%% file: main.tex
\documentclass[aos]{imsart}

\usepackage[commenters={TC}]{shortex}
\usepackage{enumitem}

\RequirePackage[numbers,sort&compress]{natbib}
\RequirePackage{cleveref}

\newcommand{\shell}{\widehat{\ell}}

\newcommand{\fstep}{f_{\mathrm{step}}}
\newcommand{\shfstep}{\shf_{\mathrm{step}}}
\newcommand{\stfstep}{\stf_{\mathrm{step}}}
\newcommand{\fdouble}{f_{\mathrm{cache}}}
\newcommand{\shfdouble}{\shf_{\mathrm{cache}}}
\newcommand{\stfdouble}{\stf_{\mathrm{cache}}}
\newcommand{\flazy}{f_{\mathrm{lazy}}}
\newcommand{\shflazy}{\shf_{\mathrm{lazy}}}
\newcommand{\stflazy}{\stf_{\mathrm{lazy}}}

\newcommand{\shwstep}{\shw_{\mathrm{step}}}
\newcommand{\stwstep}{\stw_{\mathrm{step}}}
\newcommand{\shwdouble}{\shw_{\mathrm{cache}}}
\newcommand{\stwdouble}{\stw_{\mathrm{cache}}}
\newcommand{\shwlazy}{\shw_{\mathrm{lazy}}}
\newcommand{\stwlazy}{\stw_{\mathrm{lazy}}}

\begin{document}

\begin{frontmatter}
\title{Optimal slice-adaptive tuning of hybrid slice sampling}
\runtitle{Optimal slice-adaptive tuning of hybrid slice sampling}

\begin{aug}
\author[UBC]{\fnms{Trevor}~\snm{Campbell}\ead[label=e1]{trevor@stat.ubc.ca}}
\address[UBC]{Department of Statistics, UBC\printead[presep={,\ }]{e1}}
\runauthor{T.~Campbell}
\end{aug}

\begin{abstract}
Slice sampling is a Markov chain Monte Carlo algorithm that draws its next state 
uniformly from a ``slice''---a super-level set of the target density function---at each iteration,
thereby providing automatic local adaptivity to the scale of the target. 
In practice the exact slice is not known, so general-purpose implementations use an approximate slice 
that is grown from a starting interval of length $w>0$, with a computational cost that depends on $w$.
This work presents an analysis of the average per-iteration number of target density evaluations, as a function of $w$, 
of hybrid slice sampling with various slice-finding schemes for targets with contiguous slices.
The paper uses the results of the analysis to develop automated, slice-adaptive 
tuning schemes along with suboptimality bounds and asymptotic convergence guarantees.
Simulations demonstrate that the tuning schemes reliably 
yield near-optimal slice-adaptive tuning with essentially
no dependence on the initial setting of $w$.
\end{abstract}

%

\end{frontmatter}

\input{intro}
\input{background}
\input{theory}

\input{tuning}

\input{simulations}

\input{conclusion}

%

\bibliographystyle{imsart-nameyear}
\bibliography{sources}

\appendix

\input{surrogates}

\input{proofs}

\input{pseudocode}

\end{document}

%% file: intro.tex
\section{Introduction}\label{sec:introduction}
Markov chain Monte Carlo is a commonly-used tool for 
computational Bayesian inference. 
Standard methods typically involve a fixed scale parameter that determines
the size of typical movements, e.g., the step size in random-walk Metropolis--Hastings \citep{Metropolis53, Hastings70},
Metropolis-adjusted Langevin algorithm \citep{Rossky78}, and Hamiltonian Monte Carlo \citep{Duane87,Neal11}.
In general, it is nontrivial to set the scale for even just a single well-behaved target,
let alone for distribution sequence targets (e.g., in annealed importance sampling \citep{Jarzynski97a,Jarzynski97b,Neal98,Neal01} and 
parallel tempering \citep{Swendsen86,Geyer91,Hukushima96}),
for multiscale targets where there may not be a single good choice throughout the whole state space
(e.g., Bayesian posteriors with scale priors \citep{PolsonScott12}), or in probabilistic programming libraries
that cannot rely on expert user input.
This challenge has motivated the development of \emph{locally-adaptive} samplers, which
select a scale value at each iteration based on the 
current state \citep[see, e.g.,][]{Tierney99,Mira01,Maire22,Girolami11,Nishimura16,Kleppe16,Modi24,BironLattes24,Turok24,Liu25,Livingstone21}.
Locally-adaptive samplers are typically more robust to the setting of their tuning parameters,
but also often have a larger, random number of target evaluations at each step due to the need to establish rough local scale before a move.
Slice sampling \citep{Swendsen87,Edwards88,Higdon96,Higdon98,Neal97,Neal00,Neal03} is one such locally-adaptive method
that has seen successful use in a wide variety of applications. The goal of this work is fully-automated tuning of slice sampling
in the interest of reducing its computational cost to the maximum extent possible.

The slice samplers considered in this work are those that operate in one dimension.
In particular, given a target density function $\pi$ (up to proportionality) with respect to the Lebesgue measure on $\reals^d$,
and a current state $x\in\reals^d$, the slice sampler first draws a slice variable $u\in[0,\pi(x)]$, 
then a direction $\rho\in\reals^d$, $\rho\neq 0$
and then finally a new updated state $x'\in\reals^d$ via
\[
u \dist \Unif[0, \pi(x)],\quad\rho \dist m(\d\rho; u), \quad x' \dist \Unif \{y \in \reals^d : y=x+\rho \alpha, \alpha\in\reals, \pi(y) \geq u\}.
\]
Because the new state $x'$ is drawn from the uniform distribution on the slice,
the scale of the move from $x$ to $x'$ adapts locally to the size of the slice.
With appropriate choice of the distribution $m(\d\rho; u)$, this sampler recovers slice sampling within
various multivariate schemes, e.g.,
the hit-and-run sampler \citep{Smith84,Belisle93} when $\rho \dist \Unif\lt(\sdS^{d-1}\rt)$, 
and the Gibbs sampler \cite{Gelfand90} when $\rho$ is a unit basis vector
chosen randomly or in deterministic sweeping order.

Drawing the slice level variable $u$ and direction $\rho$ at each iteration is straightforward,
but the slice itself is typically unknown, so the ideal slice sampler cannot be implemented.
A computational method to draw  from the slice is required. Two
popular methods developed by \citet{Neal97,Neal00,Neal03} begin
with an initial window of width $w>0$ around the current iterate and grow the window until the slice is (approximately) bounded. 
The first is stepping out, where the window is grown in increments of size $w$, and the second is doubling, where the window length is progressively doubled.
The choice of $w$ influences the long-run cost of the sampler. A precise characterization of this cost
and tuning methods for $w$ are the focus of this work.

The paper begins with a set of pseudocode methods that establish the precise
count of the number of target evaluations in each iteration. The pseudocode includes 
two novel variants of the doubling scheme, both of which improve upon the original method by \citet{Neal03}:
a simpler method that involves caching, and a more efficient but more complex method that involves both caching and lazy evaluation.
\cref{thm:stepcost,thm:doublecost} present an exact characterization of the cost
of both stepping out and cached doubling for targets with contiguous slices, while \cref{thm:lazycost}
establishes coarse properties of the efficient lazy doubling scheme.
These results are used to show that the optimal slice-adaptive choice of $w$ at each iteration 
is proportional to the slice width $\lambda$. In particular, the oracle optimal setting is
$w= 1.358\lambda$ for stepping out with cost $4.715$,
$w = 3.211\lambda$ for cached doubling with cost $5.901$, and
$w = 2.277\lambda$ for lazy cached doubling with cost $5.410$,
where the reported cost is the average per-iteration number of target density evaluations.

These optima establish lower bounds on the cost of slice sampling, but are typically not achievable in practice because 
the slice width $\lambda$ in general has a distribution given each slice value $u$, as opposed to being a fixed value.
The paper therefore leverages the earlier cost characterizations to develop 
one-shot and gradient-based tuning schemes for $w$ as a function of $u$ for each sampler in \cref{sec:tuning}.
The one-shot schemes, provided in \cref{eq:ostuning}, have closed-form formulae in terms of 
the mean/quantiles of the conditional distribution of the slice width $\lambda$ given the slice variable $u$. 
The gradient-based schemes, provided in \cref{eq:gtuning}, are designed have convex and smooth objectives
and hence are tractable. Each method comes with theoretical bounds on suboptimality in \cref{thm:tunequality}.
When tuned using draws from the Markov chain in practice, 
all algorithms have tuning cost that vanishes compared to the base cost of slice sampling.
Under mild additional assumptions, \cref{lem:tuningconvergence} shows that 
all schemes converge in probability to the optimum for their respective surrogate cost functions.
None of these schemes involve user input, enabling their use in an automated probabilistic programming libraries.
Simulations demonstrate that these tuning schemes in practice yield near-optimal tuning reliably with no noticeable
dependence on the initial setting of $w$.

Proofs for all results in this work are presented in the appendices.

\paragraph*{Related Work}
Markov chain Monte Carlo with auxiliary slice augmentations was first 
developed by \citet{Swendsen87}, with a flurry of work soon afterward that generalized and applied
the technique more broadly \citep{Edwards88,Higdon96,Higdon98,Damien99},
established its convergence properties \citep{Mira97,Roberts99,Mira02},
and developed the practical hybrid schemes \citep{Neal97,Neal00,Neal03} that are the focus
of the present work. Over the past two decades since those original contributions,
theoretical understanding of slice sampling has significantly 
improved (see, e.g., \citep{Rudolf13,Rudolf18,Natarovskii21,Schar25}),
and numerous generalizations of the original hybrid scheme 
have been developed, including novel hybrid slice approximations 
for multivariate targets \citep{Murray10,Thompson11,Tibbits14,Nishihara14,Karamanis21,Schar23,Schar24,Heiner24,Habeck25,Durmus26,Marco26}.
Particularly germane to the present work are those that involve adaptation, e.g., 
adaptation to the mean and covariance of multivariate targets \citep{Thompson11,Tibbits14,Marco26},
adaptation of the initial window scale using heuristic Robbins-Monro stochastic approximation \citep{Karamanis21},
and adaptation to density shape using a target approximation \citep{Heiner24}.
The present work is the first
to develop an exact characterization of the optimal slice-adaptive cost
of stepping out and doubling hybrid slice samplers, to develop tractable surrogate tuning objectives with
suboptimality guarantees, and to leverage those results to develop a 
fully-automated slice-adaptive tuning scheme. This work is orthogonal to much of the preceding 
literature on adaptive slice sampling, and thus could likely be profitably 
combined with many of those earlier developments.

%% file: background.tex
\section{Hybrid slice sampling with lazy cached doubling}\label{sec:slicesampling}
This section reviews the stepping out and doubling slice sampling algorithms by \citet{Neal03}, 
including some efficiency improvements.
For each method, detailed pseudocode in \cref{sec:pseudocode} establishes notation
and the precise count of the number of calls to $\pi(\cdot)$ for subsequent analysis. The implementation of stepping out (\cref{alg:steppingout})
and shrinkage (\cref{alg:shrinkage}) are essentially identical to the implementation by \citet{Neal03}.
For doubling and the acceptance check, there are two novel variants: one with caching (\cref{alg:doubling,alg:accept})
and one with caching and lazy evaluation (\cref{alg:lazydoubling,alg:lazyaccept})
that removes all unnecessary target density evaluations.
All other algorithms in \cref{sec:pseudocode} are related to tuning and will be discussed in \cref{sec:tuning}.

Slice sampling begins in \cref{alg:slicesampling} by drawing
$u \dist \Unif[0,\pi(x)]$ given the current state $x\in\reals^d$, and $\rho \dist m(\rho; u)$. 
While this na\"ively requires a call to $\pi(x)$, 
it can be treated as an input to \cref{alg:slicesampling} because it is obtained from the
previous iteration's shrinkage step (\cref{alg:shrinkage}).
Therefore this call does not appear in the long-run average cost.

Next, the algorithm finds the approximate slice using stepping out (\cref{alg:steppingout}),
cached doubling (\cref{alg:doubling}), or lazy cached doubling (\cref{alg:lazydoubling}) along the line $x + \rho y$, $y\in\reals$, 
starting with a window size $w$ computed as a function of $u$.
For both doubling methods, as the slice is grown the density values are stored in an array cache $\scC$ (assumed to be 1-indexed). 
This cache is used to avoid recomputing known densities during the subsequent acceptance check.
\cref{alg:lazydoubling} avoids further density computation by short-circuiting the decision to keep expanding the window.

Finally, the shrinkage algorithm (\cref{alg:shrinkage}) runs an adaptive rejection sampler
that determines whether proposals are acceptable if necessary using a call to the accept (\cref{alg:accept}) or lazy accept (\cref{alg:lazyaccept}) function. 
The (lazy) accept function traces out what would have been the sequence of evaluations by the 
doubling scheme from the proposed point, and rejects the proposal if doubling could not have created
the approximate slice. \cref{alg:accept,alg:lazyaccept} both use the cache,
and \cref{alg:lazyaccept} uses short-circuit evaluation to avoid 
computing density values that are already known or unnecessary to compute.
For the remainder of the paper, the term ``cached doubling'' refers to using only the cache (\cref{alg:doubling,alg:accept}),
and ``lazy cached doubling'' refers to using both the cache and lazy evaluation (\cref{alg:lazydoubling,alg:lazyaccept}).
In practice one should always prefer lazy cached doubling, but simple cached doubling is retained for study because it is more amenable
to analysis.

\cref{fig:acceptcomparison} displays the improvement in target density evaluations for the proposed two 
doubling schemes versus the original scheme proposed by \citet{Neal03}. The lazy caching scheme in particular saves a substantial number of evaluations
for small $w$, and around 1-2 evaluations for well-tuned $w$, which amounts to around a $15$--$25$\% overall cost reduction in that regime.

\bfig[t]
\bsubfig{0.5\textwidth}
\includegraphics[width=\textwidth]{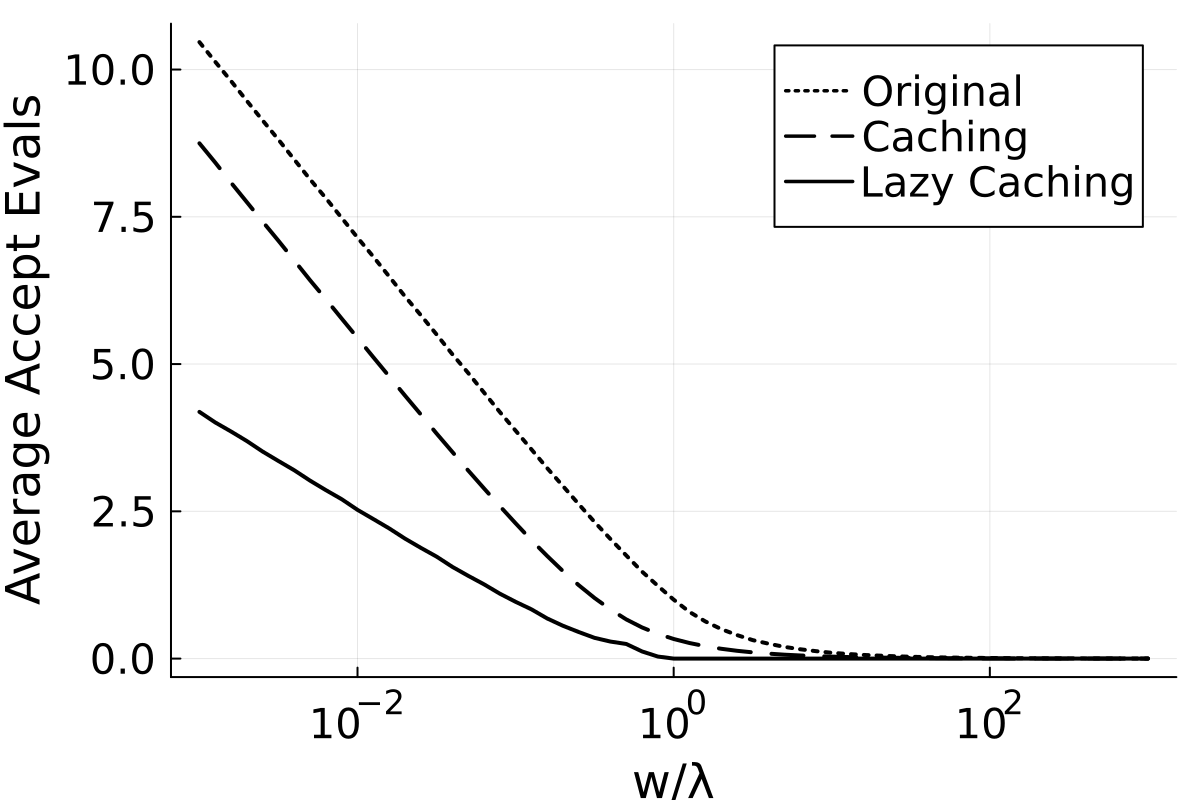}
\caption{Accept}
\esubfig
\bsubfig{0.5\textwidth}
\includegraphics[width=\textwidth]{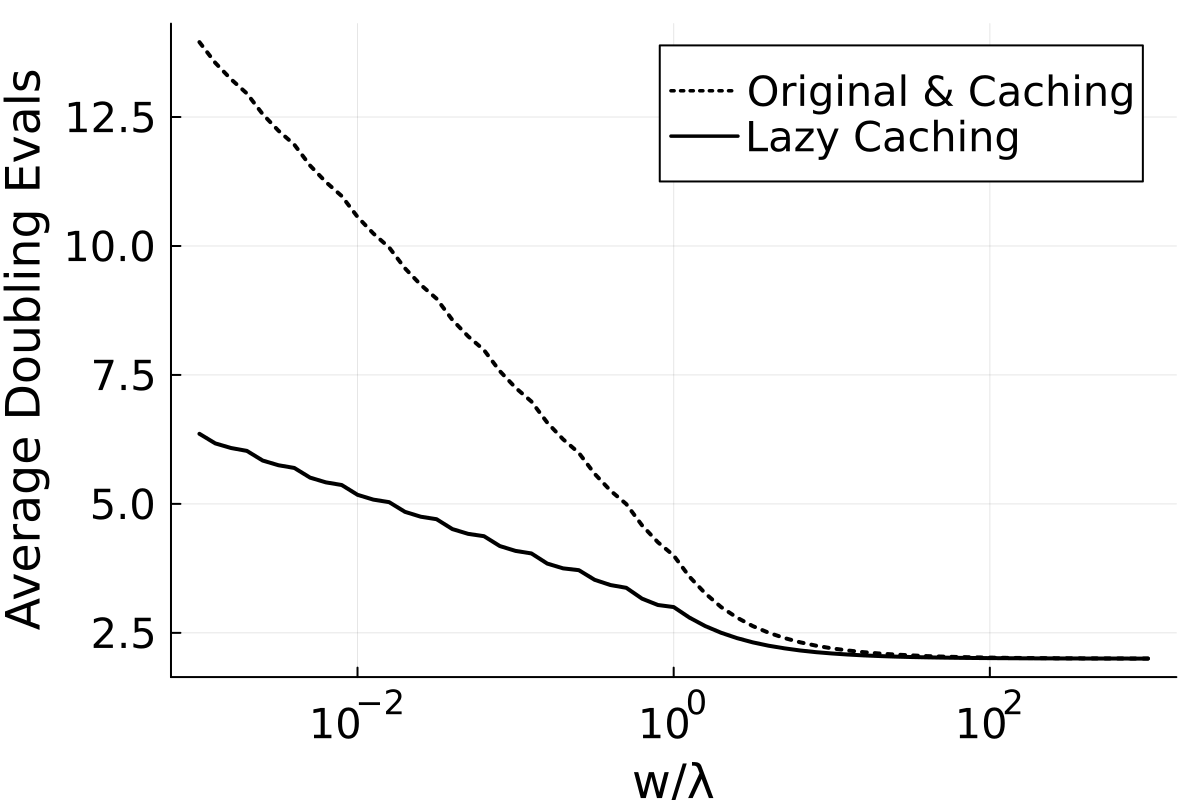}
\caption{Doubling}
\esubfig
\caption{The average number of target evaluations in the accept and doubling phase of each slice sampling step, as a function of the ratio of the window width $w$ and slice width $\lambda$, 
for the original scheme by \citet{Neal03}, the basic caching scheme,
and the lazy caching scheme. 
}\label{fig:acceptcomparison}
\efig

%% file: theory.tex
\section{Analysis}\label{sec:analysis}
This section presents an analysis of the long-run cost of both stepping-out and doubling slice sampling.
Assuming that the cost of each iteration is dominated by target density evaluation, the per-iteration cost can be
characterized by the number of calls to $\pi(\cdot)$.

The target $\pi$ is assumed to be unimodal and univariate for simplicity.
There are two notes to make about this choice.
First, the analysis suffices for broad class of scenarios:
the main results in \cref{thm:stepcost,thm:doublecost,thm:lazycost} as stated apply, without modification, to any multivariate target
such that the one-dimensional slice drawn in each iteration is guaranteed to be a contiguous interval.
For example, slice sampling within hit-and-run \citep{Smith84,Belisle93} and Gibbs \citep{Gelfand90} samplers on unimodal multivariate targets both adhere
to the below theory, with the only difference being that the slice width at each level is random as opposed to fixed.
In practice, the tuning recommendations in this work are also fruitful beyond unimodality (see \cref{sec:simulations}).
Second, unimodality technically enables further reduction of cost
by skipping the acceptability check in \cref{alg:shrinkage} \citep[Section 4.4]{Neal03}.
However, practical implementations of slice sampling cannot assume unimodality and cannot skip this check;
the analysis below corresponds to such practical implementations to ensure useful tuning advice in general.
The particular technical assumption that will be made for all theoretical results in this section is as follows.
\bassum\label{assum:unimodal}
The target $\pi$ is a probability distribution on $\reals$ with density $\pi(x)$ with respect to the Lebesgue measure
that satisfies the following:
for all $u> 0$ such that $\{x : \pi(x)\geq u\} \neq \emptyset$,
there exist $\ell,r\in\reals$, $\ell\leq r$ such that 
\[
	\{x : \pi(x)\geq u\} = [\ell, r],
\]
and for all $z$ such that $0 < \pi(z) < \infty$,
\[
	\P\lt(\lt| \{x : \pi(x) \geq u\}\rt| \leq 1\rt)  = 0, \qquad u\dist \Unif[0,\pi(z)].
\]
Furthermore, the Markov chain initial state $x_0$ satisfies $0 < \pi(x_0) < \infty$.
\eassum

\cref{assum:unimodal} stipulates weak conditions under which the iterates of stepping or doubling can be computed in finite time
and have the same distribution as those of ideal slice sampling.
The first condition is that the slices are contiguous intervals, and the second is roughly that there are no removable discontinuities
in the density function that can yield pathological initializations with slice width 0 where the shrinkage algorithm never terminates.
Because the iterates are equal in distribution to ideal slice sampling, 
the optimality of doubling and stepping samplers in this regime is determined entirely by the cost per iteration,
and does not depend on the mixing behaviour of the chain.
Furthermore, to analyze the long-run cost of stepping and doubling, it suffices to consider the expected
cost of one step starting at a state $x\dist \pi$, since the chain obeys the Markov chain law of large numbers.
Both of these results are presented precisely in \cref{lem:exactandlln}.
The law of large numbers in particular is a straightforward combination of \citet[Theorem 17.0.1]{MeynTweedie} and \citet[Corollary 1]{Tierney94},
and was originally obtained by \citet{Mira97,Mira02}, although it was not stated explicitly as such in that work.
\blem\label{lem:exactandlln}
The iterates of hybrid slice sampling with either doubling or stepping
are equal in distribution to those produced by ideal slice sampling, and require only finitely many evaluations of $\pi(\cdot)$ almost surely.
Furthermore,
for any function $h(x)$ such that $\pi(|h|) < \infty$,
\[
\frac{1}{T}\sum_{t=1}^T h(x_t) \convas \E h(X) \quad X\dist \pi.
\]
\elem

The random cost $C$ of each iteration of slice sampling can be decomposed into 3 terms: one
for finding the approximate slice, one for making proposals and shrinking the approximate slice, and
one for checking acceptability of proposals:
$C = C_{\text{slice}} + C_{\text{shrink}} + C_{\text{accept}}$.
The following sections characterize the distribution and/or expectation of each as needed.
Throughout, let $\reals_+$ denote the positive reals,
$x$ denote the current state state, $u$ denote the slice variable,
$w$ the initial window width,
$\shell, \shr$ denote the approximate slice boundaries,
$\ell, r$ denote the exact slice boundaries, and
$\lambda=r-\ell$ denote the exact slice width.
Note that in the setting of \cref{assum:unimodal}, $w$, $\ell$, $r$, and $\lambda$ are all deterministic functions of $u$,
and $\shell < \ell < r < \shr$ almost surely.

\input{shrinkage}

\input{steppingout}

\input{doubling}

\input{comparison}

%% file: shrinkage.tex
\subsection{Shrinkage}\label{sec:shrinkage}
The shrinkage algorithm takes as input a state $x$, slice variable $u$,
approximate slice from $\shell$ to $\shr$, and a stream of \iid uniform random variables $V_n \distiid \Unif[0,1]$.
Define the \emph{left exceedance} $\lambda_\ell = \ell - \shell$ and \emph{right exceedance} $\lambda_r = \shr - r$,
and recall the exact slice width is $\lambda = r-\ell$.
The number $N$ of evaluations of $\pi(\cdot)$ invoked by shrinkage (not including the accept call)  
is a function that depends only on $(\lambda_\ell, \lambda, \lambda_r)$ 
and the stream of uniform variables $(V_n)_{n=1}^\infty$.
$N$ is most naturally formulated
recursively: the algorithm proposes a point $y=(1-V_1)\shell + V_1 \shr$ and invokes one evaluation of $\pi(\cdot)$.
If the proposed point is to the left of $\ell$ (respectively, to the right of $r$)
the shrinkage algorithm is called again with $\shell$ (respectively, $\shr$) replaced by $y$.
Therefore for $W\dist\Unif[0,1]$ independent of $(V_n)_{n=1}^\infty$,
\[
N\lt(\lambda, \lambda_\ell, \lambda_r,(V_n)_{n=1}^\infty\rt)
&\eqd 1+ \lt\{\begin{array}{ll}
N\lt(\lambda, W\lambda_\ell, \lambda_r,(V_n)_{n=2}^\infty\rt) \,\, & V_1 \leq \frac{\lambda_\ell}{\lambda_\ell+\lambda_r+\lambda}\\
N\lt(\lambda, \lambda_\ell, W\lambda_r, (V_n)_{n=2}^\infty\rt) \,\, & 1-V_1 <\frac{\lambda_r}{\lambda_\ell+\lambda_r+\lambda}
\end{array}\rt. .\label{eq:recursiveN}
\] 
Using the above formula and the fact that $(V_n)_{n=1}^\infty \eqd (V_n)_{n=2}^\infty$,
one can obtain a characterization of the complementary CDF of $N$ in terms of the solution of a partial differential equation. 
The true slice width $\lambda$ is suppressed in the function arguments below as it is held constant throughout the shrinkage procedure.
\blem\label{lem:GPDEeqn}
Let $F_n(\lambda_\ell,\lambda_r)$ denote the probability that $N>n$ given $\lambda_\ell,\lambda_r,\lambda$.
Then
\[
\forall n\in\nats\cup\{0\}, \,\, F_n(\lambda_\ell,\lambda_r) = \frac{1}{n!}\lt.\pder[n]{G}{z}(\lambda_\ell,\lambda_r,z)\rt|_{z=0}\quad\text{and}\quad
\E[N|\lambda_\ell,\lambda_r,\lambda] = G(\lambda_\ell,\lambda_r,1),
\]
where $G(x,y,z)$ is the unique solution to the following system
on $x,y\geq 0$, $z\in[-1,1]$:
\[
0 &= (1-z)\lt(\pder{G}{x} + \pder{G}{y}\rt) + (x+y+\lambda)\hes{G}{x}{y}\label{eq:GPDE}\\
G(x,y,z) &= G(y,x,z), \quad 
G(x,0,z) = \lt\{\begin{array}{ll}
\lt(1-z\lt(\frac{x}{\lambda}+1\rt)^{z-1}\rt)/(1-z)& \,\,z\neq 1\\
1+\log\lt(\frac{x}{\lambda}+1\rt) & \,\,z = 1
\end{array}\rt. . 
\]
\elem
The variable-coefficient partial differential equation \cref{eq:GPDE}
can be solved, and yields a remarkably simple closed-form expression 
for the expected shrinkage cost in \cref{prop:shrinkcost}.
\bprop\label{prop:shrinkcost}
$\E C_{\text{shrink}} = 1 + 2\E\lt[\log\lt(\frac{\shr - r}{\lambda} + 1\rt)\rt]$.
\eprop
The expectation in \cref{prop:shrinkcost} is over $x\dist \pi$ and $u\dist\Unif[0,\pi(x)]$---which 
then determine the slice width $\lambda$ and right edge $r$---and all randomness in the approximate slice bounding algorithm,
which determines the distribution of $\shr-r$ conditioned on $u$. 
To use this result, then, one requires a characterization of the distribution of the right exceedance $\shr-r$
 given the slice variable $u$.
The below two sections on stepping out and doubling provide characterizations of the required
conditional distribution of $\shr-r$.

%% file: steppingout.tex
\subsection{Stepping Out}\label{sec:steppingout}
Given the current state $x, u$, 
stepping out begins by drawing $V\dist\Unif[0,1]$ and evaluating $\pi(\cdot)$
at the initial window boundaries $x-Vw$ and $x+(1-V)w$, and then proceeds
by expanding rightwards and leftwards by units of $w$ until both $\ell > \shell$ and $\shr > r$.
\cref{prop:stepout} characterizes the conditional distributions of the number of stepping out iterations
and the right exceedance given the slice variable $u$.
\blem\label{prop:stepout}
Conditioned on $u$, the number $N$ of stepping out iterations has distribution 
\[
N \eqd \lt\lfloor \frac{\lambda}{w}\rt\rfloor + B, \qquad B\dist\Bern\lt(\frac{\lambda}{w} - \lt\lfloor \frac{\lambda}{w}\rt\rfloor\rt),
\]
and conditioned on $u$, the right exceedance has distribution
\[
\shr - r \eqd w Z, \quad Z \dist \Unif[0,1].
\]
\elem
The slice-finding cost is $C_{\text{slice}} = 2+N$,
the cost to accept a proposal is identically $C_{\text{accept}}=0$,
and the expected shrinkage cost $\E C_{\text{shrink}}$ can be obtained by combining the right exceedance distribution from \cref{prop:stepout}
 with the result of \cref{prop:shrinkcost}. \cref{thm:stepcost}
uses these three facts to obtain 
the overall expected per-iteration cost of slice sampling with stepping
out.
Define the function 
\[
\fstep : \reals_+ \to \reals_+, \qquad
\fstep(x) = 1 + x + 2(x+1)\log\lt(1+1/x\rt). \label{eq:fstep}
\]
\vspace{-20pt}
\bthm\label{thm:stepcost}
Slice sampling with stepping out has expected cost per iteration
\[
\E C
&= \E\lt[\fstep\lt(\frac{\lambda}{w}\rt)\rt].\label{eq:stepcost}
\]
The cost is finite if and only if  both
$\E\lt[\frac{\lambda}{w}\rt] < \infty$ and $\E\lt[-\log\frac{\lambda}{w}\rt] < \infty$.
The optimal slice-adaptive window size and corresponding cost is
\[
w^\star = \alpha \lambda, \quad \alpha = -1 - W_{-1}\lt(-\exp(-3/2)\rt) \approx 1.358,
\quad \E C \approx 4.715,
\]
 where $W_{-1}$ is the lower branch of the Lambert $W$ function.
\ethm

\bfig[t]
\bsubfig{0.5\textwidth}
\includegraphics[width=\textwidth]{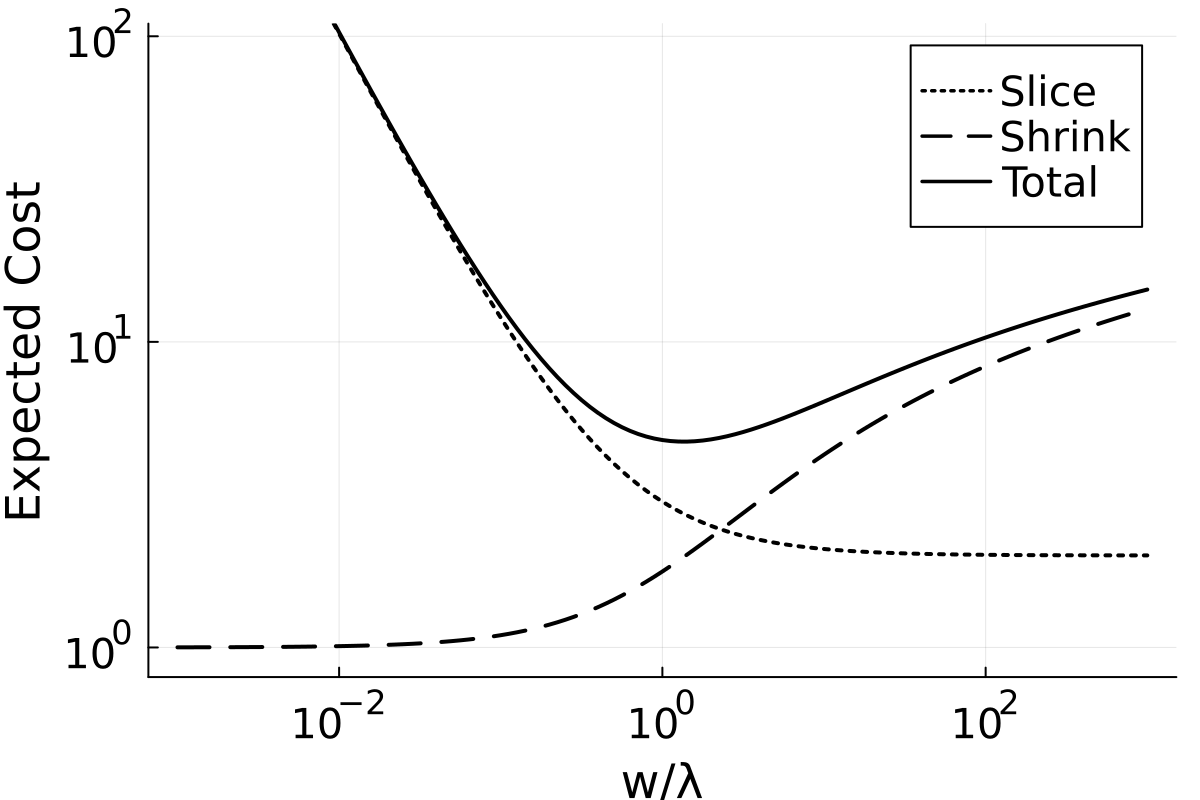}
\caption{Stepping}\label{fig:costbreakdown_stepping}
\esubfig
\bsubfig{0.5\textwidth}
\includegraphics[width=\textwidth]{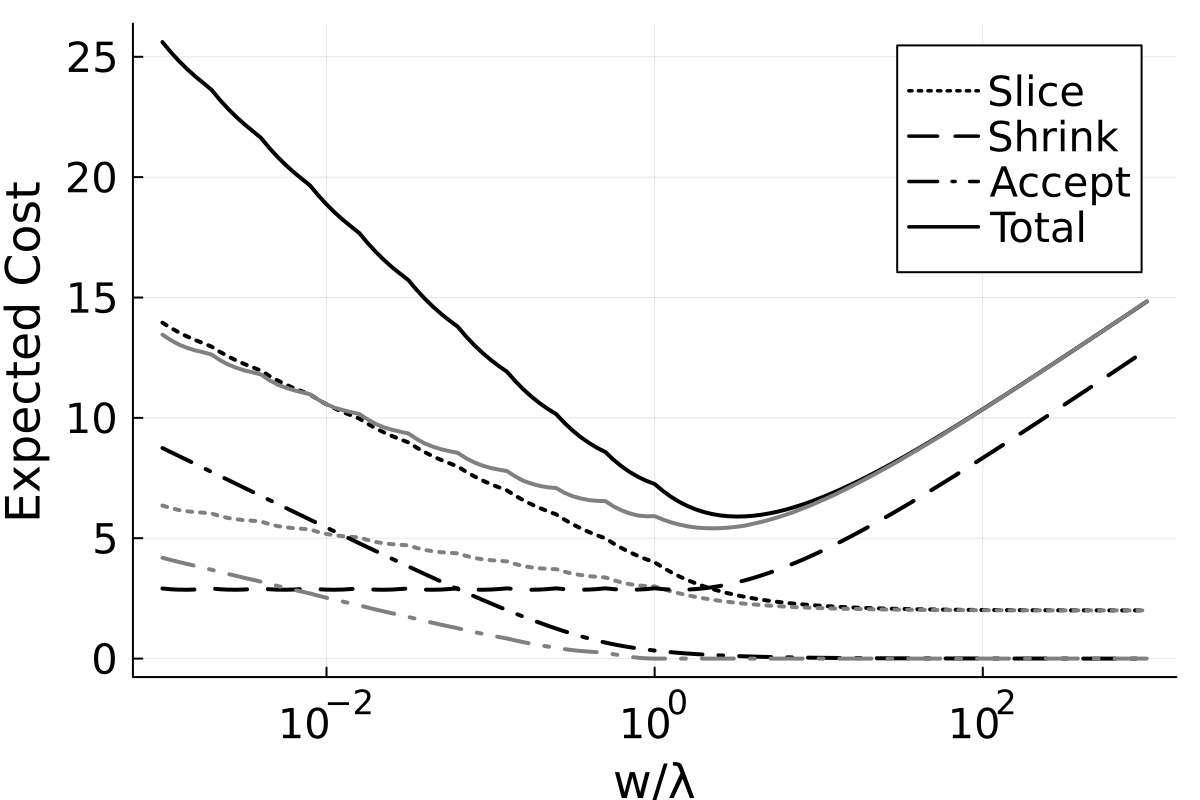}
\caption{Doubling}\label{fig:costbreakdown_doubling}
\esubfig
\caption{Cost in terms of the average number of target evaluations for one iteration of each slice sampling algorithm as a function of $w/\lambda$,
broken down by algorithmic step (slice finding, shrinkage, and accept). 
\cref{fig:costbreakdown_stepping} shows the costs for stepping, and \cref{fig:costbreakdown_doubling} shows the costs for doubling,
where black depicts cached doubling and grey depicts lazy cached doubling. For shrinkage,
only black is shown because the cost is the same for both schemes.
}\label{fig:costbreakdown}
\efig

%% file: doubling.tex
\subsection{Doubling}\label{sec:doubling}
Given the current state $x, u$, 
doubling begins by drawing $V\dist\Unif[0,1]$ and setting the initial window boundaries $x-Vw$ and $x+(1-V)w$,
and then proceeds by expanding rightwards or leftwards, each with probability $1/2$, in doubling rounds
until both $\ell > \shell$ and $\shr > r$.
\cref{prop:doubling} characterizes the joint distribution
of the number of doubling steps $N$ and right exceedance $\shr-r$ of the approximate slice conditioned on $u$.
These distributions apply to both the cached and lazy cached doubling schemes.
Define the functions $n, c : \reals_+ \to \reals$ and random values $n_0,c_0\in\reals$
\[
n(x) = \lt\lceil \log_2(x)\rt\rceil \vee 0, \quad c(x) = \frac{x}{2^{n(x)}}, \quad
n_0 = n(\lambda/w), \quad c_0 = c(\lambda/w), \label{eq:n0c0}
\]
and let $\Geom$ denote the geometric distribution with support $k\in\{1,2,\dots\}$.
\blem\label{prop:doubling}
Conditioned on $u$, the number $N$ of doubling iterations has distribution
\[
 N \eqd n_0 + BG, \qquad  B \dist \Bern\lt(c_0\rt), \quad G\dist \Geom(1/2),
\]
and conditioned on $N, u$, the right exceedance has distribution
\[
\shr-r &\dist \lt\{\begin{array}{ll}
\Unif[0, 2^Nw-\lambda] & \quad N=n_0\\
\Unif[2^{N-1}w-\lambda, 2^{N-1}w] & \quad N>n_0
\end{array}\rt. .
\]
\elem
For the simple cached doubling scheme, the slice cost is $C_{\text{slice}} = 2+N$,
the shrinkage cost $\E C_{\text{shrink}}$  is derived by combining the right exceedance distribution in \cref{prop:doubling}
and the result from \cref{prop:shrinkcost},
and \cref{lem:cacheaccept} characterizes the expected number $\E C_{\text{accept}}$ of 
evaluations in the cached acceptance check. 
The sum of these results provides the
total expected cost per iteration of slice sampling with cached doubling.
Define the function 
\[
\fdouble : \reals_+\to\reals_+, \qquad \fdouble(x) = f(n(x), c(x)),\label{eq:fdouble}
\]
where $f : \nats\cup\{0\} \times (0, 1] \to \reals$ is given by
\[
\hspace{.5cm}f(n,c) &= 1+n+((7/3)+2\log 2)c-2(1+c)\log c+ \sum_{n=0}^\infty \lt(1+c2^{-n}\rt)\log\lt(1+c2^{-n}\rt)\label{eq:fnc}\\
& + 0\vee\!\lt(\frac{c}{3}\!+\!n\!-\!1\!-\!\frac{1\!-\!2^{1-n}}{c}\!+\!\frac{1\!-\!4^{1-n}}{9c^2}\rt).
\]
\bthm\label{thm:doublecost}
Slice sampling with cached doubling has expected cost per iteration
\[
\E C = \E\lt[\fdouble\lt(\frac{\lambda}{w}\rt)\rt]. \label{eq:doublecost}
\]
The cost is finite if and only if $\E\lt|\log\frac{\lambda}{w}\rt| < \infty$.
The optimal slice-adaptive window size and corresponding cost is
\[
w^\star = \alpha \lambda, \quad \alpha = \lt(\argmin_{x\in(0,1)} f_0(0,x)+f_1(0,x)\rt)^{-1} \approx 3.211,
\quad \E C \approx 5.901.
\]
\ethm
The analysis of the cost of lazy cached doubling is significantly 
more challenging. While the cost of shrinkage is the same as for simple cached doubling
and is available in closed form, the cost of doubling and acceptance depend on what is known at each iteration
as each endpoint of the approximate slice grows/shrinks.
Rather than developing an exact cost formula,
it will suffice to understand coarser continuity and asymptotic properties,
given by \cref{lem:lipschitz,lem:lazyasymptoticslargew,lem:lazyasymptoticssmallw}.
\cref{lem:lipschitz} shows that the costs of slice-finding and acceptance
are locally Lipschitz continuous as a function of $\lambda/w$,
and \cref{lem:lazyasymptoticslargew,lem:lazyasymptoticssmallw} reveal their asymptotic
behaviour for large and small $\lambda/w$.
\cref{thm:lazycost} combines these results with the exact cost of shrinkage
to provide continuity and asymptotic results pertaining to the overall cost of lazy cached doubling,
as well as a computational guarantee about the optimum.

\bthm\label{thm:lazycost}
There exists a function $\flazy :\reals_+\to\reals_+$ such that slice sampling with lazy cached doubling has expected cost per iteration
\[
\E C = \E\lt[ \flazy\lt(\frac{\lambda}{w}\rt)\rt],\label{eq:lazycost}
\]
where the cost is finite if and only if $\E\lt|\log\frac{\lambda}{w}\rt| < \infty$.
The function $\flazy$ is locally Lipschitz,
\[
\forall |x\!-\!x'|\leq 1, \,\, |\flazy(x)\!-\!\flazy(x')| \!\leq\! 4\lt|x\!-\!x'\rt|\lt(5\!+\!\max_{y\in\{x,x'\}} n(y) \!+\! 2^{1-n(y)}(2\!+\!1/c(y))\rt),
\]
has minimum on the interval $[2^{-4}, 2^6]$,
and has asymptotic limiting behaviour
\[
\lim_{x\to 0} \frac{\flazy(x)}{-2\log x} = \lim_{x\to\infty} \frac{\flazy(x)}{(5/6)\log_2 x} = 1.
\]
\ethm
\cref{thm:lazycost} states that the cost can be minimized by setting $w\propto\lambda$
with a proportionality constant that can be found to any desired precision 
by evaluating $\flazy$ on a sufficiently fine grid on the interval $[0.0625,64]$ 
determined by the local Lipschitz constant from \cref{thm:lazycost}.
Numerical simulation yields the optimal slice-adaptive window size and corresponding cost 
for slice sampling with lazy cached doubling,
\[
w^\star = \alpha \lambda, \quad \alpha \approx 2.277,
\quad \E C \approx 5.410.
\]

%% file: comparison.tex
\subsection{Summary and comparison of hybrid methods}\label{sec:comparison}
\cref{fig:costbreakdown} displays a breakdown of the expected per-iteration cost of
slice sampling with stepping out, cached doubling, and lazy cached doubling conditioned on $u$.
Due to the results of \cref{thm:stepcost,thm:doublecost,thm:lazycost}, these costs can be compared
 in a problem-independent manner as a function of $w/\lambda$.
\cref{fig:costcomparison} displays a comparison of the total expected per-iteration cost of
slice sampling with stepping out, cached doubling, and lazy cached doubling. It demonstrates that stepping out has a $\approx 15$\% 
advantage when all samplers are individually optimally tuned, that all samplers have nearly the same cost when $w/\lambda$ is too large,
and that the doubling methods are significantly less expensive when $w/\lambda$ is too small.
In the small-$w/\lambda$ asymptotic regime, the cost of each method is dominated by slice-finding and acceptance,
where stepping has cost $\approx\lambda/w$,
cached doubling has cost $\approx 2\log_2(\lambda/w)$, and lazy cached doubling has cost $\approx (5/6)\log_2(\lambda/w)$.
In the large-$w/\lambda$ asymptotic regime, all methods have asymptotic cost $\approx 2\log(w/\lambda)$, dominated by the cost of shrinkage.
\cref{fig:costcomparison_ratio} shows that lazy cached doubling is at most $\approx 25$\% more expensive than stepping out in the worst case, 
while stepping out can be arbitrarily worse than either doubling method in a relative sense.
Lazy cached doubling is therefore recommended in general, and especially for scenarios involving automated tuning 
where its robustness prevents severe performance degradation due to mistuning.

\bfig[t]
\bsubfig{0.5\textwidth}
\includegraphics[width=\textwidth]{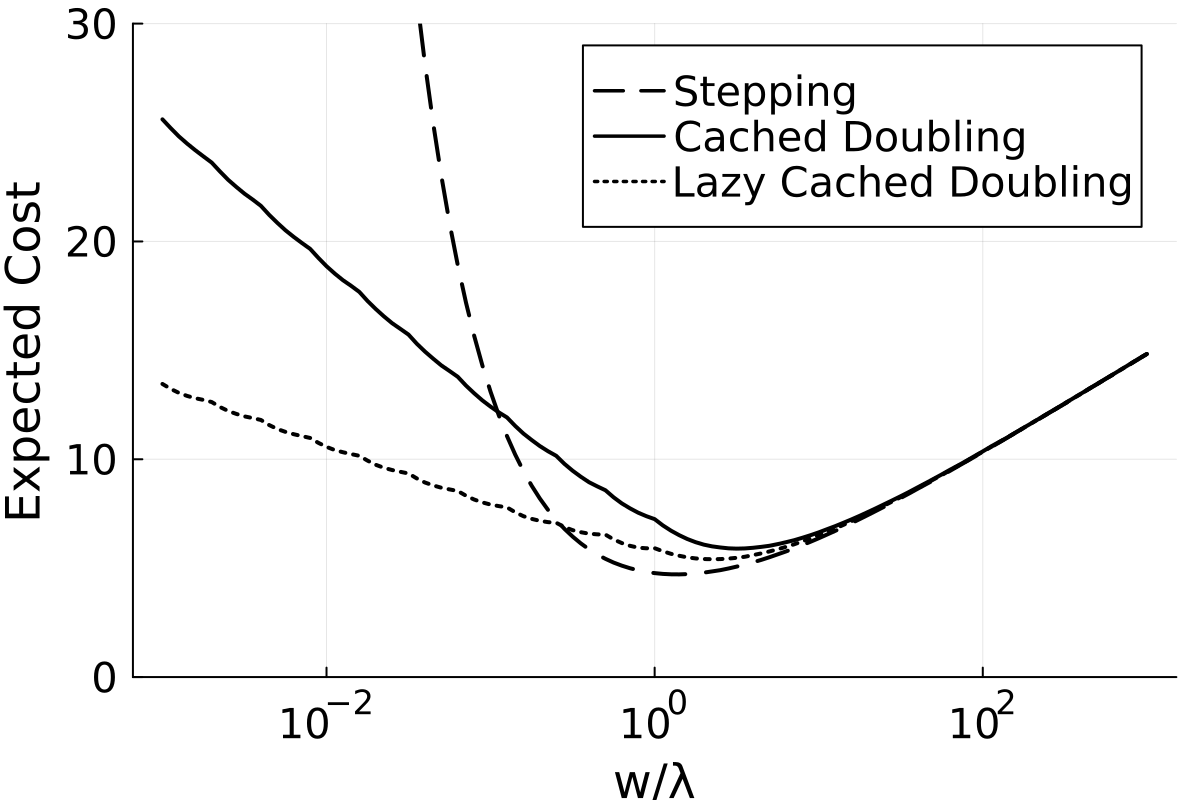}
\caption{Individual costs}\label{fig:costcomparison_separate}
\esubfig
\bsubfig{0.5\textwidth}
\includegraphics[width=\textwidth]{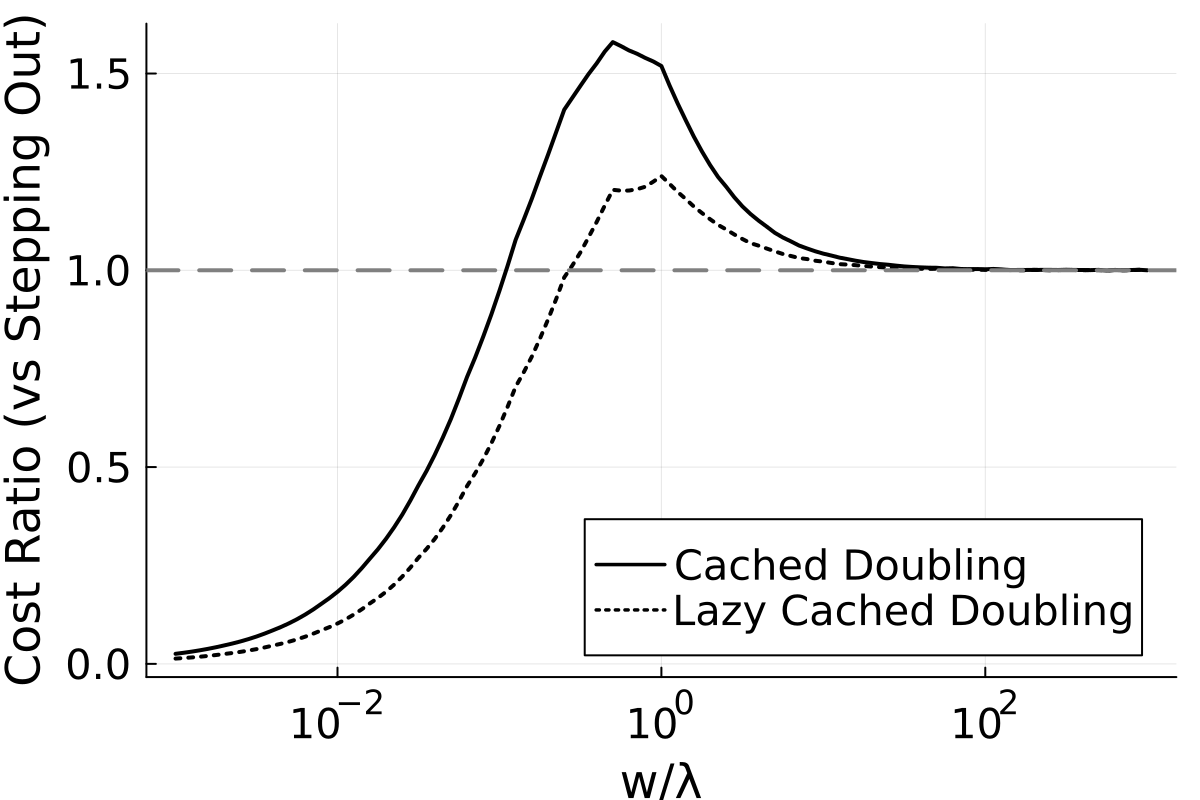}
\caption{Cost ratio}\label{fig:costcomparison_ratio}
\esubfig
\caption{
Comparison of the average cost per iteration of stepping out, 
cached doubling, and lazy cached doubling as a function of $w/\lambda$.
\cref{fig:costcomparison_separate} shows the costs plotted together, 
while \cref{fig:costcomparison_ratio} displays the ratios compared to stepping out.
}\label{fig:costcomparison}
\efig

%% file: tuning.tex
\section{Slice-adaptive tuning}\label{sec:tuning}
\cref{thm:stepcost,thm:doublecost,thm:lazycost} state that the lowest possible
per-iteration cost for slice sampling is obtained by setting
$w\propto \lambda$, with a constant that depends on whether one is using
stepping out, cached doubling, or lazy cached doubling. While that serves as an
idealized lower bound on the cost of slice sampling, those recommendations are
not implementable in practice because $\lambda$ will generally have a distribution 
conditioned on $u$ rather than being a
fixed value, e.g., for multivariate targets with the hit-and-run or
Gibbs samplers.
That being said, the expected cost functions 
in \cref{thm:stepcost,thm:doublecost,thm:lazycost} are still valid for any situation
in which the slice is guaranteed to be contiguous and the Markov chain law
of large numbers holds. In these situations, the optimal slice-adaptive tuning scheme
sets the initial width $w$ to be a function $w(u)$ of the slice variable $u$ 
given by the minimizer of the expectations in \cref{eq:stepcost,eq:doublecost,eq:lazycost} 
conditioned on $u$,
\[
w^\star_s(u) = \argmin_{w>0} \E\lt[f_{s}\lt(\frac{\lambda}{w}\rt) | u\rt], \quad s\in\{\text{step},\text{cache},\text{lazy}\}. \label{eq:optimalw}
\]
However, the minimization problem in \cref{eq:optimalw} is not tractable as it is nonconvex and nonsmooth in general
(and moreover $\flazy$ is not known in closed form).
Instead, this section uses tractable surrogate approximations of each of $\fstep$, $\fdouble$, and $\flazy$
to develop two tuning schemes for each slice sampling algorithm.
The first is a set of ``one-shot'' tuning schemes 
based only on the properties of the distribution of $\lambda$ given $u$, and the second is 
a gradient-descent-based scheme, where the optimization is initialized
at the former scheme's output. Both methods come with bounds on suboptimality
compared with the minimizer of expected cost conditioned on $u$.

The surrogate cost functions $\shfstep, \shfdouble, \shflazy, \stfstep, \stfdouble,\stflazy$ 
used to develop these tuning schemes are displayed in \cref{fig:approximation}.
The functions $\shfstep,\shfdouble,\shflazy$ are designed such that the minimum
of $\E[\shf_{(\cdot)}(\lambda/w) | u]$ is available in closed form in terms of properties of the distribution of $\lambda$
given $u$.
The functions $\stfstep, \stfdouble, \stflazy$ are designed to be
convex and locally smooth (\cref{defn:locallysmooth}) to enable gradient optimization.
The precise formulae for these surrogates and related approximation error guarantees are deferred to \cref{sec:surrogates}.

\subsection{Tuning with known slice width distribution}\label{sec:tuningsurrogates}
For each $u > 0$ with nonempty slice, let 
$q_p(u)$ be the $p$-quantile of $\lambda$ conditional on $u$.
Define the one-shot tuning schemes
\[
\shwstep(u) &= \frac{6}{5}\E[\lambda | u], & 
\label{eq:ostuning}\shwdouble(u) &= \frac{10}{3}q_{\frac{1}{1+\log 2}}(u), &
\shwlazy(u) &= 3q_{\frac{5}{12\log 2+5}}(u), 
\]
and the gradient-optimization tuning scheme
\[
\label{eq:gtuning} \stw_{s}(u) &= \exp\lt(\argmin_{x\in\reals} \E\lt[\stf_{s}\lt(\lambda \exp(-x)\rt) | u\rt]\rt), \quad s\in\{\text{step},\text{cache},\text{lazy}\}.
\]
\cref{thm:tunequality} provides suboptimality guarantees for these tuning schemes.
For stepping out, $\shwstep$ is optimal up to a multiplicative factor of $3$, while $\stwstep$ finds the slice-adaptive optimal tuning.
For both doubling methods, $\shwdouble$,$\shwlazy$ incur at most $4$ extra evaluations,
while $\stwdouble,\stwlazy$ incur at most $0.7$.
\cref{thm:tunequality} also asserts that the minimization problems in \cref{eq:gtuning}
are all tractable in the sense that they can be solved to any desired precision by 
gradient descent with backtracking line search.
\bdefn
A point $x$ is $(\epsilon,\delta)$-optimal for function $h\geq 0$ if
$h(x) \leq \epsilon \inf_x h(x) + \delta$.
\edefn
\bthm\label{thm:tunequality}
The following results hold:
\bitem
\item $\shwstep$ is $(3,0)$-optimal and $\stwstep$ is $(1,0)$-optimal for the cost of stepping out.
\item $\shwdouble$ is $(1,4)$-optimal and $\stwdouble$ is $(1,0.7)$-optimal for the cost of cached doubling.
\item $\shwlazy$ is $(1,4)$-optimal and $\stwlazy$ is $(1,0.7)$-optimal for the cost of lazy cached doubling.
\eitem
Furthermore, the optimization problems in $\stwstep,\stwdouble,\stwlazy$ are all tractable.
\ethm

\bfig[t]
\bsubfig{0.33\textwidth}
\includegraphics[width=\textwidth]{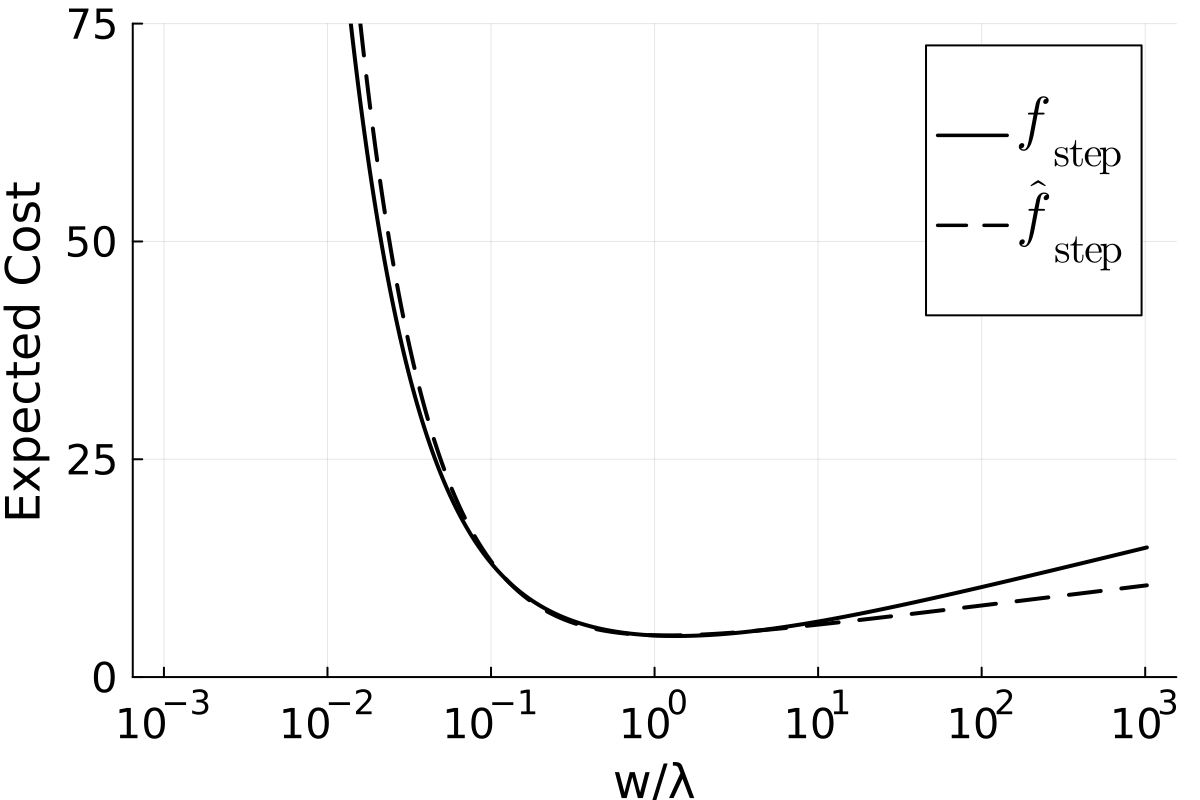}
\caption{Stepping}\label{fig:approx_stepping}
\esubfig
\bsubfig{0.33\textwidth}
\includegraphics[width=\textwidth]{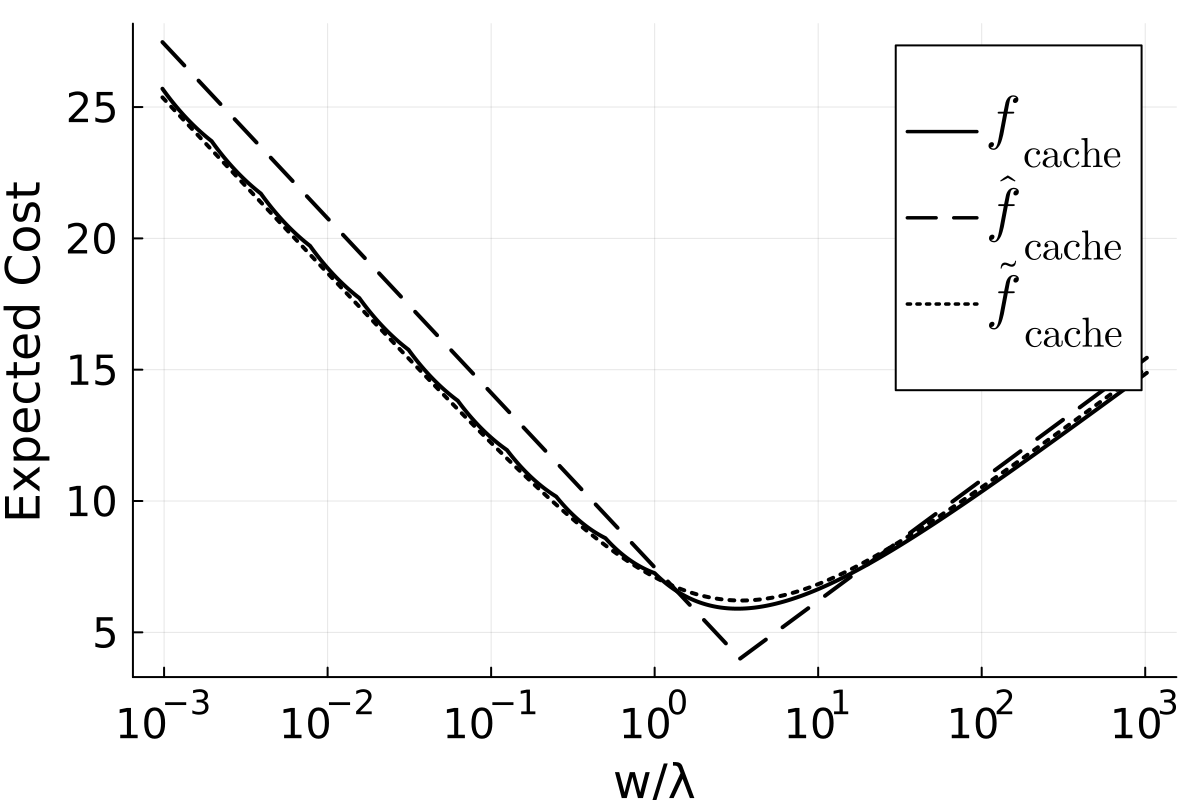}
\caption{Cached Doubling}\label{fig:approx_doubling}
\esubfig
\bsubfig{0.33\textwidth}
\includegraphics[width=\textwidth]{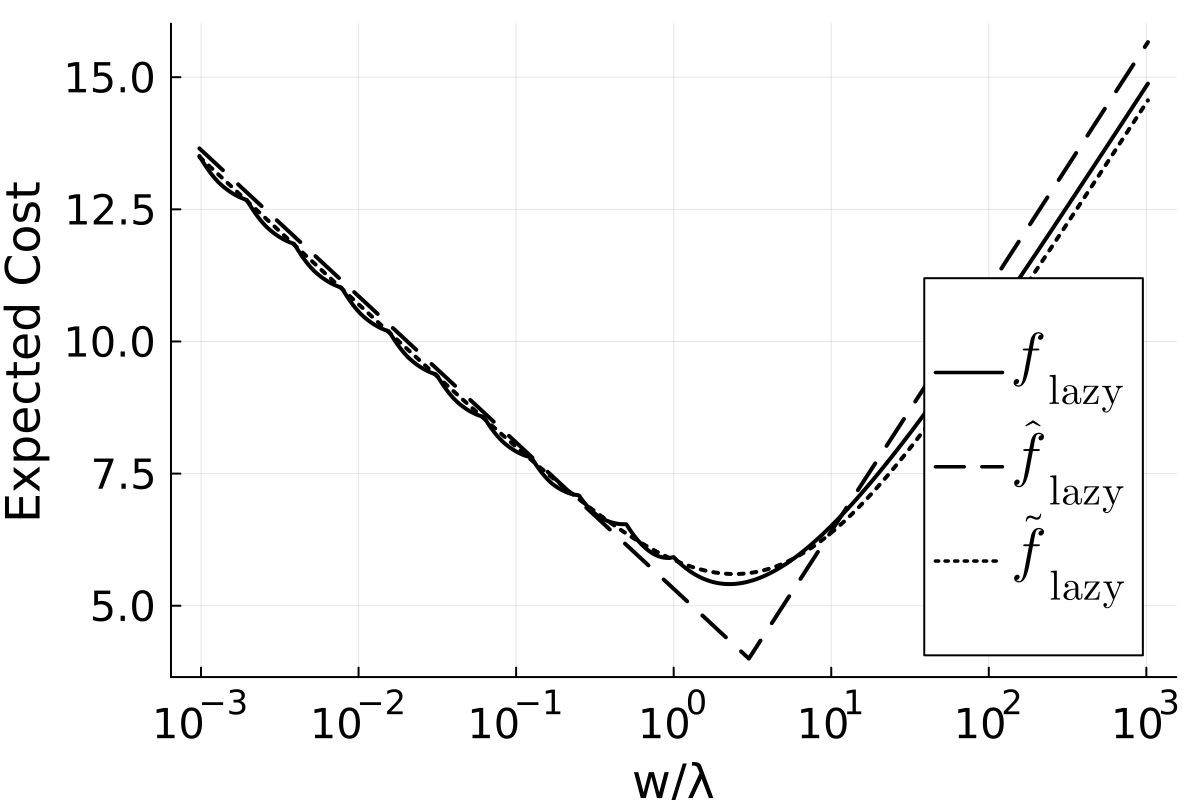}
\caption{Lazy Cached Doubling}\label{fig:approx_lazy}
\esubfig
\caption{Surrogate approximations of $\fstep$, $\fdouble$, and $\flazy$ used to develop the tuning algorithms.
See \cref{sec:surrogates} for specific formulae and approximation
error results (\cref{lem:fapprox}). Note that $\stfstep=\fstep$, so it is not shown separately.}\label{fig:approximation}
\efig

\subsection{Tuning with MCMC draws}
\cref{eq:ostuning,eq:gtuning} provide tractable near-optimal tuning of slice sampling, but require
knowing the distribution of $\lambda$ conditioned on $u$.
This section develops an algorithm that tunes using estimated conditional distributions
using draws produced by the Markov chain.
Key goals are convergence of the method to the optimal tuning (for the respective surrogate cost),
and to ensure that the additional computational cost of tuning and evaluating the initial width function $w(u)$ 
is insignificant compared to the cost of sampling with a fixed initial width.

The algorithm proceeds as follows. Suppose at some iteration $2t$, 
there is a record of the $t$ previous draws $(\lambda_j, u_j)_{j=t+1}^{2t}$.
For $\eta\in(0,1)$, $\beta\in(0.5,1)$ (in practice, $\eta=0.95$, $\beta=0.51$) set
\[
\tau = \lceil t^\eta \rceil, \qquad k = \lfloor \tau^\beta \rfloor, \qquad n = \lfloor \tau/k\rfloor. \label{eq:taukn}
\]
Then take an evenly-spaced subsequence of $(\lambda_j, u_j)$ of length $\tau$ and
sort them in order of increasing $u$, resulting in $(\tilde\lambda_j, \stu_j)_{j=1}^\tau$. 
This thinning ensures that the order statistics 
are asymptotically indistinguishable from those produced by \iid draws under mild conditions (\cref{lem:thinning}).
Then for each block of $k$ draws with slice values between $\stu_{jk}$ and $\stu_{(j+1)k}$, $j\in\nats$,
approximate the conditional distribution of $\lambda$ given $u$ using the empirical distribution of draws 
within that block. More precisely, approximate 
\[
\P(\lambda \in \cdot | u) &\approx \shmu_t(\cdot , u) =
\sum_{j=0}^{n} \1[\stu_{jk} < u \leq \stu_{(j+1)k}]\shmu_{tj}(\cdot)\qquad
\shmu_{tj} \propto \sum_{i=jk+1}^{((j+1)k)\wedge \tau} \delta_{\tilde\lambda_j}, \label{eq:approxlambdagivenu}
\]
where $\stu_0 = 0$ and $\stu_{(n+1)k} = \infty$ by convention, and each $\shmu_{tj}$ is normalized appropriately.
Then set the tuned initial window $w_t(u)$ using the results from \cref{eq:ostuning,eq:gtuning}, except
that the distribution of $\lambda$ given $u$ is replaced with the 
piecewise-constant approximation in \cref{eq:approxlambdagivenu}.
This procedure results in a piecewise-constant window function $w_t(u)$ that refines as more draws are obtained.

\cref{lem:tuningconvergence} shows that under mild technical assumptions, the above procedure results in 
a per-iteration cost that converges to the slice-adaptive optimal cost for each of the surrogate cost functions.
In particular, the result confirms that any $\eta\in(0,1)$ and $\beta\in(0.5,1)$ in \cref{eq:taukn}
suffices to ensure that the number of draws used to tune each region in the piecewise $w_t(u)$ increases
quickly enough to guarantee convergence. 
One simplification made for \cref{lem:tuningconvergence} is that within each bin, 
$w$ is selected from a finite set $\scW$ in \cref{eq:finiteminw} rather than the entirety of $\reals_+$.
This simplification does not meaningfully change the result but avoids significant unnecessary additional technicality.

\bthm\label{lem:tuningconvergence}
Let $\scW$ be a finite subset of $(0, \infty)$, and let
$h:\reals_+\to\reals$ be a function such that for all $w\in\scW$, $\E[h(\lambda/w) | u]$ is continuous in $u$
and $h(\lambda/w)$ is locally uniformly subexponential conditioned on $u$ (\cref{def:localunifsubexp}).
Suppose ideal slice sampling is geometrically ergodic for target $\pi$,
$\shmu$ is the kernel given by \cref{eq:approxlambdagivenu},
and
\[
w_t(u) = \argmin_{w\in\scW} \int h\lt(\frac{\lambda}{w}\rt) \shmu_t\lt(\d\lambda, u\rt). \label{eq:finiteminw}
\]
Then
\[
\E\lt[ h\lt(\frac{\lambda}{w_t(u)}\rt) | (u_j,\lambda_j)_{j=t+1}^{2t}\rt] \convp
\E\lt[\min_{w\in\scW}\E\lt[h\lt(\frac{\lambda}{w}\rt) | u\rt] \rt] \qquad t\to\infty.
\]
\ethm

The computational complexity of tuning $w_t$ vanishes compared to the cost of sampling.
Sorting $(\lambda_j, u_j)$ has complexity $O(\tau\log\tau) = O(t^\eta \log t) = o(t)$.
Once sorted, both one-shot and gradient tuning within each block of draws requires $O(k)$ computation, and so 
for $n$ blocks tuning requires $O(kn) = O(\tau) = o(t)$ computation. 
However, due to the need to perform a $O(\log n) = O(\log \tau^{1-\beta}) = O(\log t)$ 
binary search for the bin in which $u$ falls to evaluate $w_t(u)$, the complexity
of sampling increases by a log factor.
In practice, this issue can be ignored; for nontrivial target distributions 
the search will usually be orders of magnitude faster than  evaluating the density once, even for very large $t$  
(when $t=10^9$, for example, the binary search will require roughly $14$ floating-point lookups and comparisons). 
The issue can also be resolved simply by putting an upper limit on the number of bins $n$.

The final fully-automated, tuned slice sampling algorithm is provided in \cref{alg:tunedslicing}. 
The tuning algorithm runs in doubling rounds, each time using the last half of the draws.
In the tuning algorithm itself, the initial window size function $w(\cdot)$ is initialized
using the one-shot estimate $\shw$, and then further refined using backtracking gradient descent
to find $\stw$. Note that in practice, the values of $\lambda$ are not directly observed, 
and moreover slices may not be contiguous,
so tuning is based on an average of upper and lower bounds on $\lambda$ that are tracked by \cref{alg:eval}.

%% file: simulations.tex
\bfig[t]
\bsubfig{0.5\textwidth}
\includegraphics[width=\textwidth]{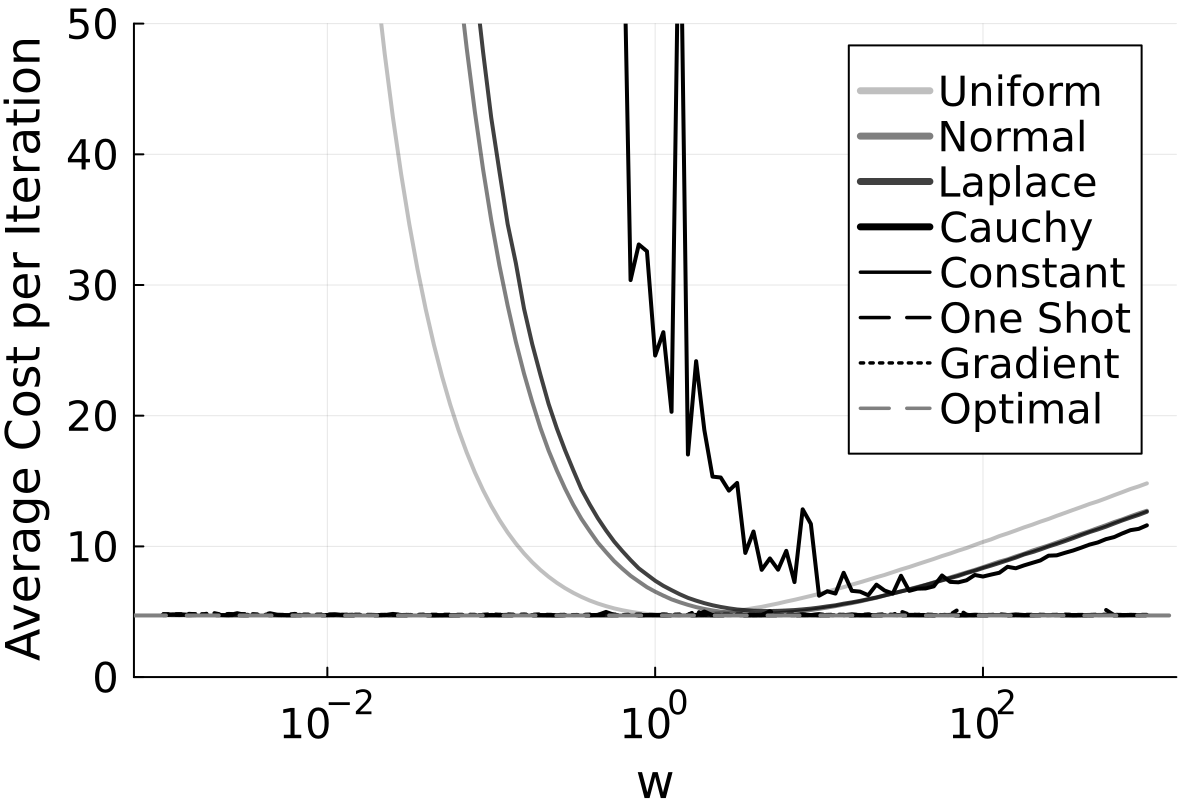}
\includegraphics[width=\textwidth]{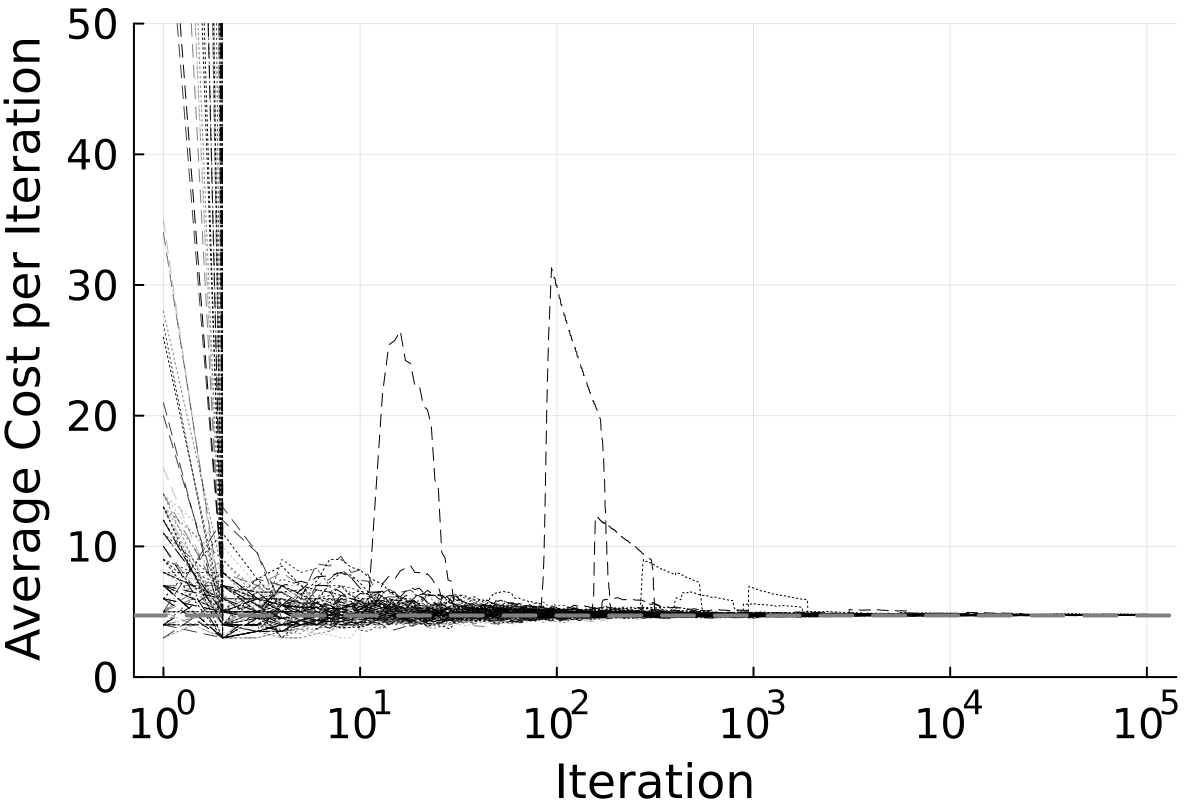}
\caption{Stepping}\label{fig:sim_cost_stepping}
\esubfig
\bsubfig{0.5\textwidth}
\includegraphics[width=\textwidth]{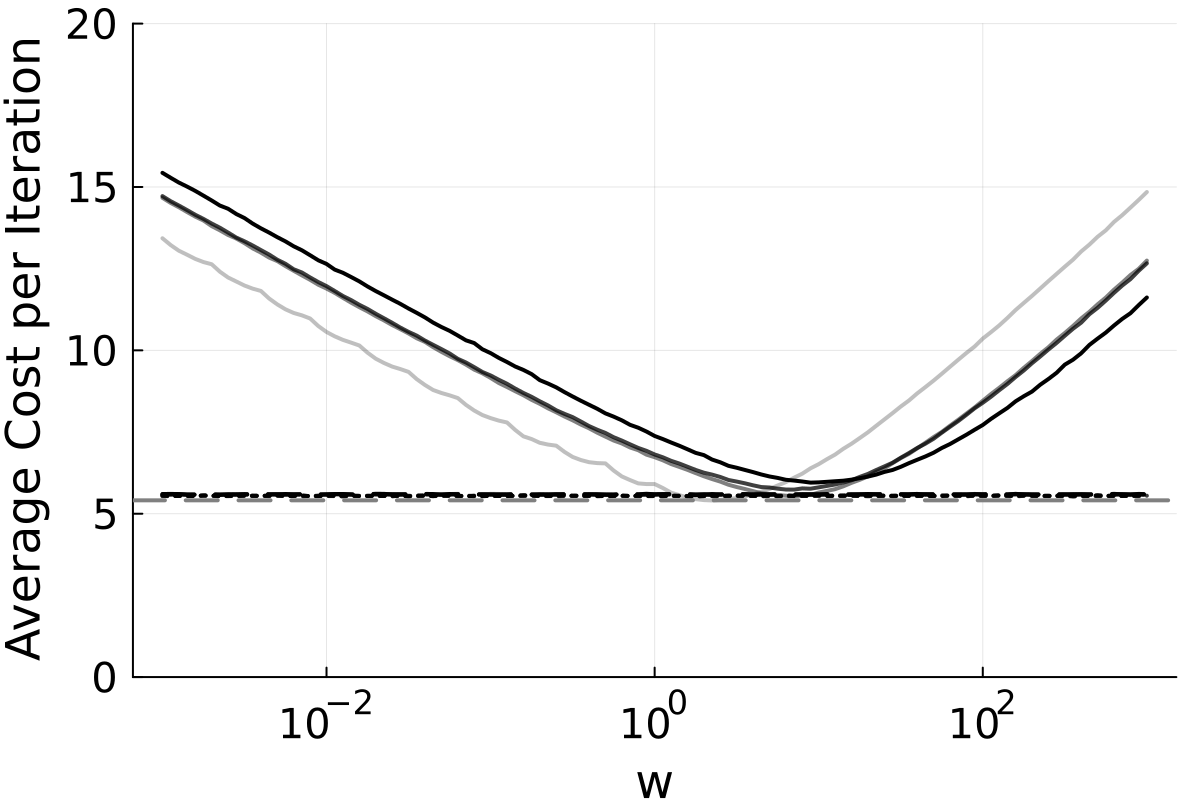}
\includegraphics[width=\textwidth]{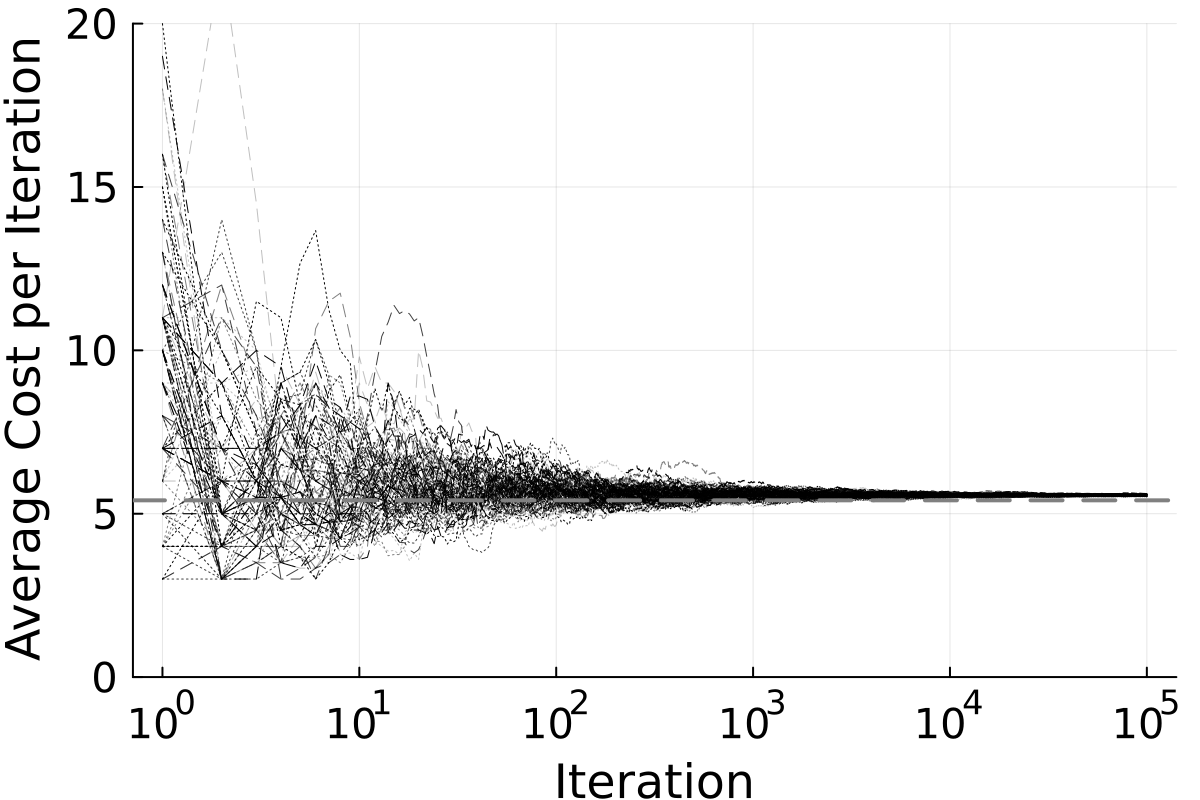}
\caption{Lazy Cached Doubling}\label{fig:sim_cost_doubling}
\esubfig
\caption{The average per iteration cost resulting from the proposed tuning schemes
over 100{,}000 simulated iterations, as a function of the initial window size $w$ (top row),
and as a function of the iteration number (bottom row). Line shade corresponds to the target
(uniform, normal, Laplace, and Cauchy), while line style corresponds to the tuning scheme
(constant, one-shot, and gradient). The bottom row figures display only the one-shot and gradient tuning
schemes, and for each tuning scheme and target, 13 traces are displayed across evenly-spaced initial choices
of $w\in(10^{-3},10^3)$.  The horizontal grey dashed line displays the 
lowest possible cost with $w\propto \alpha \lambda$ and optimal $\alpha$ given by \cref{thm:stepcost,thm:lazycost}.
All of the proposed methods for both stepping out and lazy cached doubling provide near-optimal cost per iteration after about 100 iterations, 
regardless  of the initial choice of $w$.
}\label{fig:sim_cost_synth}
\efig

\bfig[t]
\bsubfig{0.5\textwidth}
\includegraphics[width=\textwidth]{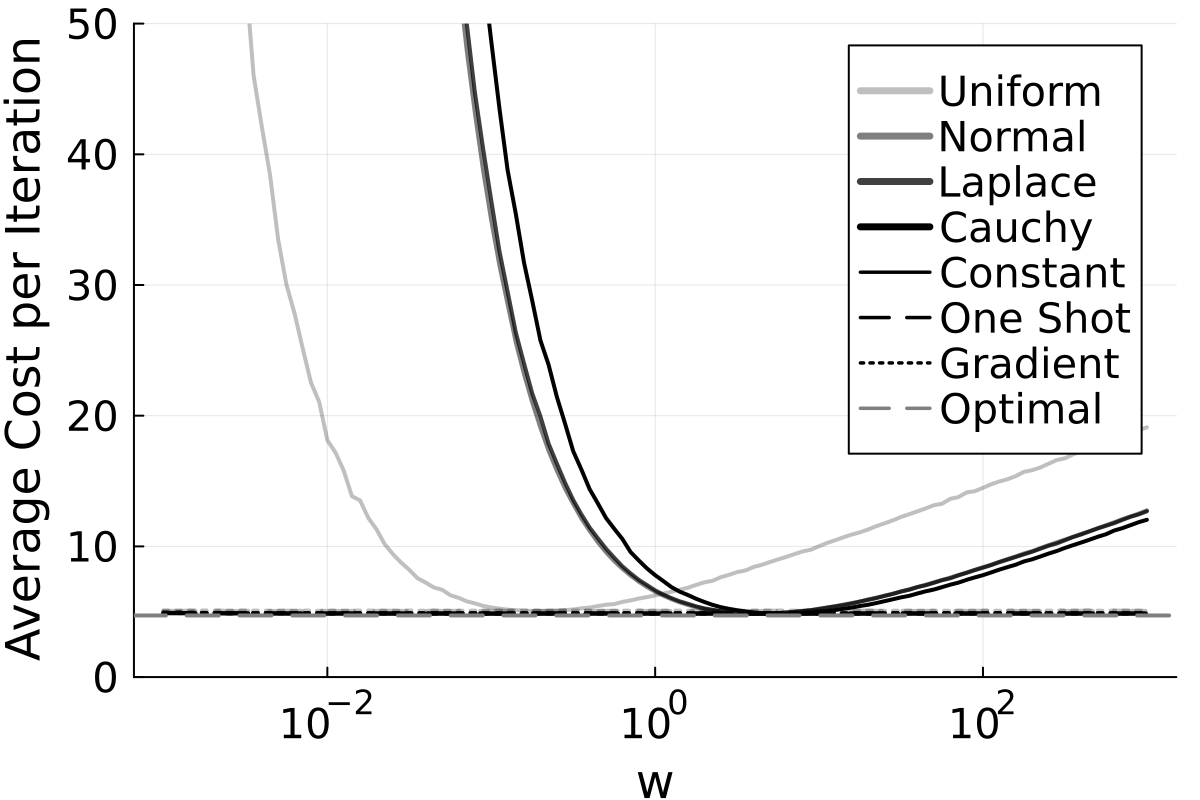}
\includegraphics[width=\textwidth]{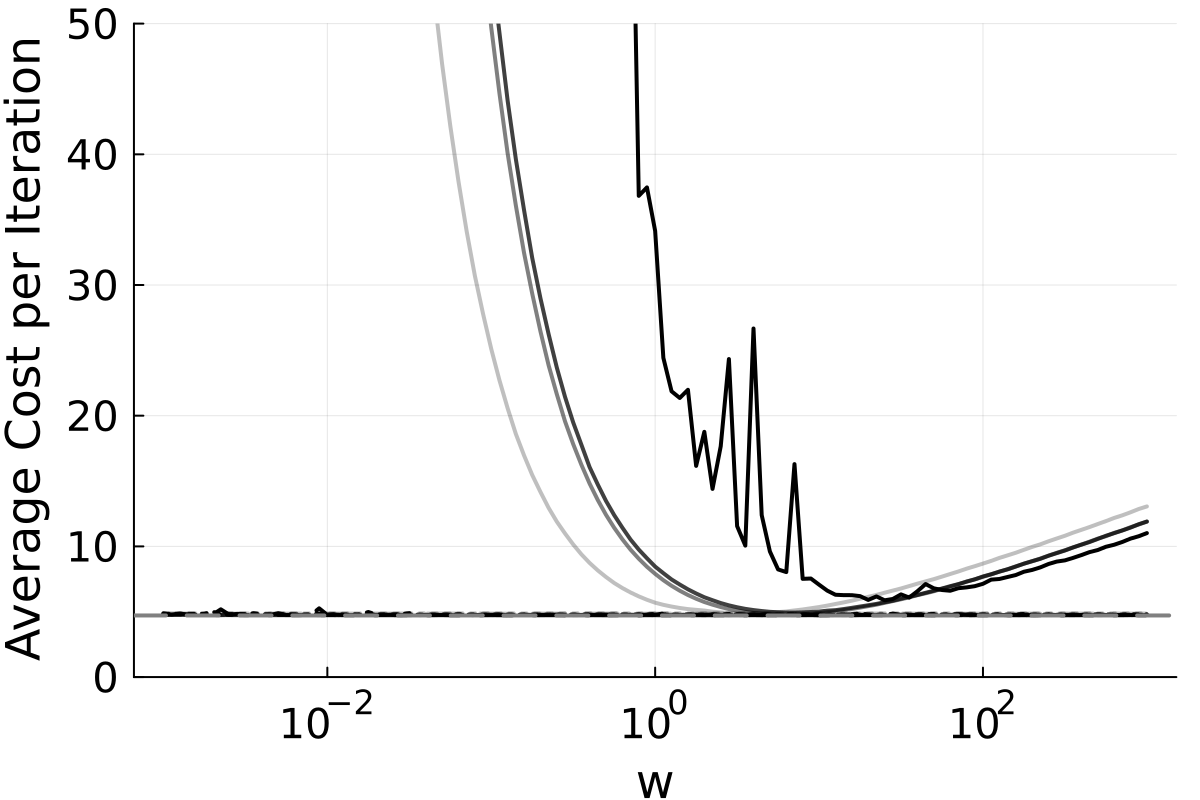}
\caption{Stepping}\label{fig:sim_cost_multi_stepping}
\esubfig
\bsubfig{0.5\textwidth}
\includegraphics[width=\textwidth]{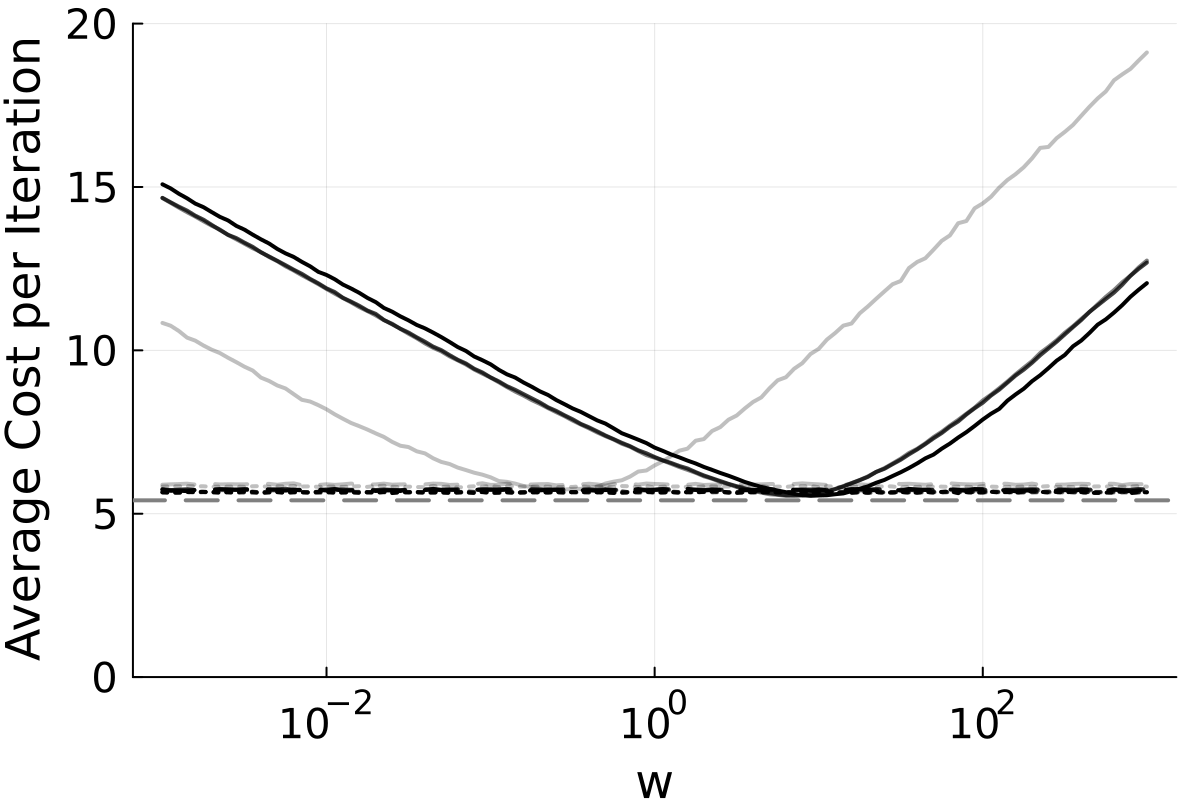}
\includegraphics[width=\textwidth]{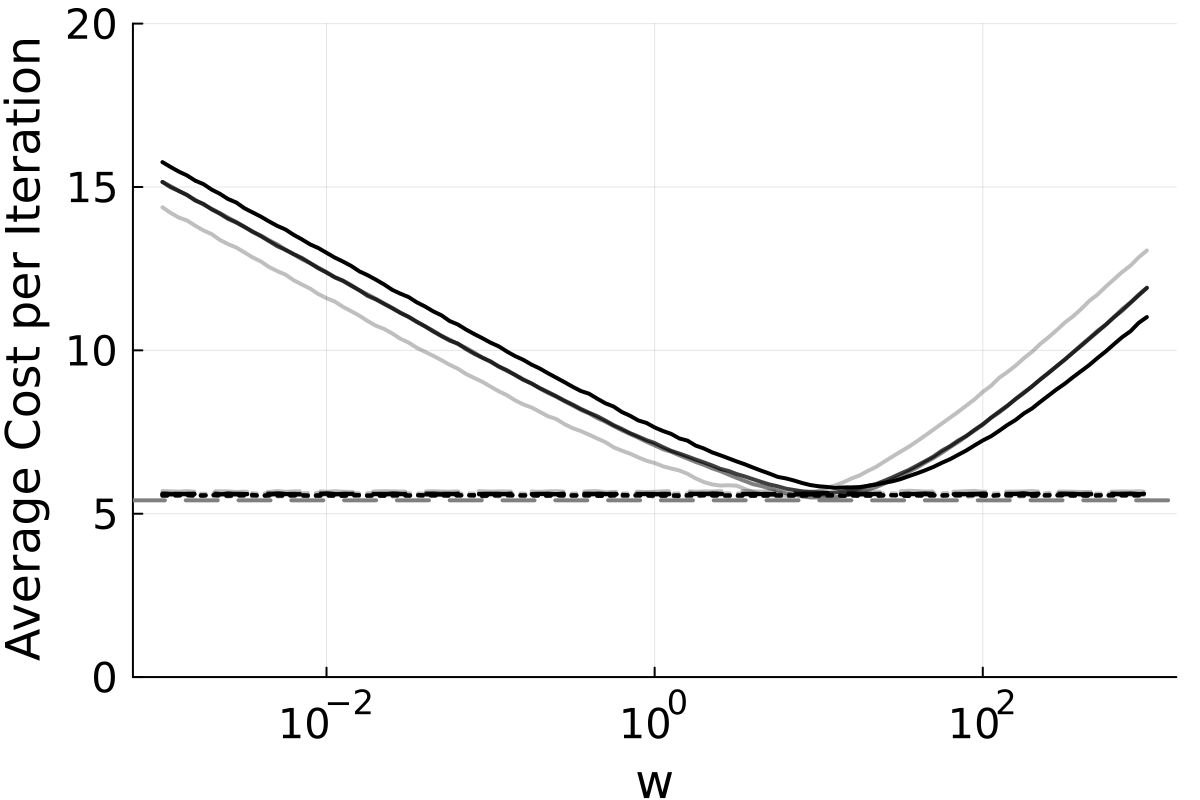}
\caption{Lazy Cached Doubling}\label{fig:sim_cost_multi_doubling}
\esubfig
\caption{The average per iteration cost resulting from the proposed tuning schemes
over 100{,}000 simulated iterations, as a function of the initial window size $w$,
for hit-and-run sampling on multidimensional targets (top row),
and univariate multimodal targets (bottom row). Line shade corresponds to the target, while line style corresponds to the tuning scheme. 
The horizontal grey dashed line displays the 
lowest possible cost with $w\propto \alpha \lambda$ and optimal $\alpha$ given by \cref{thm:stepcost,thm:lazycost}.
All of the proposed methods for both stepping out and lazy cached doubling provide near-optimal cost per iteration
regardless of the initial choice of $w$.
}\label{fig:sim_cost_multi_synth}
\efig

\bfig[t]
\bsubfig{0.5\textwidth}
\includegraphics[width=\textwidth]{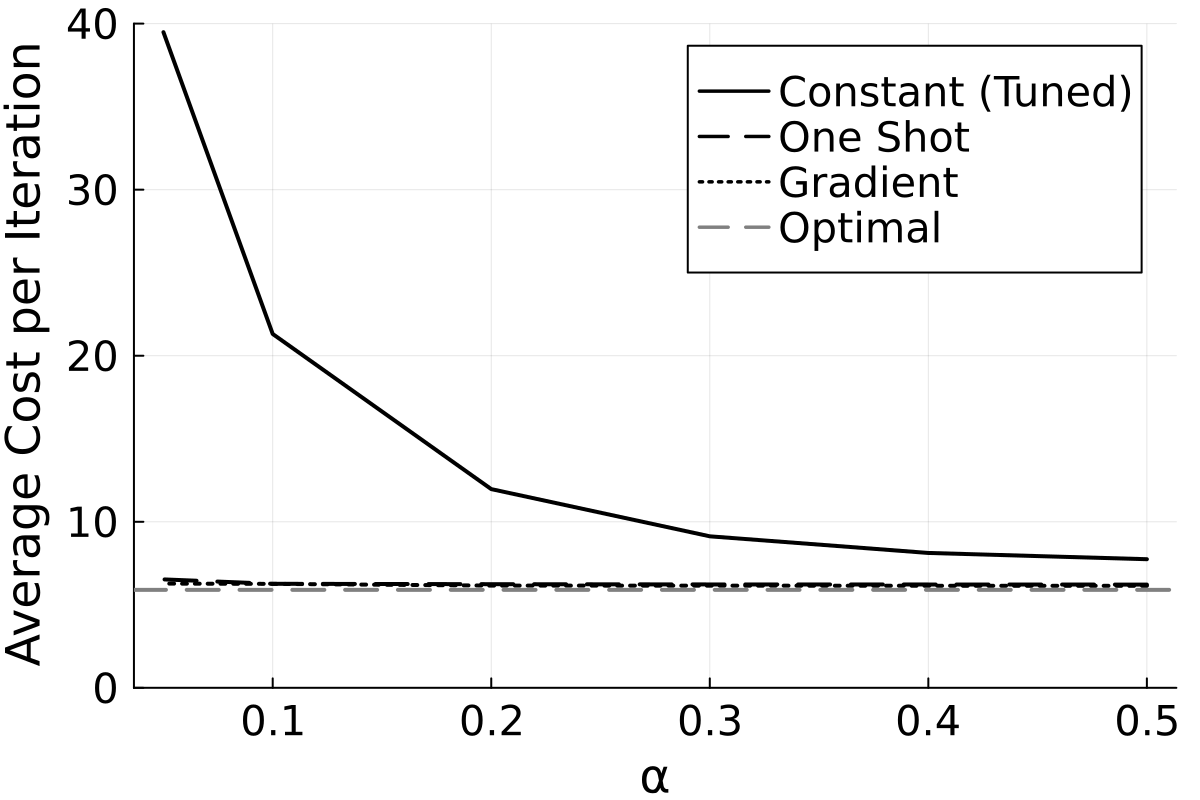}
\caption{Gamma$(\alpha,1)$}\label{fig:costvtail_gamma}
\esubfig
\bsubfig{0.5\textwidth}
\includegraphics[width=\textwidth]{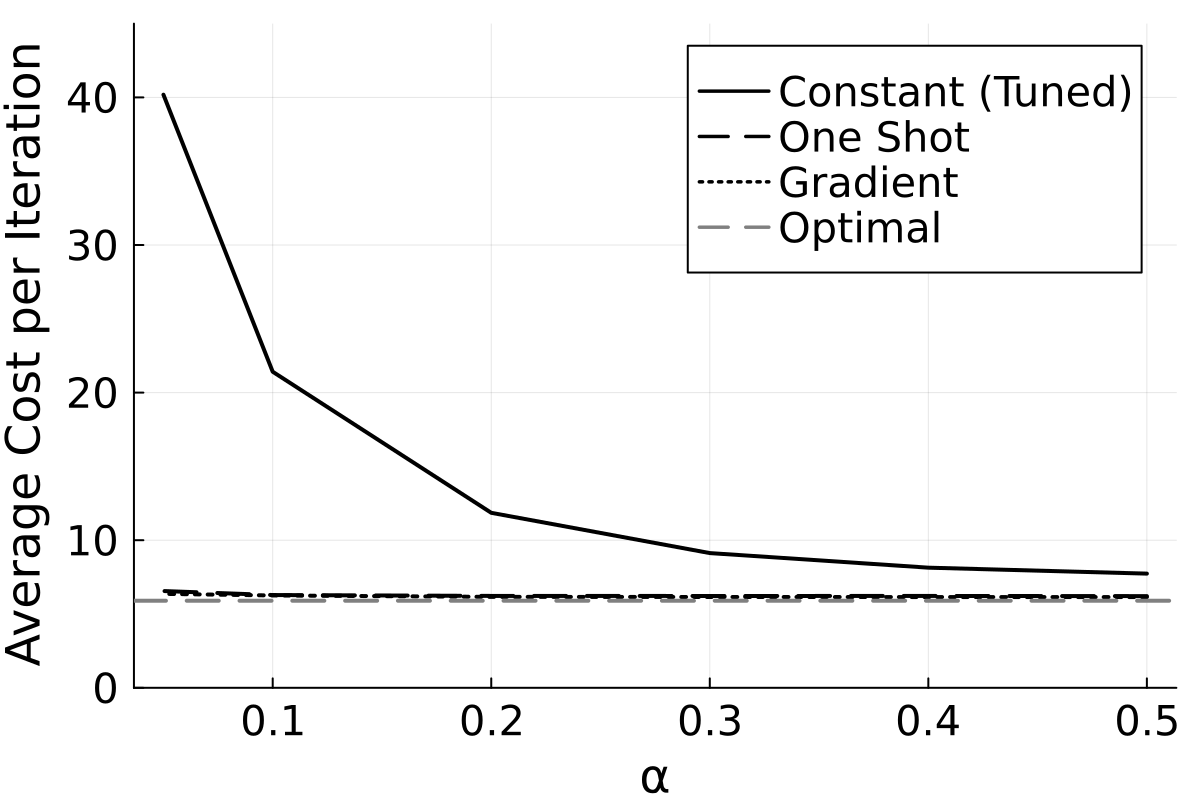}
\caption{T$(\alpha)$}\label{fig:costvtail_student}
\esubfig
\caption{The average per iteration cost for slice-adaptive lazy cached doubling versus
constant tuning over 1{,}000{,}000 simulated iterations.
\cref{fig:costvtail_gamma} displays the result for the gamma distribution
with shape parameter $\alpha$, and \cref{fig:costvtail_student} displays
the result for the T distribution with $\alpha$ degrees of freedom.
Line style corresponds to the tuning scheme (tuned constant, one-shot, and gradient).
The proposed slice-adaptive methods provide increasing benefit the more the slice
width $\lambda$ varies from iteration to iteration (smaller $\alpha$ in each case).
}\label{fig:sim_tails}
\efig

\subsection{Simulations}\label{sec:simulations}
This section presents simulation results for the proposed tuning schemes, including 
both the one-shot window $\shw$ and gradient-based window $\stw$ initialized using the one-shot scheme.
These simulations are designed to be a demonstration that the tuning schemes reliably work well without 
user input across a range of target tail behaviour, dimension, and uni/multimodality.
Results are shown only for stepping out and lazy cached doubling; simple cached doubling is omitted as 
it is dominated by lazy cached doubling.

\cref{fig:sim_cost_synth} displays 
the average cost per iteration of tuned slice sampling
for unimodal, univariate $\Unif[0,1]$, $\Norm(0,1)$, $\Laplace(0,1)$, and $\Cauchy(0,1)$ targets.
The top row figures show the average cost over the last 50{,}000 draws produced by 100{,}000 total iterations
as a function of the initial setting of window size $w \in [10^{-3}, 10^3]$.
Every proposed tuning method nearly matches the oracle optimal costs from \cref{thm:stepcost,thm:lazycost},
regardless of the initial choice of $w$, while the untuned slice sampler has performance that varies significantly depending
on the choice of $w$. The bottom row figures show the cost as a function of iteration number, averaged over the most recent half of
the iterations. Tuning traces are shown for 13 log-evenly-spaced starting values of $w\in[10^{-3}, 10^3]$ for each method, with
the same legend convention as in the top row of plots. 
These figures demonstrate that the tuning converges after about 100 iterations for both stepping out and lazy cached doubling,
with some infrequent jumps in cost for stepping out, likely caused by temporary mistuning of $w$ due to stochasticity in $\shmu_t$.
\cref{fig:sim_cost_multi_synth} repeats these experiments for 256-dimensional \iid products of the same four univariate targets,
and for univariate multimodal mixtures
\[
\text{Uniform Mix: }&0.4\Unif[-3/2,-1/2]+0.2\Unif[-1/2,1/2] + 0.4\Unif[1/2,3/2]\\
 \text{Normal Mix: }&0.5\Norm(-1,1)+0.5\Norm(1,1)\\
 \text{Laplace Mix: }&0.5\Laplace(-1,1)+0.5\Laplace(1,1)\\
 \text{Cauchy Mix: }&0.5\Cauchy(-1,1)+0.5\Cauchy(1,1).
\]
The results have the same qualitative characteristics as in the previous results; the tuned methods reliably
achieve an average cost near the oracle optimal for each method, regardless of the initialization of $w$.
Taken together, these results suggest that the proposed tuning methods are very robust to the initial choice of $w$,
and initializing $w(u)=1$ as suggested in \cref{alg:tunedslicing} is likely reasonable for most problems.

Finally, \cref{fig:sim_tails} displays a comparison of slice sampling with the proposed slice-adaptive tuning schemes
versus using a tuned but constant, non-slice-adaptive
choice of $w$. The results show that when the slice lengths tend to vary more---e.g., for targets with heavy tails or 
unbounded density functions---slice-adaptivity becomes increasingly important to obtaining a near oracle-optimal per-iteration cost.
Results are shown only for lazy cached doubling, which is much more robust to mistuning; stepping 
out often became too expensive to run during early rounds before the tuning methods had stabilized. 
In practice, lazy cached doubling should essentially always be preferred over the other two methods.

%% file: conclusion.tex
\section{Conclusion}\label{sec:conclusion}
This paper presented an analysis of the cost of slice sampling along unidimensional
manifolds, and a fully-automated, near-optimal implementation.
Key contributions include an improvement to the doubling slice finding scheme via lazy caching (\cref{alg:lazydoubling,alg:lazyaccept}),
characterizations of the per-iteration cost as a function of initial window size
and slice width (\cref{thm:stepcost,thm:doublecost,thm:lazycost}),
simple one-shot and tractable gradient-based tuning schemes along with suboptimality
guarantees (\cref{thm:tunequality}),
and tuning methods  using draws from Markov chain Monte Carlo (\cref{alg:tunedslicing})
that provide asymptotic convergence in probability to the surrogate optimal cost (\cref{lem:tuningconvergence}).

This work focused on initial widths as a function of only the slice variable $w(u)$.
In multivariate settings, a natural and useful extension to this work would be to extend $w$ to be a function of both the slice variable $u$
and the direction $\rho$. 
All of the theory and tuning methods in this work
extend without much effort to this setting by using the conditional distribution of slice width $\lambda$
given $u, \rho$ in place of the conditional distribution of $\lambda$ given $u$.
For slice sampling within Gibbs sampling, for example,
one should use the proposed procedure in this paper to tune a separate function $w_i(u)$ for each coordinate direction $i$ of motion, 
because different variables in a model will often exhibit different posterior scale.

One limitation of the present work is that tuning is based on bounds on the slice width
at each iteration (\cref{alg:eval}) rather than the value of $\lambda$ itself.
A possible avenue for future work is therefore to rigorously handle the lack of observability 
of the slice width at each iteration in a manner that does not significantly increase
computational cost. As mentioned earlier, there are also many avenues for follow-up work
in combining earlier developments in adaptive slice sampling with the initial window tuning from this work, e.g.,
covariance adaptation to improve condition number dependence.

%% file: surrogates.tex
\section{Surrogate approximations}\label{sec:surrogates}
The tuning methods in this work rely on surrogate approximations of $\fstep$, $\fdouble$, and $\flazy$, 
displayed in \cref{fig:approximation},
that yield tractable optimization problems given a known 
distribution of $\lambda$ conditioned on $u$.
For stepping out, define the functions 
\[
\shfstep(x) &= 3.6 + 1.2x-\log (x),  \quad \stfstep(x) = \fstep(x).
\]
For cached doubling, define the functions
\[
\shfdouble(x) &= 4 + \lt\{\begin{array}{ll}
\frac{2}{\log 2}\log\lt(\frac{x}{0.3}\rt) & x > 0.3\\
-2\log\lt(\frac{x}{0.3}\rt) & x \leq 0.3
\end{array}\rt. \\
\stfdouble(x) &= \lt(\frac{2}{\log 2}+2\rt)\log\lt(1+\lt(2+\frac{\log 2}{2}\rt)x\rt) - 2\log x + 1.2.
\]
Finally, for lazy cached doubling, define the functions
\[
\shflazy(x) &= 4 + \lt\{\begin{array}{ll}
\frac{5}{6\log 2}\log\lt(3x\rt) & x > 1/3\\
-2\log\lt(3x\rt) & x \leq 1/3,
\end{array}\rt. \\
\stflazy(x) &= \lt(\frac{5}{6\log 2}+2\rt)\log\lt(1 + \frac{3(2+ \log(2))}{2}x\rt) - 2\log\lt(\frac{3}{2}x\rt) + (3/2).
\]
The approximations $\stfstep$, $\stfdouble$, $\stflazy$ are designed to be convex and locally smooth (\cref{defn:locallysmooth})
on log-scale to enable gradient optimization with guarantees.
The approximations $\shfstep$, $\shfdouble$, $\shflazy$ are designed to have 
a closed-form
optimum of $\E[\shf(\lambda/w) | u]$ over $w$ for the purpose of one-shot tuning
and initialization of the aforementioned optimization.

As these are approximations of the original cost functions, it is important to quantify their 
downstream suboptimality when used for tuning. This work uses the notion of an $(\epsilon,\delta)$-approximation
given by \cref{defn:approx}.
\bdefn\label{defn:approx}
Fix $\epsilon,\delta\geq 0$. A function $\shh$ is an $(\epsilon,\delta)$-approximation of $h\geq0$ if
\[
\sup_x \frac{|h(x)-\shh(x)|}{\epsilon h(x) + \delta} \leq 1. \label{eq:epsdeltaapprox}
\]
\edefn
\cref{lem:fapprox} shows that each of the surrogate cost functions is an $(\epsilon,\delta)$-approximation
of its respective exact cost function. \cref{fig:approximation_error} validates these theoretical results via simulation;
each plot shows the relative-additive error \cref{eq:epsdeltaapprox} for each surrogate approximation and its respective exact cost,
with all lines falling below 1.
\blem\label{lem:fapprox}
The following statements hold:
\bitem
\item $\shfstep$ is a $(0.5, 0)$-approximation of $\fstep$.
\item $\shfdouble$ is a $(0, 2)$-approximation of $\fdouble$.
\item $\stfdouble$ is a $(0, 0.35)$-approximation of $\fdouble$.
\item $\shflazy$ is a $(0, 2)$-approximation of $\flazy$
\item $\stflazy$ is a $(0, 0.35)$-approximation of $\flazy$
\item $\fstep\circ \exp$, $\stfdouble\circ \exp$, and $\stflazy\circ \exp$ are each convex and locally smooth.
\eitem
\elem

\bfig[t]
\bsubfig{0.33\textwidth}
\includegraphics[width=\textwidth]{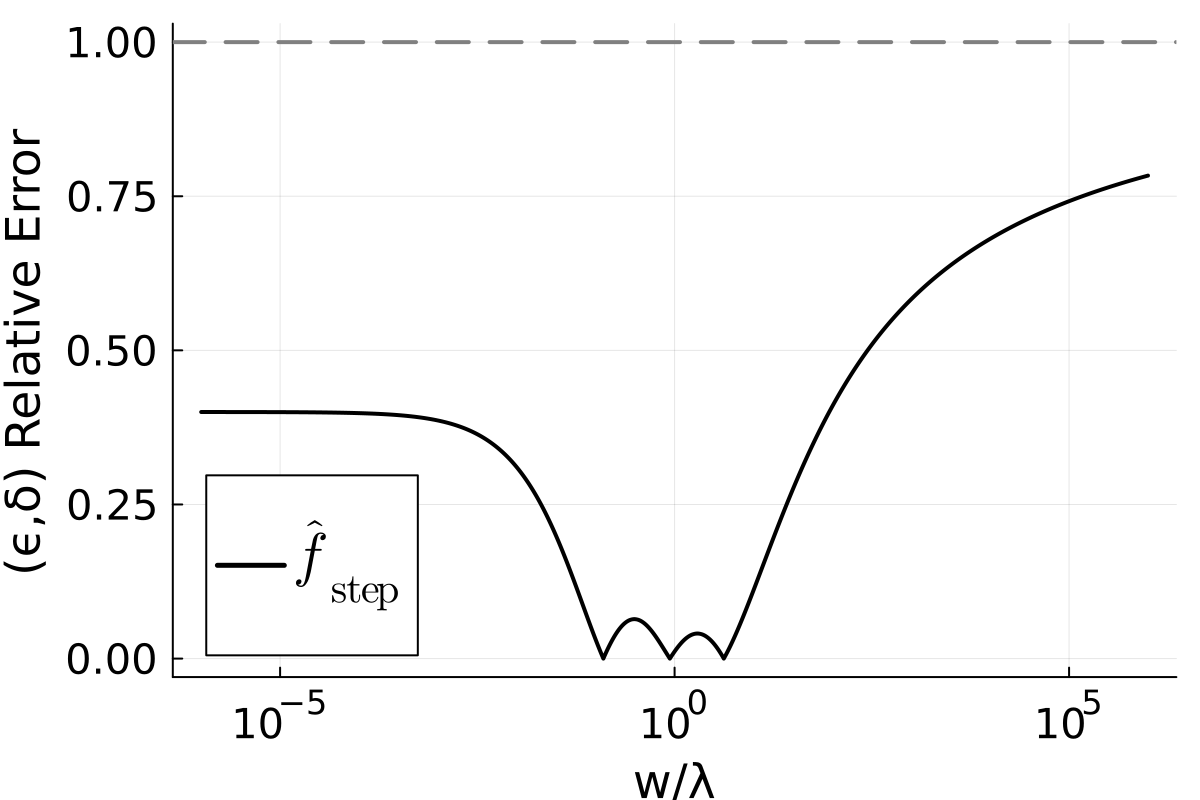}
\caption{Stepping}\label{fig:approx_error_step}
\esubfig
\bsubfig{0.33\textwidth}
\includegraphics[width=\textwidth]{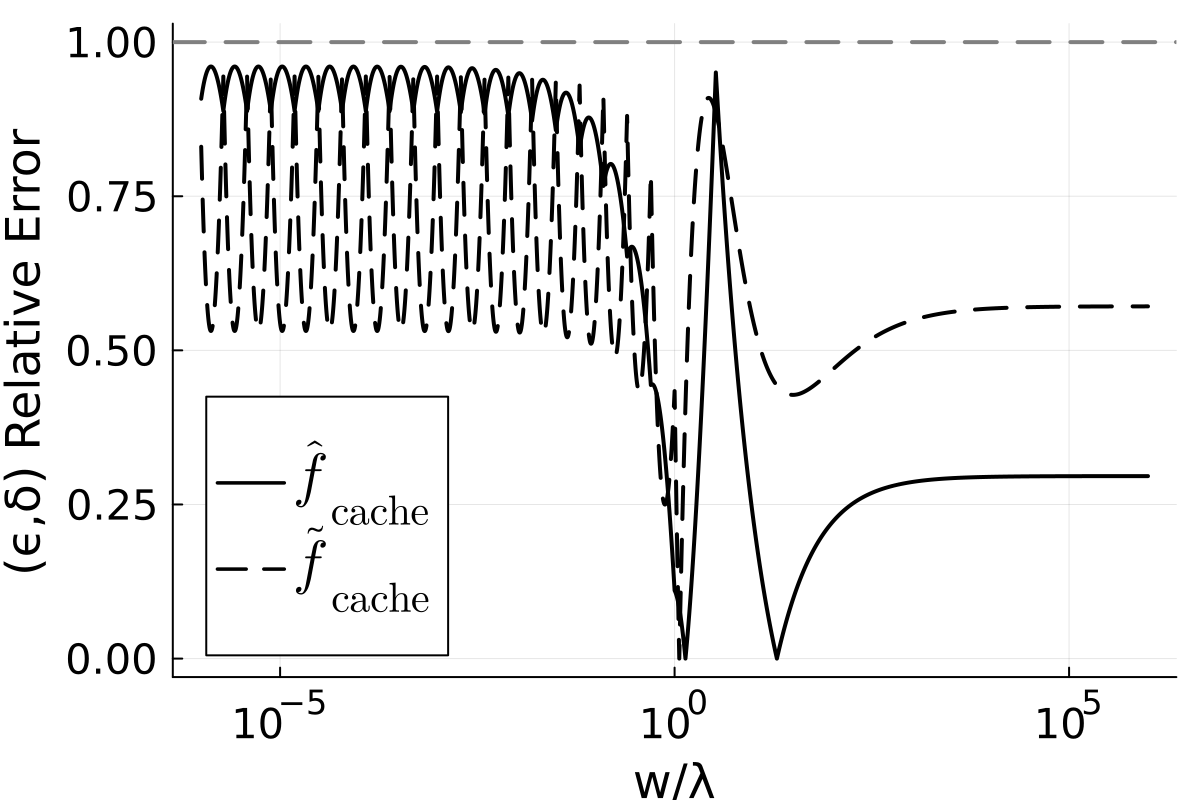}
\caption{Cached Doubling}\label{fig:approx_error_cache}
\esubfig
\bsubfig{0.33\textwidth}
\includegraphics[width=\textwidth]{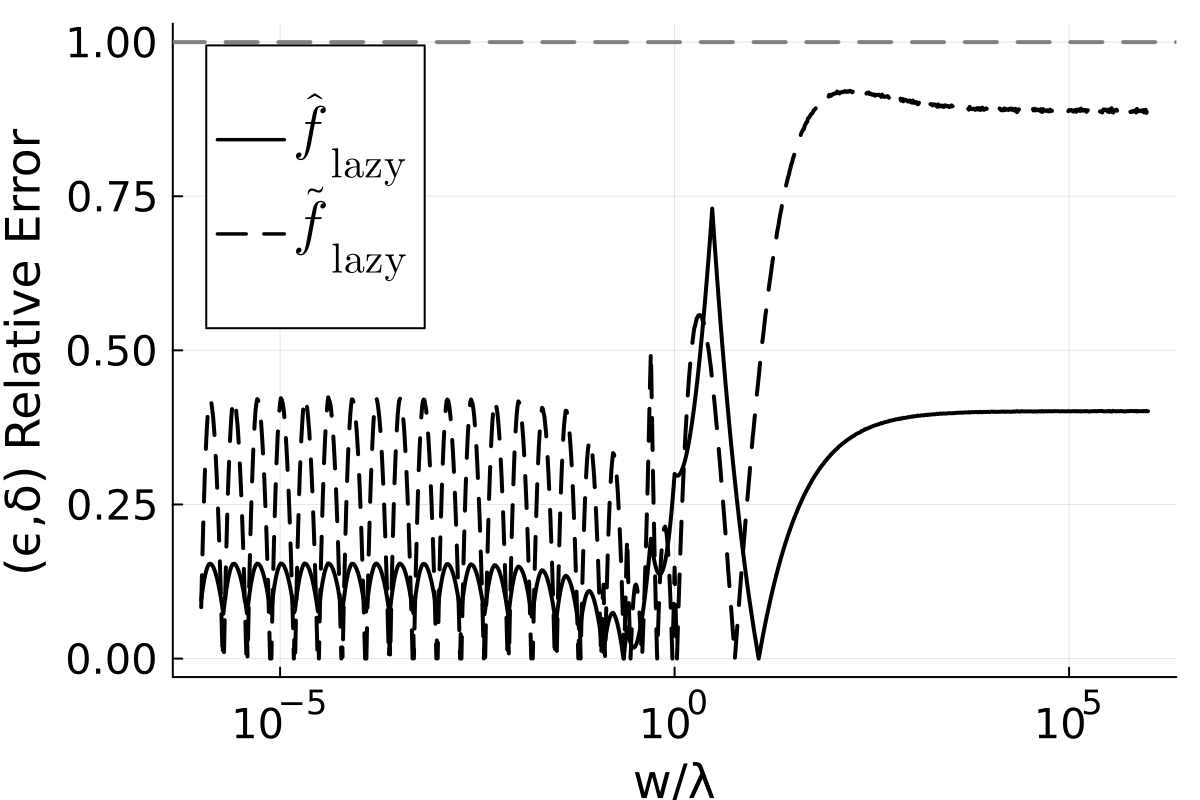}
\caption{Lazy Cached Doubling}\label{fig:approx_error_lazy}
\esubfig
\caption{Lines displaying the error of each $(\epsilon,\delta)$-approximation of $f$, 
normalized by $\epsilon f + \delta$ using the 
values of $\epsilon$ and $\delta$ obtained from \cref{lem:fapprox}. All lines are uniformly
bounded above by 1, validating the theory in \cref{lem:fapprox}.}\label{fig:approximation_error}
\efig

Optimization of each surrogate cost function yields the corresponding tuned slice-adaptive initial window function.
For each $s\in\{\text{step},\text{cache},\text{lazy}\}$,
\[
\shw_{s}(u) \!&=\! \exp\lt(\argmin_{x\in\reals} \E\lt[\shf_{s}\lt(\lambda \exp(-x)\rt) | u\rt]\rt),\,\,
\stw_{s}(u) \!=\! \exp\lt(\argmin_{x\in\reals} \E\lt[\stf_{s}\lt(\lambda \exp(-x)\rt) | u\rt]\rt).
\]

In general, optimizing an $(\epsilon,\delta)$-approximation of a cost function 
results in a suboptimality guarantee for the original cost, as stated by
\cref{lem:subopt}. This result is elementary and well-known in the 
$\epsilon$-approximation literature \citep{Li01,Langberg10} \citep[e.g.,][Theorem 2.1]{Bachem17}.
We use \cref{lem:fapprox,lem:subopt} to analyze the suboptimality
of the one-shot 
and gradient-based slice-adaptive tuning strategies for slice sampling
in \cref{eq:ostuning,eq:gtuning}.

\blem\label{lem:subopt}
Suppose $\shh(x,z)$ is an $(\epsilon,\delta)$-approximation of $h(x,z)\geq 0$,
$Z$ is a random variable, and $\shx^\star = \argmin_x \E \shh(x,Z)$ exists.
Then $\shx^\star$ is $\lt(\frac{1+\epsilon}{1-\epsilon}, \frac{2\delta}{1-\epsilon}\rt)$-optimal
for $\E h(x,Z)$.
\elem

%% file: proofs.tex
\section{Proofs}\label{sec:proofs}

\bprfof{\cref{lem:exactandlln}}
If the current state $x \in \{x : 0<\pi(x)<\infty\}$, then the slice variable $u \in (0, \pi(x)]$ almost surely.
The slice defined by $u$ is nonempty (in particular, $x$ is in the slice)
so the slice is a contiguous interval of finite length by \cref{assum:unimodal}.
The approximate slice found by either doubling (\cref{alg:doubling}) or 
stepping (\cref{alg:steppingout}) is guaranteed to contain the exact slice by design,
and both slice bounding algorithms terminate in finite time because the slice interval has a finite length.
The shrinkage algorithm (\cref{alg:shrinkage}) is an adaptive rejection sampler
that removes only invalid next states, and hence the next state $x'$ is drawn uniformly from the slice
conditioned on $u$. Because the slice has a nonzero length almost surely by \cref{assum:unimodal},
the shrinkage algorithm terminates in finite time almost surely.
Note also that all states $x'$ on the slice satisfy $\pi(x') \geq u > 0$, 
and since $\pi$ is a distribution it must be the case that $\pi(\{x : \pi(x) = \infty\}) = 0$, 
so therefore $x' \in \{x : 0 < \pi(x) < \infty\}$ almost surely.
By induction, the sequence of states have the same distribution as those produced by the ideal slice sampler
and each step terminates in finite time almost surely.

For the LLN, by \citet[Theorem 17.0.1]{MeynTweedie} and by the equivalence to ideal slice sampling, 
it suffices to show that ideal slice sampling is positive Harris recurrent.
By \citet[Corollary 1]{Tierney94}, it therefore suffices to show that ideal slice sampling is $\pi$-irreducible
and dominated by $\pi$. Let the slice sampler kernel be denoted $P(x, A)$.
Consider a set $A$ with $\pi(A) = 0$. Then the density $\pi(\cdot) = 0$ Lebesgue-almost everywhere on $A$. 
For any current state $x$, we have that $0 < \pi(x) < \infty$ almost surely by the earlier argument,
so the slice variable $u\dist \Unif[0,\pi(x)]$ satisfies $u>0$ almost surely, 
and so at most a null set in $A$ lies in the slice; hence $P(x, A) = 0$ and the slice sampler kernel is dominated by $\pi$.
Now consider a set $A$ with $\pi(A) > 0$; there must be some $\epsilon > 0$ such that
$A \cap \{\pi(x) \geq \epsilon\}$ has nonzero Lebesgue measure.
Therefore
\[
P(x, A) &= \int_0^{\pi(x)} \int \1[\pi(x')\geq u] \1[x' \in A] \d x'\d u\\
&\geq \int_0^{\epsilon} \text{Leb}(A \cap \{ \pi(x')\geq u\})\d u\\
&\geq \epsilon \text{Leb}(A \cap \{ \pi(x')\geq \epsilon\})\\
&> 0.
\]
Therefore the slice sampler can transition to any set $A$ with $\pi(A)>0$ in one step; hence the sampler is irreducible.
\eprfof

\bprfof{\cref{lem:GPDEeqn}}
By \cref{eq:recursiveN} and the fact that $(V_n)_{n=1}^\infty \eqd (V_n)_{n=2}^\infty$,
we have that $F_0(x,y) = 1$ and for all $n\in\nats$,
\[
F_n(x,y) &= \frac{x}{x+y+\lambda}\int_0^1 F_{n-1}(vx,y,z)\d v + \frac{y}{x+y+\lambda}\int_0^1F_{n-1}(x,vy,z)\d v.
\]
A transformation of variables in each integral yields
\[
F_n(x,y) &= \frac{1}{x+y+\lambda}\lt(\int_0^x F_{n-1}(v,y,z)\d v+ \int_0^yF_{n-1}(x,v,z)\d v\rt).\label{eq:intermediateintegraleqn}
\]
Define the generator $G(x,y,z) = \sum_{n=0}^\infty z^nF_n(x,y)$. 
Then $G$ is well-defined on $x,y\geq 0$ and $z\in[-1,1]$,
since $F_n \geq 0$ and $\sum_{n=0}^\infty F_n(x,y) < \infty$
due to the fact that $N$ is dominated by a geometric random variable (corresponding to a less efficient shrinkage algorithm where we
keep $\lambda_\ell, \lambda_r$ fixed after each attempt). 
By similar dominated convergence arguments, both
$\frac{1}{n!}\lt.\der[n]{G}{z}\rt|_{z=0} = F_n$
and $\E N = G(x,y,1) = \sum_{n=0}^\infty F_n(x,y)$ as required.
Furthermore, 
by symmetry of the shrinkage algorithm, $F_n(x,y) = F_n(y,x)$, and hence $G$ preserves this symmetry.
Multiplying both sides of \cref{eq:intermediateintegraleqn} by $z^n$, summing over $n\in\nats$, and interchanging
the sum and integrals yields
\[
G(x,y,z) &= 1+ \frac{z}{x+y+\lambda}\lt(\int_0^x G(v,y,z)\d v + \int_0^y G(x,v,z)\d v\rt).
\]
The interchange of sum and integral follows again by dominated convergence.
Finally multiplying both sides by $x+y+\lambda$ and differentiating in $x$ and $y$
yields the PDE
\[
(1-z)\lt(G_x + G_y\rt) + (x+y+\lambda)G_{xy}&= 0.
\]
For the boundary, we fix $z$, set $y=0$ and denote $g(x) = G(x,0,z)$. Multiplying both sides 
of the integral equation by $x+\lambda$ and 
taking the derivative in $x$ yields the first order differential equation
\[
(1-z)g + (x+\lambda)g_x&= 1 \qquad g(0) = 1.
\]
The solution to this equation when $z\neq 1$ is
\[
g(x) &= \frac{1-z\lt(\frac{x}{\lambda}+1\rt)^{z-1}}{1-z},
\]
and when $z=1$ the solution is
\[
g(x) = 1+\log\lt(\frac{x}{\lambda}+1\rt).
\]
Finally, for uniqueness, 
suppose there were two solutions $G,H$ to the above PDE with symmetry and its boundary condition.
Then if we let $u(x,y) = x+y$, $v(x,y) = x-y$, the reparametrized 
difference function $K(u(x,y), v(x,y)) = G(x,y) - H(x,y)$ on the domain $u\geq 0$, $v\in[-u,u]$ 
satisfies
\[
K_{uu} - K_{vv} + \frac{2(1-z)}{u+\lambda}K_u = 0,\qquad
K(u,v) = K(u, -v),\qquad
K(x,x) = 0.
\]
Consider the energy function
\[
E(u) = \frac{1}{2}\int_{-u}^u K_u^2 + K_v^2 \d v. 
\]
Note that $E(0) = 0$. 
Differentiating in $u$ yields
\[
E'(u) &= \frac{1}{2}\lt(K_u^2+K_v^2\rt)_{v=-u}^{v=u} + \int_{-u}^u K_u K_{uu}\d v + \int_{-u}^u K_v K_{uv}\d v\\
&= \frac{1}{2}\lt(K_u^2+K_v^2\rt)_{v=-u}^{v=u} + \int_{-u}^u K_u (K_{uu}-K_{vv})\d v + \lt(K_uK_v\rt)_{v=-u}^{v=u}\\
&= \frac{1}{2}\lt(K_u^2+K_v^2\rt)_{v=-u}^{v=u}+ \lt(K_uK_v\rt)_{v=-u}^{v=u} - \frac{2(1-z)}{u+\lambda}\int_{-u}^u K^2_u \d v\\
&= - \frac{2(1-z)}{u+\lambda}\int_{-u}^u K^2_u \d v \leq 0,
\]
where the second equation follows by integration by parts, 
the third follows by the PDE, and the last follows by $K(u,v) = K(u,-v)$.
Therefore $E'(u) \leq 0$, $E(0) = 0$, and $E(u) \geq 0$, which implies that $E(u) = 0$ identically.
Therefore $K_u = K_v = 0$ everywhere, which combined with the boundary again implies $K=0$ identically.
Therefore $G=H$ and the solution is unique.
\eprfof

\bprfof{\cref{prop:shrinkcost}}
For $z=1$, \cref{eq:GPDE} becomes
\[
\hes{G}{x}{y} &= 0 \implies G(x,y) = g(x) + h(y).
\]
By symmetry, $h = g$. Given the boundary condition and symmetry,
\[
g(x)= 1+\log\lt(\frac{x}{\lambda}+1\rt) - g(0).
\]
Setting $x=0$ yields
$2g(0) = 1 \implies g(0) = 1/2$.
Therefore
\[
\E[N|\lambda_\ell, \lambda_r] = G(\lambda_\ell,\lambda_r) = 1 + \log\lt(\frac{\lambda_r}{\lambda}+1\rt) + \log\lt(\frac{\lambda_\ell}{\lambda}+1\rt).
\]
In slice sampling, $\lambda_\ell = \ell - \shell$, $\lambda_r = \shr - r$, and $\lambda = r-\ell$.
Therefore
\[
\E[C_{\text{shrink}}|\shr,\shell,u] &= 1 + \log\lt(\frac{\shr-r}{\lambda}+1\rt) + \log\lt(\frac{\ell-\shell}{\lambda}+1\rt).
\]
The result follows by the symmetry of the distributions of $\shr-r$ and $\ell-\shell$.
\eprfof

\bprfof{\cref{prop:stepout}}
Stepping out begins by evaluating the target 
at the initial window boundaries $x-Vw$ and $x+(1-V)w$, and then proceeds
by expanding rightwards for $N_r$ iterations and leftwards for $N_\ell$ iterations.
Conditioned on $u$, $x\dist\Unif[\ell, r]$. Therefore, expressing $x$ as the convex combination $x = Y \ell + (1-Y)r$ for $Y\dist\Unif[0,1]$,
we have that
\[
N_r &= \lt\lceil0\vee \frac{r - (x+(1-V)w)}{w}\rt\rceil = \lt\lceil0\vee Y\frac{\lambda}{w}-(1-V)\rt\rceil\\
N_\ell &= \lt\lceil0\vee\frac{(x-Vw) - \ell}{w}\rt\rceil = \lt\lceil0\vee(1-Y)\frac{\lambda}{w}-V\rt\rceil.
\]
The right exceedance is therefore
\[
\shr - r &= x+(1-V)w + N_rw - r\\
 &= \lt(\lt\lceil\frac{\lambda Y}{w}-(1-V)\rt\rceil-\lt(\frac{\lambda Y}{w}-(1-V)\rt)\rt)w.
\]
For any $a\in\reals$, $V\dist\Unif[0,1]$, $\lceil a+V\rceil -(a+V) \dist \Unif[0,1]$.
Therefore $\shr - r \eqd w Z$ for $Z\dist\Unif[0,1]$.
Next, the cost of stepping out is
\[
C_{\text{slice}} &= 2 + N_r + N_\ell\\
&= 2 + \lt\lceil0\vee Y\frac{\lambda}{w}-(1-V)\rt\rceil + \lt\lceil0\vee(1-Y)\frac{\lambda}{w}-V\rt\rceil\\
&= 1 + \lt\lceil Y\frac{\lambda}{w}+V\rt\rceil + \lt\lceil(1-Y)\frac{\lambda}{w}-V\rt\rceil\\
&= 1 +\lt\lfloor \frac{\lambda}{w}\rt\rfloor+ \lt\lceil Y\frac{\lambda}{w}+V\rt\rceil - \lt\lfloor Y\frac{\lambda}{w}+V\rt\rfloor
+ \lt\lceil \text{frac}\lt(\frac{\lambda}{w}\rt) - \text{frac}\lt(Y\frac{\lambda}{w}+V\rt)\rt\rceil\\
&\eqas 2 +\lt\lfloor \frac{\lambda}{w}\rt\rfloor + \lt\lceil \text{frac}\lt(\frac{\lambda}{w}\rt) - \text{frac}\lt(Y\frac{\lambda}{w}+V\rt)\rt\rceil.
\]
For any fixed $a\in\reals$, $\text{frac}(a+V) \dist\Unif[0,1]$;
therefore
\[
C_{\text{slice}} \eqd 2 +\lt\lfloor \frac{\lambda}{w}\rt\rfloor + B
\qquad B\dist\Bern\lt( \text{frac}\lt(\frac{\lambda}{w}\rt)\rt).
\]
\eprfof

\bprfof{\cref{thm:stepcost}}
Combining the results of \cref{prop:shrinkcost,prop:stepout} yields 
\[
\E[C] &= 3 + \E\lt[\frac{\lambda}{w}\rt] + 2\E \int_0^1 \log\lt(\frac{w v}{\lambda}+1\rt)\d v\\
&= 3 + \E\lt[\frac{\lambda}{w}\rt] + 2\E \frac{\lambda}{w}\int_0^{\frac{w}{\lambda}} \log\lt(s+1\rt)\d s\\
&= 3 + \E\lt[\frac{\lambda}{w}\rt] + 2\E \lt[\frac{\lambda}{w}\lt((1+s)\log(1+s)-(1+s)\rt)_0^{\frac{w}{\lambda}}\rt] \\
&= 1 + \E\lt[\frac{\lambda}{w} + 2\lt(\frac{\lambda}{w}+1\rt)\log\lt(1+\frac{w}{\lambda}\rt)\rt].
\]
This expectation is finite if and only if $\E[\lambda/w] < \infty$ and $\E[-\log(\lambda/w)] < \infty$ by inspection.
To find the height-adaptive optimal initial window size $w^\star$, we take the derivative of the integrand and set it to 0.
\[
0 &= -\frac{\lambda}{w^2} -2\frac{\lambda}{w^2}\log\lt(1+\frac{w}{\lambda}\rt) + 2\frac{\frac{\lambda}{w}+1}{1+\frac{w}{\lambda}} \frac{1}{\lambda}\\
2\frac{w}{\lambda} &= 1 +2\log\lt(1+\frac{w}{\lambda}\rt),
\]
and therefore
$w^\star = \alpha \lambda$, $\alpha = -1 - W_{-1}\lt(-\exp(-3/2)\rt) \approx 1.358$, where $W_{-1}$ is the lower branch of the Lambert $W$ function.
This yields expected cost
\[
\E C &= 1 + \alpha^{-1} + 2\lt(\alpha^{-1}+1\rt)\log\lt(1+\alpha\rt) \approx 4.715.
\]
\eprfof

\blem\label{lem:cacheaccept}
Cached doubling has expected accept cost 
\[
 \E C_{\text{accept}} &=  \E\lt[\frac{c_0}{3} + 0\vee\lt(\frac{c_0}{3}+n_0-1-\frac{1-2^{1-n_0}}{c_0} + \frac{1-4^{1-n_0}}{9c_0^2}\rt)\rt].
\]
\elem
\bprfof{\cref{prop:doubling} and \cref{lem:cacheaccept}}
We prove both \cref{prop:doubling} and \cref{lem:cacheaccept} simultaneously 
to avoid repeating proof techniques across multiple results.
Let $N$ be the number of doubling iterations, such that $C_{\text{slice}} = 2+N$.
Let $K$ be the number of additional evaluations in the call to \texttt{accept} for cached doubling.
Since these two variables are dependent, we analyze them jointly.
Let $Z_i$ be the direction choice Bernoullis drawn as per \cref{alg:doubling}.
Then the event $N \leq n, K\leq k-1$ is equivalent to
the event where the approximate slice covers the exact slice,
\[
x + (1-V) w + \sum_{i=0}^{n-1} w 2^i (1-Z_i) &\geq r \quad \text{and}\quad x - V w - \sum_{i=0}^{n-1} w 2^i Z_i \leq \ell,
\]
and the proposal $x'$ falls in the approximate slice at $k$ iterations,
\[
x + (1-V) w + \sum_{i=0}^{k-1} w 2^i (1-Z_i) &\geq x' \quad \text{and}\quad x - V w - \sum_{i=0}^{k-1} w 2^i Z_i \leq x',
\]
where we use the convention $\sum_{i=0}^{-1} = 0$, 
and $Z_i = 1$ (respectively $0$) denotes a leftward (respectively rightward) doubling iteration $i$.
Rearranging these inequalities, this is equivalent to the event
\[
\frac{\ell-x}{w} + \sum_{i=0}^{n-1}2^iZ_i + V \geq 0 
\quad \frac{x-r}{w} + 2^n - \sum_{i=0}^{n-1}2^iZ_i - V \geq 0\\
\frac{x'-x}{w} + \sum_{i=0}^{k-1}2^iZ_i + V \geq 0 
\quad \frac{x-x'}{w} + 2^k - \sum_{i=0}^{k-1}2^iZ_i - V \geq 0
\]
Note that conditioned on $u,w$, we have that $x,x'\distiid \Unif[\ell, r]$, i.e.,
$x = Y\ell+(1-Y)r$ and $x' = X\ell+(1-X)r$, where $X,Y\distiid \Unif[0,1]$.
Also note that we need only consider the case where $k\leq n$, since the event with $k> n$ is equivalent to the event with $k=n$.
So we can write
\[
\sum_{i=0}^{n-1}2^iZ_i+V = \sum_{i=0}^{k-1}2^iZ_i+V + \sum_{i=k}^{n-1}2^iZ_i  = \sum_{i=0}^{k-1}2^iZ_i+V + 2^k\sum_{i=0}^{n-k-1}2^i Z_{i+k}
\eqd 2^kZ + 2^kW_{n-k},
\]
where $Z\dist\Unif[0,1]$ and $W_j\dist\Unif\{0,1,\dots,2^{j}-1\}$ are independent.
Combining all of these facts, the above event is equivalent to
\[
 \frac{(1-Y)\lambda}{2^kw} \leq Z + W_{n-k}\leq 2^{n-k}-\frac{Y\lambda}{2^kw} \\ 
 \frac{(X-Y)\lambda}{2^kw} \leq Z \leq \frac{(X-Y)\lambda}{2^kw} + 1.
\]
Since we require only the marginal distributions of $N$ and $K$,
we can use continuity of probability to find each event $N\leq n$ and $K\leq k-1$ separately.
In particular, the event $N\leq n$ is equivalent to substituting $k=n$ in the above inequalities, which reduce to
\[
 \frac{(1-Y)\lambda}{2^nw} \leq Z \leq 1-\frac{Y\lambda}{2^nw}.
\]
Therefore
\[
\P(N\leq n | u) = 1 - 1\wedge \frac{\lambda}{2^nw},
\]
and so the conditional PMF of $N$ given $u$ is given by
\[
\P(N=n|u) &= \lt\{\begin{array}{ll}
0 & n < n_0\\
1-\frac{\lambda}{2^nw} & n = n_0\\
\frac{\lambda}{2^nw} & n > n_0
\end{array}\rt. .
\]
In other words, to draw $N$, we first
 draw a Bernoulli with probability $\frac{\lambda}{2^{n_0}w}$; if the result is 0, we set $N=n_0$,
 and otherwise set $N=n_0 + G$, where $G\dist\Geom(1/2)$.

Next, the marginal event $K\leq k-1$ can be obtained by taking the limit $n\to \infty$, yielding
\[
 \frac{(X-Y)\lambda}{2^kw} \leq Z \leq \frac{(X-Y)\lambda}{2^kw} + 1.
\]
Therefore for $k\geq 0$,
\[
\P\lt(K\leq k|u\rt) 
 &= \E\lt[ 0\vee \lt(1 - \frac{|X-Y|\lambda}{2^{k+1}w}\rt) | u\rt].
\]
From here we could characterize the distribution of $K$; however, it takes a somewhat complicated piecewise form.
As we ultimately require only the expected value of $K$, we
use the identity $\E[K|u] = \sum_{k=0}^\infty \P(K> k|u)$, 
\[
\E[K | u] &= \sum_{k=0}^\infty \E\lt[1 \wedge \frac{|X-Y|\lambda}{2^{k+1}w} | u\rt]\\
&= \lt\{\begin{array}{ll}
\frac{\lambda}{3w} & n_0 \leq 1\\
\frac{2}{3}c_0 + n_0-1-\frac{1-2^{1-n_0}}{c_0} + \frac{1-4^{1-n_0}}{9c_0^2} & n_0 > 1
\end{array}\rt.\\
&=\frac{c_0}{3} + 0\vee\lt(\frac{c_0}{3}+n_0-1-\frac{1-2^{1-n_0}}{c_0} + \frac{1-4^{1-n_0}}{9c_0^2}\rt).
\]

Finally, for the right exceedance, let $J_n(r) = \P\lt(N=n, \shr-r\geq r|u\rt)$.
Once again using the fact that $x\dist\Unif[\ell,r]$ given $u$,
we have for $V,Y\distiid \Unif[0,1]$, $Z_i\distiid \Bern(1/2)$, $Z = 2^{-n}\lt(\sum_{i=0}^{n-1}2^iZ_i + V\rt)\dist\Unif[0,1]$,
\[
J_n(r)
&= \P\lt(N=n,Z \leq 1 - \frac{Y\lambda + r}{2^nw} | u\rt)\\
&= \P\lt(N\leq n,Z \leq 1 - \frac{Y\lambda + r}{2^nw} | u\rt)-\P\lt(N\leq n-1,Z \leq 1 - \frac{Y\lambda + r}{2^nw} | u\rt).
\]
For $n\geq 0$, the event $N\leq n$ is equivalent to
\[
\lt(1-Y\rt)\frac{\lambda}{2^nw} \leq Z \leq 1 - Y\frac{\lambda}{2^nw}.
\]
So for the case $n=0$,
\[
J_n(r)
&= \P\lt(\lt(1-Y\rt)\frac{\lambda}{w} \leq Z \leq 1 - \frac{Y\lambda+r}{w}|u\rt) = 0\vee\lt(1-\frac{r+\lambda}{w}|u\rt).
\]
And when $n>0$, if $Z' = 2^{-(n-1)}\lt(\sum_{i=0}^{n-2}2^iZ_i+V\rt) = 2Z-Z_{n-1}$,
\[
J_n(r)
&= \P\lt(\lt(1-Y\rt)\frac{\lambda}{2^nw} \leq Z \leq 1 - \frac{Y\lambda+r}{2^nw}| u\rt)\\
&-\P\lt(\lt(1-Y\rt)\frac{\lambda}{2^{n-1}w} \leq Z' \leq 1 - Y\frac{\lambda}{2^{n-1}w},Z' \leq 2 - Z_{n-1} - \frac{Y\lambda + r}{2^{n-1}w} | u\rt)\\
&= 0\vee\lt(1 - \frac{r+\lambda}{2^nw}\rt) - \E\lt[ 0\vee\lt(1\wedge \lt(2 - Z_{n-1} - \frac{r}{2^{n-1}w}\rt) - \frac{\lambda}{2^{n-1}w}\rt) | u\rt]\\
&= 0\vee\lt(1 - \frac{r+\lambda}{2^nw}\rt) - \frac{1}{2} 0\vee\lt(1\wedge \lt(2 - \frac{r}{2^{n-1}w}\rt) - \frac{\lambda}{2^{n-1}w}\rt)- \frac{1}{2} 0\vee\lt(1 - \frac{r+\lambda}{2^{n-1}w} \rt).
\]
Taking the negative derivative of $J_n(r)$ yields the joint probability mass/density function of $N,r$ given $u$.
When $n=0$,
\[
f_n(r) &= \frac{1}{2^nw}\1[r+\lambda \leq 2^nw],
\]
and when $n>0$,
\[
f_n(r) &= \frac{1}{2^nw}\1[r+\lambda\leq 2^nw, r< 2^{n-1}w]
-\frac{1}{2^nw}\1[r+\lambda\leq 2^{n-1}w]\\
&= \frac{1}{2^nw}\lt\{\begin{array}{ll}
0 & \lambda > 2^nw\\
\1[r \leq 2^nw-\lambda] & 2^{n-1}w < \lambda \leq 2^nw\\
\1[r< 2^{n-1}w]
-\1[r\leq 2^{n-1}w-\lambda]
& \lambda \leq 2^{n-1}w
 \end{array}\rt. .
\]
By inspection, conditioned on $u$, $w$, and $N=n$, 
if $n=0$,
\[
r\dist \Unif[0, w-\lambda],
\]
and if $n>0$,
\[
2^{n-1}w < \lambda \leq 2^nw &\implies r \dist \Unif[0, 2^nw-\lambda]\\
\lambda \leq 2^{n-1}w &\implies r\dist \Unif[2^{n-1}w-\lambda, 2^{n-1}w].
\]
This can be written compactly as
\[
r \dist \lt\{\begin{array}{ll}
\Unif[0, 2^nw-\lambda] & \quad n=n_0\\
\Unif[2^{n-1}w-\lambda, 2^{n-1}w] & \quad n>n_0
\end{array}\rt. .
\]
\eprfof

\bprfof{\cref{thm:doublecost}}
Using the distribution of the number of doubling steps $N$ given $u$ and the right exceedance $\shr-r$ distribution given $u,N$ from \cref{prop:doubling},
\[
&\E\lt[\log\lt(\frac{\shr-r}{\lambda}+1\rt)|u\rt]\\
&= \frac{1}{2^{n_0}w}\int_0^{2^{n_0}w-\lambda} \log\lt(\frac{x}{\lambda}+1\rt)\d x
+ \sum_{n=n_0+1}^\infty \frac{1}{2^nw} \int_{2^{n-1}w-\lambda}^{2^{n-1}w} \log\lt(\frac{x}{\lambda}+1\rt)\d x\\
&= \log(2) c_0 -(1+c_0)\log\lt(c_0\rt) - 1 + \frac{1}{2}\sum_{n=0}^\infty \lt(1+c_02^{-n}\rt)\log\lt(1+c_02^{-n}\rt).
\]
Combined with the other results from \cref{prop:doubling}, we have that
\[
&\E[C|u] = \E\lt[C_{\text{slice}} + C_{\text{accept}}+C_{\text{shrink}}|u\rt]= f\lt(n_0, c_0\rt),
\]
where $f$ is given in \cref{eq:fnc}.
By inspection of this function, the cost is finite if and only if $\E[n(\lambda/w)] < \infty$ and $\E[c(\lambda/w)]<\infty$,
which together are equivalent to $\E|\log(\lambda/w)| < \infty$.
To minimize the expected cost, recall the definition
\[
n_0 = \lt\lceil \log_2\lt(\frac{\lambda}{w}\rt) \rt\rceil\vee 0, \qquad c_0 = \frac{\lambda}{2^{n_0}w}.
\]
For all $k=0,1,\dots$,
 if $2^k w < \lambda \leq 2^{k+1}w$, then $n_0 = k+1$, and $c_0 = \frac{\lambda}{2^{k+1}w}$, so $c_0 \in (1/2, 1]$. 
For all $n_0>0$, the objective function is monotone increasing in $c_0$ for $c_0\in(1/2,1]$, 
and the infimum therefore occurs at the limit as $c_0 \to 1/2$ from the right.
This yields expected cost
\[
\E[C|u] = 2n_0 - 4(1-2^{-n_0}) + 16(1-4^{-n_0})/9 + D, \quad D \approx 6.086.
\]
For $n_0\geq 1$, has positive derivative in $n_0$ so is increasing, so
the minimum is at $n_0=1$ with cost $\approx 7.42$.
On the other hand, if $w \geq \lambda$, then $n_0 = 0$, so
\[
\E[C|u] = f(0,c_0),
\]
which has an $\argmin$ on $(0,1]$ of $c_0 \approx 0.3115$, and hence
\[
w^\star \approx 3.211\lambda,
\]
with cost $\approx 5.901$. This cost is lower than the optimum on the range $n_0\geq 1$, and hence is optimal.
\eprfof

\blem\label{lem:lipschitz}
Fix a value of $u>0$. Let $S(w)$ and $A(w)$ denote the 
conditional expectations of $C_{\text{slice}}$, $C_{\text{accept}}$ given $u$ with 
initial window width $w$ for lazy cached doubling.
For all $w,w' > 0$ such that $|\lambda/w - \lambda/w'| \leq 1$, 
\[
|S(w) - S(w')| \vee |A(w)-A(w')|
&\leq \lt|\frac{\lambda}{w}-\frac{\lambda}{w'}\rt|\lt(10+n\lt(\frac{\lambda}{w}\rt)+n\lt(\frac{\lambda}{w'}\rt)\rt).
\]
\elem
\bprf
Consider running two coupled copies of the slice sampling algorithm
with a shared slice variable $u$. 
The slice width $\lambda$ and the two initial windows $w$ and $w'$ are determined,
and all other variables between the two copies can be coupled 
(the current state $x$, the first accepted proposal $x'$, the window shift $V\dist\Unif[0,1]$, and the doubling directions $Z_i$).
Each algorithm will only evaluate the density on at most the grid of points $x+(1-V+k)w$ and $x+(1-V+k)w'$, respectively, for $k\in\ints$.
Therefore as long as these grids of points have density values that fall above/below the slice value $u$ in the same way,
the algorithms will behave identically, i.e., if the grids satisfy
\[
\forall k\in\ints, \quad \pi(x+(1-V+k)w) \geq u \iff \pi(x+(1-V+k)w') \geq u.
\]
Let $B$ be the event where the grids satisfy the above property,
and write $C_{\text{slice}} = F(x,V,(Z_i)_i, \lambda, w)$ as an explicit deterministic function of its arguments.
Then
\[
&\lt|\E\lt[F(x,V,(Z_i)_i,\lambda,w)|u\rt] 
-\E\lt[F(x,V,(Z_i)_i,\lambda,w')|u\rt] 
\rt|\\
&\leq \E\lt[\1[B^c]\lt|F(x,V,(Z_i)_i,\lambda,w) - F(x,V,(Z_i)_i,\lambda,w')\rt||u\rt]\\ 
&\leq \P(B^c|u)\E\lt[\sup_{x\in[\ell,r],v\in[0,1]} F\lt(x,v,(Z_i), \lambda,w\rt)+ \sup_{x\in[\ell,r],v\in[0,1]}F\lt(x,v,(Z_i), \lambda,w'\rt)|u\rt].
\]
Note that the number of evaluations for lazy doubling is bounded above by $2+N$, where $N$ is the number of doubling rounds that occur.
Regardless of the value of $x$ or $v$, that number is bounded above by $n(\lambda/w)$ plus the time it takes for at least one $Z_i = 0$ and $Z_{i'} = 1$
for $i, i' > n(\lambda/w)$ (at which point the slice is guaranteed to be covered). 
Because the $Z_i\distiid \Bern(1/2)$, this is equivalent to $n(\lambda/w) + 1 + G$, where $G\dist\Geom(1/2)$.
Therefore
\[
&\lt|\E\lt[F(x,V,(Z_i)_i,\lambda,w)|u\rt] 
-\E\lt[F(x,V,(Z_i)_i,\lambda,w')|u\rt] \rt|\\
&\leq \P(B^c|u)\lt(\E\lt[2+n(\lambda/w) + 1 + G|u\rt] + \E\lt[2+n(\lambda/w') + 1 + G | u\rt]\rt)\\
&= \P(B^c|u)\lt(10+n(\lambda/w)+n(\lambda/w')\rt).
\]
The probability that both grids to the right of $x$ have density values that lie above/below the slice value in the same way is
equal to the probability that the index of the first grid point for $w$ beyond the right edge is the same as that for $w'$.
Combined with the fact that $x \eqd Y\ell+(1-Y)r$ given $u$ for $Y\dist\Unif[0,1]$,
\[
&\P\lt(\lt\lceil \frac{r - (x+(1-V)w)}{w}\rt\rceil = \lt\lceil \frac{r - (x+(1-V)w')}{w'}\rt\rceil | u\rt)\\
&=\P\lt(\lt\lceil \frac{r - (Y\ell + (1-Y)r)}{w}-(1-V)\rt\rceil = \lt\lceil \frac{r - (Y\ell+(1-Y)r)}{w'}-(1-V)\rt\rceil | u\rt)\\
&=\P\lt(\lt\lceil Y\frac{\lambda}{w}+V\rt\rceil = \lt\lceil Y\frac{\lambda}{w'}+V\rt\rceil | u\rt),
\]
and if $|\lambda/w - \lambda/w'| \leq 1$, this probability is equal to
\[
\P\lt(\lt\lceil \frac{r - (x+(1-V)w)}{w}\rt\rceil = \lt\lceil \frac{r - (x+(1-V)w')}{w'}\rt\rceil | u\rt)
&= 1 - \frac{\lt|\frac{\lambda}{w} - \frac{\lambda}{w'}\rt|}{2}.
\]
The result follows by symmetry for the left edge and the union bound,
and by repeating the proof identically for $C_{\text{accept}}$.
\eprf
\blem\label{lem:lazyasymptoticslargew}
Fix a value of $u>0$. Let $S(w)$ and $A(w)$ denote the 
conditional expectations of $C_{\text{slice}}$, $C_{\text{accept}}$ given $u$ with 
initial window width $w$ for lazy cached doubling.
For all $w \geq \lambda$,
\[
2 \leq S(w) \leq 2 + 5(\lambda/w), \qquad 0 \leq A(w) \leq 5(\lambda/w).
\]
\elem
\bprf
Fix a slice value $u$. 
If both $x-Vw < \ell$ and $x + (1-V)w > r$, the doubling algorithm requires precisely 2 evaluations. Otherwise,
the number of doubling iterations is bounded above by the waiting time for at least one leftward and one rightward expansion,
at which point doubling will terminate (since $w\geq\lambda$). Furthermore, doubling always requires at least 2 evaluations.
Therefore for $G\dist\Geom(1/2)$, and noting that $x \eqd Y\ell + (1-Y)r$, $Y\dist\Unif[0,1]$ given $u$,
\[
2 \leq \E[C_{\text{slice}}|u] &\leq 2\P\lt(x-Vw < \ell, x+(1-V)w > r | u\rt) \\
&+(2+\E\lt[1+G\rt])\P\lt(x-Vw \geq \ell \text{ or } x+(1-V)w \leq r | u\rt)\\
&\leq 2 + 5\lt(\P\lt(x-Vw \geq \ell|u\rt) + \P\lt(x+(1-V)w \leq r | u\rt)\rt)\\
&= 2 + 10\P\lt(Y\lambda \geq Vw |u\rt)\\
&= 2 + 5\frac{\lambda}{w}.
\]
The same logic applies to the accept algorithm, except that it requires 0 additional evaluations when the initial window exceeds the slice.
Therefore
\[
0 \leq \E[C_{\text{accept}}|u] &\leq 5\frac{\lambda}{w}.
\]
\eprf

\blem\label{lem:asympaccept}
Fix a value of $u>0$. Let $A(w)$ denote the 
conditional expectation of $C_{\text{accept}}$ given $u$ with 
initial window width $w$ for lazy cached doubling.
Then for all $w<\lambda/2$,
\[
\lt| A(w) - (A(2w) + 1/2)\rt| \leq 24w/\lambda.
\]
\elem
\bprf
Fix the slice value $u>0$. 
The accept cost is a function of the original state $x$, proposal $x'$, initial window size $w$,
true slice bounds $\ell, r$, and approximate slice bounds $\shell, \shr$.
If $\min\{|x-x'|, |\ell-x'|, |r-x'|\} > 4w$---in other words, if the proposed new point $x'$ is not close to either slice boundary or the old state $x$---then 
when the halving procedure reaches a slice size of $2w$, the slice is guaranteed to have
precisely one known end density value and one unknown. Because the proposal $x'$ is uniform on $[\ell, r]$ conditioned
on $u,\shell,\shr$---and hence uniform within the remaining halved slice---with probability 1/2 the last halving will move the known edge, with
resulting cost 1, and with probability 1/2 the unknown edge will move with cost 0.
Otherwise, the cost with width $w$ is no less than the cost with $2w$, and no more than 2 additional.
Denote $B$ to be the event where $\min\{|x-x'|, |\ell-x'|, |r-x'|\} > 4w$,
and write $C_{\text{accept}} = F(x,x',\shell,\shr,w)$ as an explicit function of its arguments.
We suppress $\ell,r$ in the notation below because they are constant throughout.
Then for some random variable $K \in [0,2]$,
\[
&\E\lt[F(x,x',\shell,\shr,w)| u,\shell,\shr\rt]\\
&= \E\lt[\1[B]F(x,x',\shell,\shr,w) + \1[B^c]F(x,x',\shell,\shr,w) | u,\shell,\shr \rt]\\
&= \E\lt[\1[B]\lt(1/2 + F(x,x',\shell,\shr,2w)\rt) | u,\shell,\shr\rt] + \E\lt[\1[B^c]\lt(K + F(x,x',\shell,\shr,2w)\rt) | u,\shell,\shr\rt],
\]
and therefore for some $K \in [-1/2, 3/2]$,
\[
&= (1/2) + K\P(B^c|u,\shell,\shr) + \E\lt[F(x,x',\shell,\shr,2w) | u,\shell,\shr\rt].
\]
Crucially, since $2w<\lambda$, $c(\lambda/(2w)) = c(\lambda/w)$ and $n(\lambda/(2w)) = n(\lambda/w)-1$, so
\cref{prop:doubling} states that the distribution of $\shell,\shr$ conditioned on $u$ for initial width $w$
is identical to that with initial width $2w$.
Therefore the law of total expectation can be used on both sides of the equation to reduce it to
\[
\E\lt[F(x,x',\shell,\shr,w)| u\rt]
&= (1/2) + K\P(B^c|u) + \E\lt[F(x,x',\shell,\shr,2w)| u\rt].
\]
By the union bound,
\[
\P(B^c | u) \leq 16w/\lambda.
\]
The result follows.
\eprf

\blem\label{lem:asympslice}
Fix a value of $u>0$. Let $S(w)$ denote the 
conditional expectation of $C_{\text{slice}}$ given $u$ with 
initial window width $w$ for lazy cached doubling.
Then for all $w<\lambda$,
\[
\lt| S(w) - (1+(1/2)(S(2w)-1) + (1/2)S(4w)) \rt| \leq 6\frac{w}{\lambda}\lt(2n\lt(\frac{\lambda}{w}\rt)+12\rt).
\]
\elem
\bprf
Fix the slice value $u>0$
and write $C_{\text{slice}} = F(x, V, (Z_i)_{i=0}^\infty,w)$ as an explicit deterministic function $F$ of its arguments;
we leave $\ell,r$ implicit as they do not change throughout. 
We break the behaviour of the algorithm for the first and second expansion into 4 cases.
If the first and second expansions are both rightward ($Z_0=Z_1=0$),
\[
F(x,V,(Z_i)_{i=0}^\infty,w) &= 1 + \1[x-Vw<\ell] 
+ (F(x,V/2,(Z_i)_{i=1}^\infty,2w) - 1).
\]
Note the $V/2$ in the argument of the slice cost with initial width $2w$; we need to
manipulate this term so that the argument once again has a $\Unif[0,1]$ distribution before proceeding.
To address this, note that we can shift $x\to x+w$ and $V/2 \to (1+V)/2$ without changing the boundaries
of the initial slice, and hence without changing the cost of doubling.
Therefore when $Z_0=Z_1=0$,
\[
&F(x,V,(Z_i)_{i=0}^\infty,w) = 1 + \1[x-Vw<\ell] \\
&+ (1/2)(F(x,V/2,(Z_i)_{i=1}^\infty,2w) - 1)\\
&+ (1/2)(F(x+w,(1+V)/2,(Z_i)_{i=1}^\infty,2w) - 1).
\]
Now take the expectation conditioned on $u$. Note that $x\dist\Unif[\ell,r]$ and $x+w\dist\Unif[\ell+w,r+w]$,
and so the two can be coupled when $x\in[\ell+w, r]$ with probability $1-w/\lambda$.
Furthermore when they are coupled, the $v$-argument takes value $V/2$ or $(1+V)/2$ with even probability;
the marginal distribution of the argument is $\Unif[0,1]$ as desired.
When they are not coupled, the slice cost is bounded above by $n(\lambda/w)$ plus the number of iterations it takes for one additional
leftward and rightward expansion, and bounded below by $0$. Therefore for some value $K\in[-1,1]$,
\[
\label{eq:F1}&\E\lt[\1[Z_0=Z_1=0]F(x,V,(Z_i)_{i=0}^\infty,w)|u\rt] = (K/2)(w/\lambda)\lt(2n\lt(\lambda/w\rt)+6\rt)  \\
& + 1/4 + (1/2)\lt(\E\lt[\1[Z_1=0]F(x,V,(Z_i)_{i=1}^\infty, 2w)|u\rt] - 1/2\rt).
\]
If the first and second expansions are both leftward ($Z_0=Z_1=1$), 
\[
&F(x,V,(Z_i)_{i=0}^\infty,w)\\
&= 1 + \1[x+(1-V)w>r] 
+ (F(x,(1+V)/2,(Z_i)_{i=1}^\infty,2w) - 1).
\]
An identical argument for this case yields, for some $K\in[-1,1]$,
\[
 \label{eq:F2}&\E\lt[\1[Z_0=Z_1=1]F(x,V,(Z_i)_{i=0}^\infty,w)|u\rt]= (K/2)(w/\lambda)\lt(2n\lt(\lambda/w\rt)+6\rt)  \\
& + 1/4 + (1/2)\lt(\E\lt[\1[Z_1=1]F(x,V,(Z_i)_{i=1}^\infty,2w)| u \rt] - 1/2\rt) .
\]
If the first expansion is leftward and the second is rightward ($Z_0=1$, $Z_1=0$),
\[
&F(x,V,(Z_i)_{i=0}^\infty,w)\\
&= \1[x+(1-V)w>r](2 + F(x,(1+V)/2,(Z_i)_{i=1}^\infty,2w))\\
&+ \1[x+(1-V)w\leq r](1+F(x,(1+V)/4,(Z_i)_{i=2}^\infty,4w)).
\]
Note that the
above cost of doubling starting from width $2w$ is at least as much as starting from $4w$,
and at most 2 additional,
\[
F(\dots,4w) \leq F(\dots,2w) \leq F(\dots,4w) + 2,
\]
and so for some $K\in[0,2]$,
\[
F(x,V,(Z_i)_{i=0}^\infty,w)
&= 1 + \1[x+(1-V)w>r](1 + K) + 
 F(x,(1+V)/4,(Z_i)_{i=2}^\infty,4w).
\]
Once again we can handle the fact that $(1+V)/4$ does not marginally have a $\Unif[0,1]$ distribution
by shifting $x$ and $V$ appropriately, noting that this does not move the boundaries of the initial slice
and hence does not change the cost of doubling.
\[
F(x,V,(Z_i)_{i=0}^\infty,w)
&= 1 + \1[x+(1-V)w>r](1 + K)\\
&+(1/4)F(x+w,V/4,(Z_i)_{i=2}^\infty,4w)\\
&+(1/4)F(x,(1+V)/4,(Z_i)_{i=2}^\infty,4w)\\
&+(1/4)F(x-w,(2+V)/4,(Z_i)_{i=2}^\infty,4w)\\
&+(1/4)F(x-2w,(3+V)/4,(Z_i)_{i=2}^\infty,4w).
\]
Once again we can couple $x+w, x, x-w, x-2w$ as long as $x\in[\ell+w, r-2w]$, which occurs
with probability $1-3w/\lambda$. When coupled, the $v$-arguments marginally have the correct $\Unif[0,1]$ distribution.
Following the same logic as before, for some $K\in[-1,1]$,
\[
&\label{eq:F3}\E\lt[\1[Z_0=1,Z_1=0]F(x,V,(Z_i)_{i=0}^\infty,w)|u\rt] = (K/2)(3w/\lambda)\lt(2n\lt(\lambda/w\rt)+12\rt)\\
& + 1/4  + (1/4)\E\lt[F(x,V,(Z_i)_{i=2}^\infty,4w)| u \rt],
\] 
and the same formula holds by symmetry for the case where $Z_0=0, Z_1=1$.
Adding \cref{eq:F1,eq:F2}, and twice \cref{eq:F3} yields, for some $K\in[-1,1]$,
\[
&\E\lt[F(x,V,(Z_i)_{i=0}^\infty,w)|u\rt] = 6K(w/\lambda)\lt(2n\lt(\lambda/w\rt)+12\rt) \\
&+1  + (1/2)\E\lt[F(x,V,(Z_i)_{i=1}^\infty,2w)-1|u\rt] + (1/2)\E\lt[F(x,V,(Z_i)_{i=2}^\infty,4w)|u\rt].
\]
Since $(Z_i)_{i=0}^\infty\eqd (Z_i)_{i=1}^\infty \eqd (Z_i)_{i=2}^\infty$, the result follows.
\eprf

\blem\label{lem:approxrecursion2}
Fix sequences of real values $(\beta_i)_{i=1}^\infty$ and $(x_i)_{i=0}^\infty$.
If 
\[
 \forall k\geq 1, \quad \lt|x_{k} - (1/2 + x_{k-1})\rt| \leq \beta_k,
\]
then for $k\geq 1$,
\[
\lt|x_k - (x_0 + k/2)\rt| \leq \sum_{j=1}^k \beta_j.
\]
\elem
\bprf
Note that both
\[
|x_0 - (x_0 + 0/2)| = 0, \quad \text{and}\quad |x_1 - (x_0+1/2)| \leq \beta_1.
\]
Suppose that for $k \geq 2$ and all $j=1,\dots,k-1$,
\[
\lt|x_j - (x_0 + j/2)\rt| \leq \alpha_j.
\]
Then
\[
&\lt|x_{k} - \lt(k/2 + x_0\rt)\rt|\\
&\leq \lt|x_{k} - 1/2 - x_{k-1}\rt|
+ \lt|1/2 + x_{k-1} - \lt(k/2 + x_0\rt)\rt|\\
&\leq \beta_{k}
+ \lt|1/2 + (k-1)/2 + x_0 - k/2 - x_0\rt| + \alpha_{k-1}\\
&\leq \beta_{k} + \alpha_{k-1}.
\]
Therefore the inductive result holds if
\[
\alpha_{k} = \beta_{k} + \alpha_{k-1} = \sum_{j=1}^k \beta_j.
\]
\eprf

\blem\label{lem:approxrecursion1}
Fix sequences of real values $(\beta_i)_{i=1}^\infty$ and $(x_i)_{i=0}^\infty$.
If
\[
 |x_1 - x_0 - 1/3| \leq \beta_1\quad\text{and}\quad\forall k\geq 2, \quad \lt|x_{k} - (1/2)(1+x_{k-1} + x_{k-2})\rt| \leq \beta_k,
\]
then for $k\geq 1$,
\[
\lt|x_k - (x_0 + k/3)\rt| \leq \frac{2}{3}\sum_{j=1}^k\beta_j\lt(1 - (-1/2)^{k-j+1}\rt).
\]
\elem
\bprf
Note that both
\[
|x_0 - (x_0 + 0/3)| = 0, \quad \text{and}\quad |x_1 - (x_0+1/3)| \leq \beta_1.
\]
Suppose that for $k \geq 2$ and all $j=1,\dots,k-1$,
\[
\lt|x_j - (x_0 + j/3)\rt| \leq \alpha_j.
\]
Then
\[
&\lt|x_{k} - \lt(k/3 + x_0\rt)\rt|\\
&\leq \lt|x_{k} - (1/2)(1+x_{k-1} + x_{k-2})\rt|
+ \lt|(1/2)(1+x_{k-1} + x_{k-2}) - \lt(k/3 + x_0\rt)\rt|\\
&\leq \beta_{k}
+ \lt|(1/2)(1+x_{k-1} + x_{k-2}) - k/3 - x_0\rt|\\
&\leq \beta_{k+1}
+ \lt|(1/2)(1+(x_0+(k-1)/3) + (x_0+(k-2)/3)) - k/3 - x_0\rt|\\
&+ (1/2)\alpha_k + (1/2)\alpha_{k-1}\\
&= \beta_{k+1} + (1/2)(\alpha_k+\alpha_{k-1}).
\]
Therefore the inductive result holds if we set
\[
\alpha_{k+1} = \beta_{k+1} + (1/2)(\alpha_k+\alpha_{k-1}), \qquad \alpha_1 = \beta_1, \quad \alpha_0 = 0.
\]
To solve this recursion for $\alpha_k$, define $d_k = \alpha_k - \alpha_{k-1}$. Then
\[
d_{k+1} &= \alpha_{k+1}-\alpha_k\\
&= \beta_{k+1} + (1/2)(\alpha_{k-1}-\alpha_k)\\
&= -(1/2)d_k + \beta_{k+1}.
\]
Therefore for $k\geq 2$,
\[
d_k = (-1/2)^{k-1}d_1 + \sum_{j=2}^k (-1/2)^{k-j} \beta_j.
\]
Since $\alpha_k = \alpha_0 + \sum_{i=1}^k d_i$, for $k\geq 2$,
\[
\alpha_k &= \alpha_0 + d_1 +  \sum_{i=2}^k\lt( (-1/2)^{i-1}d_1 + \sum_{j=2}^i (-1/2)^{i-j} \beta_j\rt)\\
&= \alpha_0 + (2/3)d_1 +\frac{1}{3}(-1/2)^{k-1}d_1 + \sum_{i=2}^k\sum_{j=2}^i (-1/2)^{i-j} \beta_j\\
&= \alpha_0 + (2/3)d_1 +\frac{1}{3}(-1/2)^{k-1}d_1 + \frac{2}{3}\sum_{j=2}^k\beta_j\lt(1 - (-1/2)^{k-j+1}\rt)\\
&= \alpha_0 + (2/3)(\alpha_1-\alpha_0) +\frac{1}{3}(-1/2)^{k-1}(\alpha_1-\alpha_0) + \frac{2}{3}\sum_{j=2}^k\beta_j\lt(1 - (-1/2)^{k-j+1}\rt).
\]
Substituting $\alpha_0 = 0$, $\alpha_1=\beta_1$,
\[
\alpha_k &= \frac{2}{3}\sum_{j=1}^k\beta_j\lt(1 - (-1/2)^{k-j+1}\rt).
\]
\eprf

\blem\label{lem:lazyasymptoticssmallw}
Fix a value of $u>0$. Let $S(w)$ and $A(w)$ denote the 
conditional expectations of $C_{\text{slice}}$, $C_{\text{accept}}$ given $u$ with 
initial window width $w$ for lazy cached doubling.
For all $w<\lambda/2$ and $k\in\nats\cup\{0\}$,
\[
|A(2^{-k}w) &- \lt(A(w)+k/2\rt)| \leq 24(w/\lambda)\\
|S(2^{-k}w) - \lt(S(w)+k/3\rt)| &\leq 
|S(w/2)-(S(w)+1/3)| + 4\frac{w}{\lambda}\lt(n\lt(\frac{\lambda}{w}\rt)+9\rt).
\]
\elem
\bprf
Let $x_k = A(2^{-k}w, u)$.
Then by \cref{lem:asympaccept}, for all $k\geq 1$,
\[
|x_k - (1/2 + x_{k-1})| \leq 2^{-k}\cdot 24w/\lambda.
\]
By \cref{lem:approxrecursion2} with $\beta_k = 2^{-k}\cdot 24w/\lambda$,
for $k\geq 1$,
\[
\lt| A(2^{-k}w) - (A(w) + k/2)\rt| \leq \sum_{j=1}^k 2^{-k}\cdot 24w/\lambda \leq 24(w/\lambda).
\]
Next let $x_k = S(2^{-k}w)$.
Then by \cref{lem:asympslice}, for all $k\geq 2$,
\[
|x_k - (1/2)(1+x_{k-1}+x_{k-2})| &\leq 2^{-k}\cdot 6 \frac{w}{\lambda}\lt(2n\lt(\frac{\lambda}{2^{-k}w}\rt)+12\rt) \\
&=  2^{-k}\cdot 6\frac{w}{\lambda}\lt(2n\lt(\frac{\lambda}{w}\rt)+2k+12\rt).
\]
By \cref{lem:approxrecursion1} with $\beta_k = 2^{-k}\cdot 6 \frac{w}{\lambda}\lt(2n\lt(\frac{\lambda}{w}\rt)+2k+12\rt)$
for $k\geq 2$,
\[
&\lt|S(2^{-k}w)-(S(w)+k/3)\rt| \leq (2/3)|S(w/2)-S(w)-1/3| \\
&+ \frac{2}{3}\sum_{j=2}^k \lt(2^{-j}\cdot 6 \frac{w}{\lambda}\lt(2n\lt(\frac{\lambda}{w}\rt)+2j+12\rt)\rt)\lt(1-(-1/2)^{k-j+1}\rt)\\
&\leq (2/3)|S(w/2)-S(w)-1/3| + \frac{24}{3}\frac{w}{\lambda}\sum_{j=2}^k 2^{-j} \lt(n\lt(\frac{\lambda}{w}\rt)+j+6\rt)\\
&\leq |S(w/2)-S(w)-1/3| + 4\frac{w}{\lambda}\lt(n\lt(\frac{\lambda}{w}\rt)+9\rt).
\]
\eprf

\bprfof{\cref{thm:lazycost}}
Conditioned on $u$, the exact slice boundaries are $\ell, r$. 
By inspection the algorithm is invariant to shifting by $-\ell$ 
and scaling by $\lambda$, resulting in slice sampling on the exact slice $[0,1]$
with initial width $w/\lambda$.
Therefore, there exists a function $f_{\text{lazy}}:\reals_+\to\reals_+$ such that
\[
\E\lt[C | u\rt] = f_{\text{lazy}}\lt(\lambda/w\rt) \implies \E[C] = \E\lt[\flazy\lt(\frac{\lambda}{w}\rt)\rt].
\]
The formula for the expected cost of shrinkage conditioned on $u$ is (see the earlier proof of \cref{thm:doublecost})
\[
\E[C_{\text{shrink}}|u]
&= -1 + 2\log(2) c_0 - 2(1+c_0)\log\lt(c_0\rt) + \sum_{j=0}^\infty \lt(1+c_02^{-j}\rt)\log\lt(1+c_02^{-j}\rt).
\]
Note that the shrinkage cost above is continuous and almost everywhere differentiable in $\lambda/w$,
with derivative
\[
\der{}{(\lambda/w)}\E[C_{\text{shrink}}|u] &=
2^{-n_0}\lt(-2(1-\log 2+\log c_0+c_0^{-1}) + \sum_{j=0}^\infty 2^{-j}\lt(1+\log(1+c_02^{-j})\rt)\rt).
\]
The absolute value of this expression is bounded above by
\[
\lt|\der{}{(\lambda/w)}\E[C_{\text{shrink}}|u] \rt| &\leq 2^{2-n_0}(c_0^{-1} + 2).
\]
This bound is monotone in $\lambda/w$; therefore for any $w,w'$ 
the derivative over the interval between $\lambda/w$ and $\lambda/w'$ is bounded by the sum of the derivatives
at the endpoints. Combined with the result of \cref{lem:lipschitz}, the function $\flazy$ is locally Lipschitz
continuous: for $x,x'$ such that $|x-x'|\leq 1$,
\[
|\flazy(x)-\flazy(x')| \leq 4\lt|x-x'\rt|\lt(5 +\max_{y\in\{x,x'\}} n(y) + 2^{1-n(y)}(2+1/c(y))\rt).
\]
Next, when $w\geq \lambda$, the shrinkage cost formula above can be bounded above and below by
\[
-1
\leq \E[C_{\text{shrink}}|u] - (-2\log(\lambda/w))&\leq 
4 - 2(\lambda/w)\log\lt(\lambda/w\rt),
\]
due to the fact that $c_0 = \lambda/w$.
Combining with the results of \cref{lem:lazyasymptoticslargew} yields, 
for all $x\leq 1$, 
\[
-1 \leq \flazy(x) - (-2\log x) \leq 6 - 10x - 2 x\log x.
\]
This shows that $\flazy(x) \sim -2\log x$ as $x\to 0$,
and also that $\flazy(x) \geq 7$ for all $x \leq 2^{-4}$.

When $w < \lambda$, the shrinkage cost can be bounded by
\[
1 \leq \E[C_{\text{shrink}}|u]\leq  5
\]
due to the fact that $c_0 \in (1/2,1]$ (see the earlier proof of \cref{thm:doublecost}). 
Because the shrinkage cost is bounded, the results of \cref{lem:lazyasymptoticssmallw} immediately
show that $\flazy(x) \sim (1/2+1/3)\log_2 x = (5/6)\log_2 x$ as $x\to\infty$.
Furthermore combining the above lower bound with the results of \cref{lem:lazyasymptoticssmallw} yields
for all $w < w_0 < \lambda/2$, the total expected cost with initial width $w>0$
has lower bound
\[
\E[C|u] &\geq (5/6)\lfloor \log_2(w_0/w)\rfloor + 1 + A(w_0) + S(w_0)\\
& - 24 w_0/\lambda - |S(w_0/2)-(S(w_0)+1/3)| - 4(w_0/\lambda)(n(\lambda/w_0)+9).
\]
Setting $w_0 = 2^{-6}\lambda$ and noting that $(5/6)\lfloor \log_2(w_0/w)\rfloor \geq 0$,
numerical evaluation yields the lower bound 
\[
\forall w < w_0, \quad \E[C|u] &\geq 6.95.
\]
Numerical evaluation at $w = \lambda$ yields a cost of $5.92$, which is less than the two lower bounds
on the ranges $w < 2^{-6}\lambda$ and $w > 2^4\lambda$. Hence the minimum of $\flazy$
occurs on the interval $[2^{-4}, 2^6]$.
\eprfof

\bprfof{\cref{lem:fapprox}}
The proof is straightforward but involves significant tedious algebra and numerical evaluation,
and so only a sketch is provided here. The $(\epsilon,\delta)$-approximation results follow by bounds on $|f-\shf|$
arising from asymptotic expansions for $x \leq 1/b$ and $x \geq b$ for sufficiently large $b>0$,
bounding the Lipschitz constant on the compact region $[1/b, b]$, and using numerical
simulation at a sufficiently fine grid that the Lipschitz constant bounds imply an overall
upper bound. 
 See \cref{fig:approximation_error} for an illustration
of the fact that the required approximation bounds hold.
Convexity and smoothness for the surrogates are verified by checking nonnegativity and boundedness 
of the second derivative.
\eprfof

\bprfof{\cref{lem:subopt}}
Since $\shh$ is a $(\epsilon,\delta)$-approximation of $h$,
we have that for all $x,z$,
\[
(1-\epsilon)h(x,z) -\delta \leq \shh(x,z) \leq (1+\epsilon)h(x,z) + \delta.
\]
Therefore
\[
\E h(\shx^\star,Z) \leq \frac{\E \shh(\shx^\star,Z) + \delta}{1-\epsilon}
\leq \frac{\inf_x \E \shh(x,Z)+\delta}{1-\epsilon}
\leq \frac{(1+\epsilon)\inf_x \E h(x,Z)+2\delta}{1-\epsilon}.
\]
\eprfof

\bdefn\label{defn:locallysmooth}
A function $g:\reals\to\reals$ is \emph{locally smooth}
if it is differentiable and its derivative is Lipschitz continuous on every compact set.
\edefn

\blem\label{lem:shiftedconvexsmooth}
Let $Z$ be a random variable on $\reals$,
and $g:\reals\to\reals_+$ be twice continuously differentiable (and hence locally smooth), 
nonnegative, and convex.
Suppose further that there exist constants $0 < \alpha, C_0, C_1 < \infty$ such that
for all $x$, 
\[
\E[g(Z-x)] < \infty, \quad\text{and}\quad \sup_{|y|\leq \alpha}\lt|g''(x+y)\rt| \leq C_0+C_1 g(x).
\]
Then $\stg(x) = \E[g(Z-x)]$ is twice continuously differentiable (and hence locally smooth), nonnegative, and convex.
\elem
\bprf
The function $\stg$ is an average of horizontally shifted copies of $g$, which is nonnegative and convex, and therefore $\stg$ is nonnegative and convex.
Let $\mu$ be the distribution of $Z$; then
\[
\der[2]{}{x} \int g(z-x)\mu(\d z) &= \lim_{h\to 0} \int \frac{g(z-(x+h)) - 2g(z-x) + g(z-(x-h))}{h^2} \mu(\d z).
\]
By the mean value theorem, for all $0 < h < \alpha$, there exists some $|y|\leq h$
such that
\[
\lt|
\frac{g(z-(x+h)) - 2g(z-x) + g(z-(x-h))}{h^2}
\rt| = \lt|g''(z-x+y)\rt| \leq C_0+C_1 g(z-x).
\]
By Lebesgue dominated convergence, the integral is twice differentiable and
\[
\stg''(x) = \der[2]{}{x} \int g(z-x)\mu(\d z) &= \int g''(z-x)\mu(\d z) = \E\lt[g''(Z-x)\rt].
\]
Finally, for $|h|<\alpha$,
\[
\lt|\stg''(x+h) - \stg''(x)\rt| &\leq \E\lt[\lt|g''(Z-(x+h)) - g''(Z-x)\rt|\rt] \leq 2C_0 + 2C_1\E[g(Z-x)] < \infty, 
\]
and so again by Lebesgue dominated convergence and the fact that $g''$ is continuous, $\stg''$ is continuous.
\eprf

\bprfof{\cref{thm:tunequality}}
The minimum of
\[
\E \shfstep(\lambda/w) &= \E\lt[3.6 + 1.2\frac{\E\lt[\lambda|u\rt]}{w} - \E\lt[\log\lambda|u\rt] + \log w \rt]
\]
over $w(u)$ can be found by setting the derivative of the integrand in $w$ to 0, resulting in
\[
w(u) = \frac{6}{5}\E\lt[\lambda|u\rt] = \shwstep(u).
\]
By \cref{lem:fapprox,lem:subopt}, this initial window is $(3,0)$-optimal.
Next, the minimum of
\[
\E\shfdouble(\lambda/w)
&= \E\lt[4 + \int_{-\infty}^{0.3w} -2\log\lt(\frac{\lambda}{0.3w}\rt)p(\lambda|u)\d \lambda
+ \int_{0.3w}^\infty \frac{2}{\log 2}\log\lt(\frac{\lambda}{0.3w}\rt)p(\lambda|u) \d\lambda \rt]
\]
over $w(u)$ can again be obtained by finding the stationary point of the integrand.
Let $F(\cdot|u) = \P(\lambda \leq \cdot | u)$.
\[
0 &= 
 2\frac{1}{w}\int_{-\infty}^{0.3w}p(\lambda|u)\d \lambda
-\frac{2}{\log 2} \frac{1}{w}\int_{0.3w}^\infty p(\lambda|u) \d\lambda\\
0 &= 2\P(\lambda \leq 0.3w | u)-\frac{2}{\log 2}\P(\lambda > 0.3w | u)\\
0 &= \lt(2+\frac{2}{\log 2}\rt)F(0.3w|u) - \frac{2}{\log 2}\\
F(0.3w|u) &=\frac{2}{2\log 2+2}\\
w &= \frac{1}{0.3} F^{-1}\lt(\frac{1}{\log 2 + 1}|u\rt) = \shwdouble(u).
\]
Therefore by \cref{lem:fapprox,lem:subopt}, this initial window is $(1,4)$-optimal.
The minimum of 
\[
\E\shflazy(\lambda/w)
&= \E\lt[4 + \int_{-\infty}^{w/3} -2\log\lt(3\frac{\lambda}{w}\rt)p(\lambda|u)\d \lambda
+ \int_{w/3}^\infty \frac{5}{6\log 2}\log\lt(3\frac{\lambda}{w}\rt)p(\lambda|u) \d\lambda \rt]
\]
can be obtained by finding the stationary point of the integrand.
Let $F(\cdot|u) = \P(\lambda \leq \cdot | u)$.
\[
0 &= 
 2\frac{1}{w}\int_{-\infty}^{w/3}p(\lambda|u)\d \lambda
-\frac{5}{6\log 2} \frac{1}{w}\int_{w/3}^\infty p(\lambda|u) \d\lambda\\
0 &= 2\P(\lambda \leq w/3 | u)-\frac{5}{6\log 2}\P(\lambda > w/3 | u)\\
0 &= \lt(2+\frac{5}{6\log 2}\rt)F(w/3|u) - \frac{5}{6\log 2}\\
F(w/3|u) &=\frac{5}{12\log 2 + 5}\\
w &= 3 F^{-1}\lt(\frac{5}{12\log 2 + 5}|u\rt).
\]
Therefore by \cref{lem:fapprox,lem:subopt}, this initial window is $(1,4)$-optimal.
By inspection, for $s\in\{\text{step},\text{cache},\text{lazy}\}$,
\[
\E\stf_s(\lambda/w) &= \E\lt[ \E\lt[ \stf_s\lt(\lambda \exp(-\log w)\rt) | u\rt] \rt]
\]
is minimized by setting
\[
w(u) = \exp\lt(\argmin_{x\in\reals}\E\lt[ \stf_s\lt(\lambda \exp(-x)\rt) | u\rt]\rt) = \stw_s(u).
\]
For $\stfstep$, this initial window is $(1,0)$-optimal by definition since $\stfstep = \fstep$,
and for $\stfdouble,\stflazy$, this initial window is $(1,0.7)$-optimal by \cref{lem:fapprox,lem:subopt}.  
The optimization is tractable if 
\[
\E\lt[ \stf_s\lt(\exp(\log\lambda -x)\rt) | u\rt]
\]
is convex and locally smooth, which both follow for each $\stf_s$ by \cref{lem:fapprox,lem:shiftedconvexsmooth}.
\eprfof

\blem\label{lem:thinning}
Let $(X_t)_{t}$ be a Markov chain on a standard Borel space $\scX$ initialized at $X_0 \dist \mu_0$, 
and $\pi$ be a probability distribution on $\scX$. 
Suppose that there exist $\xi : \nats \to \reals_+$ and $C: \scX \to \reals_+$ with
 $\pi(C) < \infty$ such that
for all $x_0\in\scX$, $t\in\nats$,
$\TV(P^t_{x_0}, \pi) \leq C(x_0)\xi(t)$, where $P^t_x$ denotes the $t$-step marginal distribution of the chain
initialized at $x\in\scX$.
Then for all $k,n\in\nats$,
\[
\TV\lt(\mu_{n,k}, \pi^{\otimes(n+1)}\rt) \leq \TV(\mu_0, \pi) + \pi(C)\xi(k)n.
\]
where $X_0, X_{k}, X_{2k},\dots, X_{nk}\dist \mu_{n,k}$ and $\pi^{\otimes(n+1)}$ denotes $n+1$ independent copies of $\pi$.
\elem
\bprf
Since $\scX$ is standard Borel, $\mu_{n,k}$ can be disintegrated
\[
\TV &= \frac{1}{2}\int \lt| \mu_{n,k}(\d x_{0,\dots,nk}) - \pi^{\otimes(n+1)}(\d x_{0,\dots,nk})\rt|\\
&= \frac{1}{2}\int \lt| p(\d x_{nk} | x_{(n-1)k}) \mu_{n-1,k}(\d x_{0,\dots,(n-1)k}) - \pi(\d x_{nk}) \pi^{\otimes n}(\d x_{0,\dots,(n-1)k})\rt|.
\]
By the triangle inequality,
\[
\TV
&\leq 
\frac{1}{2}\int \lt| p(\d x_{nk} | x_{(n-1)k}) - \pi(\d x_{nk})\rt| \pi^{\otimes n}(\d x_{0,\dots,(n-1)k})\\
&+ \frac{1}{2}\int p(\d x_{nk} | x_{(n-1)k}) \lt|\mu_{n-1,k}(\d x_{0,\dots,(n-1)k} - \pi^{\otimes n}(\d x_{0,\dots,(n-1)k})\rt|\\
&\leq 
\xi(k) \pi(C)+ \TV\lt(\mu_{n-1,k}, \pi^{\otimes n}\rt).
\]
Continuing this process recursively yields the result.
\eprf

\blem\label{lem:quantileapprox}
Suppose $(X_n)_{n=1}^N$ are random variables drawn \iid from a distribution on $\reals$ with quantile function 
\[
q:[0,1]\to\reals, \quad q(p) = \inf \lt\{x \in \reals : \P(X\leq x) \geq p\rt\}.
\]
Let $(\stX_n)_{n=1}^N$ be the order statistics of $(X_n)_{n=1}^N$.
Then for all $j,k,\ell\in\nats$, $0<j<k<\ell<N$,
\[
\P\lt( q(j/N) \leq \stX_{k} \leq q(\ell/N)\rt) &\geq 1- 
\lt(e^{-2\frac{(\ell-k)^2}{N}} + e^{-2\frac{(k-j)^2}{N}}\rt).
\]
\elem
\bprf
\[
\P\lt( \stX_k > q(\ell/N)\text{ or }\stX_k < q(j/N)\rt)
&= \P\lt(\stX_{k} > q(\ell/N)\rt) + \P\lt(\stX_{k} < q(j/N) \rt).
\]
The probability that $\stX_{k} > q(\ell/N)$ is equal to the probability
that at most $k-1$ draws are less than or equal to $q(\ell/N)$.
Similarly, the probability that $\stX_{k} < q(j/N)$
is equal to the probability that at least $k$ draws are strictly less than $q(j/N)$.
Therefore
\[
\P\lt( \stX_k > q(\ell/N)\text{ or }\stX_k < q(j/N)\rt)
&= \P(B \leq k-1) + \P(B' \geq k)\\
&= \P(B \leq k-1) + \P(N-B' \leq N-k),
\]
where
\[
B \dist \Binom\lt(N, \P\lt(X_1 \leq q(\ell/N)\rt)\rt)\qquad
N - B' \dist \Binom\lt(N, \P\lt(X_1 \geq q(j/N)\rt)\rt).
\]
The Hoeffding inequality asserts that for $Z\dist\Binom(N,p)$, $z \leq Np$,
\[
\P(Z\leq z) \leq e^{-2N\lt(p-z/N\rt)^2}.
\]
By the definition of the quantile function,
\[
N\P(X_1\leq q(\ell/N)) \geq \ell > k-1,
\]
and
\[
N\P(X_1 < q(j/N)) = \lim_{\epsilon\downarrow 0}N\P(X_1 \leq q(j/N)-\epsilon) \leq j < k,
\]
verifying that both of the above binomial tail probabilities satisfy $z \leq Np$. Therefore
\[
&\P\lt( \stX_k > q(\ell/N)\text{ or }\stX_k < q(j/N)\rt)\\
&\leq e^{-2N\lt(\P\lt(X_1 \leq q(\ell/N)\rt) - (k-1)/N\rt)^2} + e^{-2N \lt(\P\lt(X_1 \geq q(j/N)\rt) - (N-k)/N\rt)^2}\\
&\leq e^{-2N\lt(\ell/N - (k-1)/N\rt)^2} + e^{-2N \lt(k/N -j/N\rt)^2}\\
&= e^{-\frac{2(\ell-k)^2}{N}} + e^{-\frac{2(k-j)^2}{N}}.
\]
\eprf

\blem\label{lem:uniformconditionalapprox}
Suppose $(u_n, \lambda_n)_{n=1}^N$ are drawn \iid from a distribution on $\reals^2$.
Let $(\stu_n, \tilde\lambda_n)_{n=1}^N$ be the same draws reordered so that $(\stu_n)_{n=1}^N$ are the order statistics of $(u_n)_{n=1}^N$.
Fix  $j,k\in\nats$, $1\leq j\leq k < N$, $k-j > 1$ and
functions $h, M:\reals \to \reals_+$ such that
\[
\forall 0\leq s \leq \frac{1}{M(u)}, \quad \E\lt[ e^{s(h(\lambda)-\E[h(\lambda)|u])} | u\rt] \leq e^{\frac{s^2M(u)}{2}}.
\] 
Finally, for all $a,b\in\reals$, $a\leq b$, and $\Delta \geq 0$, let
\[
L(a,b,\Delta) &= \sup_{\begin{subarray}{c}a\leq u_1,u_2\leq b\\ |u_1-u_2|\leq \Delta\end{subarray}} 
\lt|\E\lt[h(\lambda) | u=u_1\rt] - \E\lt[h(\lambda) | u=u_2\rt]\rt| \qquad M(a,b) = \sup_{a\leq u\leq b} M(u).
\]
Then conditioned on $\stu_j,\stu_k$, for all $0<\epsilon \leq 1$,
\[
\sup_{\stu_{j} \leq u\leq \stu_{k}} \lt| \E\lt[h\lt(\lambda\rt) | u\rt] - \frac{1}{k-j-1}\sum_{i=j+1}^{k-1}h\lt(\tilde\lambda_i\rt)\rt|  > \epsilon
\]
with probability at most
\[
2\exp\lt(-\frac{(k-j-1)((\epsilon-L(\stu_j ,\stu_k,\stu_k-\stu_j))\vee 0)^2}{2M(\stu_j,\stu_k)}\rt).
\]
\elem
\bprf
Conditioned on the order statistics $\stu_j, \stu_k$, any symmetric function of $(\stu_i,\tilde\lambda_i)_{i=j+1}^{k-1}$
is equal in distribution to that same function of independent copies
$(u'_i, \lambda'_i)_{i=1}^{k-j-1}$ conditioned on the event that for each $i$, $u'_i \in (\stu_j, \stu_k)$.
Therefore
\[
&\P\lt(
\sup_{\stu_{j} \leq u\leq \stu_{k}}
\lt| \E\lt[h\lt(\lambda\rt) | u\rt] - \frac{1}{k-j-1}\sum_{i=j+1}^{k-1}h\lt(\tilde\lambda_i\rt)\rt| > \epsilon \m| \stu_j, \stu_k \rt)\\
=&\P\lt(
\sup_{\stu_{j} \leq u\leq \stu_{k}}
\lt| \E\lt[h\lt(\lambda\rt) | u\rt] - \frac{1}{k-j-1}\sum_{i=1}^{k-j-1}h\lt(\lambda'_i\rt) \rt|>\epsilon \m| \stu_j,\stu_k,\,\,\forall i,\,\,u'_i \in (\stu_j, \stu_k) \rt)\\
\leq&\P\lt(\lt| \E\lt[h(\lambda)|u\in(\stu_j,\stu_k)\rt] - \frac{1}{k-j-1}\sum_{i=1}^{k-j-1}h\lt(\lambda'_i\rt) \rt| > \epsilon - L(\stu_j,\stu_k,\stu_k-\stu_j)\m| \dots\rt)\\
\leq& 2\exp\lt(-\frac{(k-j-1)((\epsilon-L(\stu_j,\stu_k,\stu_k-\stu_j))\vee 0)^2}{2M(\stu_j,\stu_k)}\rt),
\]
where the last inequality follows because $\epsilon - L(\stu_j,\stu_k,\stu_k-\stu_j) \leq 1$.
\eprf

\bdefn\label{def:localunifsubexp}
A random variable $X\in\reals$ is \emph{locally uniformly subexponential conditioned on $Y\in\reals$}
if there exists a locally bounded function $M:\reals \to \reals_+$ such that
\[
\forall\, 0\leq s \leq \frac{1}{M(Y)}, \quad \E\lt[ e^{s(X-\E[X|Y])} | Y\rt] \leq e^{\frac{s^2M(Y)}{2}}.
\] 
\edefn
\bprfof{\cref{lem:tuningconvergence}}
Since the goal is to demonstrate convergence in probability of some function of $(\stu_j,\tilde\lambda_j)_{j=1}^\tau$ as $\tau\to\infty$,
we can replace the distribution of $(\stu_j, \tilde\lambda_j)_{j=1}^\tau$ in the analysis with any other distribution
that converges to it in total variation. In particular, by geometric ergodicity, \cref{lem:thinning},
the fact that for any $\xi\in[0,1)$ and $k,n$ defined in \cref{eq:taukn},  $n \xi^k\to 0$ as $\tau\to\infty$,
and the fact that we only use the latter half of draws from iterations $t,\dots, 2t$,
throughout this proof we can treat the distribution of $(\stu_j,\tilde\lambda_j)_{j=1}^\tau$ as if these are 
$u$-ordered draws produced \iid from the target, rather than from a Markov chain.

We begin by obtaining a lower bound on the cost:
\[
\E\lt[ h\lt(\frac{\lambda}{w_t(u)}\rt) | (\stu_j,\tilde\lambda_j)_{j=1}^{\tau}\rt]
&= \E\lt[ \E\lt[ h\lt(\frac{\lambda}{w_t(u)}\rt) | u, (\stu_j,\tilde\lambda_j)_{j=1}^{\tau}\rt] | (\stu_j,\tilde\lambda_j)_{j=1}^{\tau}\rt]\\
&\geq \E\lt[ \min_{w\in\scW} \E\lt[ h\lt(\frac{\lambda}{w}\rt) | u, (\stu_j,\tilde\lambda_j)_{j=1}^{\tau}\rt] | (\stu_j,\tilde\lambda_j)_{j=1}^{\tau}\rt]\\
&= \E\lt[ \min_{w\in\scW} \E\lt[ h\lt(\frac{\lambda}{w}\rt) | u\rt] \rt].
\]
Next, we obtain an upper bound.
Let $A_\tau$ be the event where $a_\tau \leq u \leq b_\tau$, where $(a_\tau, b_\tau)$ are 
any deterministic sequence of intervals such that $\P(A_\tau)\to 1$ as $\tau\to\infty$.
\[
&\E\lt[ h\lt(\frac{\lambda}{w_t(u)}\rt) | (\stu_j,\tilde\lambda_j)_{j=1}^{\tau}\rt]\\
&=\E\lt[\1_{A_\tau}(u) h\lt(\frac{\lambda}{w_t(u)}\rt) | (\stu_j,\tilde\lambda_j)_{j=1}^{\tau}\rt]
+\E\lt[\1_{A_\tau^c}(u) h\lt(\frac{\lambda}{w_t(u)}\rt) | (\stu_j,\tilde\lambda_j)_{j=1}^{\tau}\rt].
\]
The second term vanishes because 
$\P\lt(A_\tau\rt) \to 1$ as $\tau\to\infty$ and
$h(\lambda/w_t(u))$ is guaranteed to be uniformly tight 
by the finiteness of $\scW$,
\[
\E\lt[\1_{A_\tau^c}(u) h\lt(\frac{\lambda}{w_t(u)}\rt) | (\stu_j,\tilde\lambda_j)_{j=1}^{\tau}\rt]
&\leq \sum_{w\in\scW}\E\lt[\1_{A_\tau^c}(u) h\lt(\frac{\lambda}{w}\rt)\rt] \to 0.
\]
For the first term, suppose for now that for some $0 < \epsilon < \infty$,
\[
\sup_{a_\tau \leq u \leq b_\tau, w\in\scW} \lt|\E\lt[h\lt(\frac{\lambda}{w}\rt)\m|u\rt] - \shmu\lt(h\lt(\frac{\lambda}{w}\rt), u\rt) \rt| \leq \epsilon.\label{eq:supassump}
\]
Then
\[
&\E\lt[\1_{A_\tau}(u) h\lt(\frac{\lambda}{w_t(u)}\rt) | (\stu_j,\tilde\lambda_j)_{j=1}^{\tau}\rt]\\
&= \E\lt[\1_{A_\tau}(u) \E\lt[ h\lt(\frac{\lambda}{w_t(u)}\rt) | u, (\stu_j,\tilde\lambda_j)_{j=1}^{\tau}\rt] | (\stu_j,\tilde\lambda_j)_{j=1}^{\tau}\rt]\\
&\leq \epsilon + \E\lt[ \1_{A_\tau}(u)\shmu\lt(h\lt(\frac{\lambda}{w_t(u)}\rt), u\rt) | (\stu_j,\tilde\lambda_j)_{j=1}^{\tau}\rt]\\
&\leq \epsilon + \E\lt[ \1_{A_\tau}(u)\shmu\lt(h\lt(\frac{\lambda}{w^\star(u)}\rt), u\rt) | (\stu_j,\tilde\lambda_j)_{j=1}^{\tau}\rt]\\
&\leq 2\epsilon + \E\lt[ \1_{A_\tau}(u)\E\lt[ h\lt(\frac{\lambda}{w^\star(u)}\rt) | u, (\stu_j,\tilde\lambda_j)_{j=1}^{\tau}\rt]| (\stu_j,\tilde\lambda_j)_{j=1}^{\tau}\rt]\\
&= 2\epsilon + \E\lt[ \1_{A_\tau}(u)\min_{w\in\scW} \E\lt[ h\lt(\frac{\lambda}{w}\rt) | u, (\stu_j,\tilde\lambda_j)_{j=1}^{\tau}\rt]| (\stu_j,\tilde\lambda_j)_{j=1}^{\tau}\rt]\\
&= 2\epsilon + \E\lt[ \1_{A_\tau}(u)\min_{w\in\scW} \E\lt[ h\lt(\frac{\lambda}{w}\rt) | u\rt]\rt]\\
&\leq 2\epsilon + \E\lt[\min_{w\in\scW} \E\lt[ h\lt(\frac{\lambda}{w}\rt) | u\rt]\rt].
\]
Therefore if we can show
that for all sufficiently small $\epsilon > 0$, \cref{eq:supassump} holds with probability increasing to 1,
the result follows. By the union bound,
\[
&\P\lt(\sup_{a_\tau \leq u \leq b_\tau, w\in\scW} \lt|\E\lt[h\lt(\frac{\lambda}{w}\rt)\m|u\rt] - \shmu\lt(h\lt(\frac{\lambda}{w}\rt), u\rt) \rt| > \epsilon\rt)\\
&\leq\sum_{w\in\scW}\sum_{j=0}^{n}\P\lt(\sup_{\begin{subarray}{c}\stu_{jk} \leq u \leq \stu_{(j+1)k}\\ a_\tau\leq u\leq b_\tau\end{subarray}} \lt|\E\lt[h\lt(\frac{\lambda}{w}\rt)\m|u\rt] - \shmu_j\lt(h\lt(\frac{\lambda}{w}\rt)\rt) \rt|>\epsilon\rt)\\
&\leq\sum_{w\in\scW}\sum_{j=0}^{n}\P\lt(\stu_{jk}\leq b_\tau, \stu_{(j+1)k}\geq a_\tau, \sup_{\stu_{jk} \leq u \leq \stu_{(j+1)k}} \lt|\E\lt[h\lt(\frac{\lambda}{w}\rt)\m|u\rt] - \shmu_j\lt(h\lt(\frac{\lambda}{w}\rt)\rt) \rt|>\epsilon\rt).
\]
By \cref{lem:uniformconditionalapprox}, for $0<\epsilon\leq1$,
\[
&\leq2\sum_{w\in\scW}\sum_{j=0}^{n}\E\lt[\1\lt[\stu_{jk}\leq b_\tau, \stu_{(j+1)k}\geq a_\tau\rt]e^{-\frac{(k-1)((\epsilon-L_w(\stu_{jk},\stu_{(j+1)k},\stu_{(j+1)k}-\stu_{jk}))\vee 0)^2}{2M_w(\stu_{jk},\stu_{(j+1)k})}}\rt],
\]
where the $w$ subscript in $L,M$ from \cref{lem:uniformconditionalapprox} indicate that there will generally be different functions $L,M$ for each value of $w$.
Let $B_\tau$ be the event where
\[
\stu_k \leq q(2k/\tau) \quad \stu_{nk} \geq q((n-1)k/\tau),
\]
and
for each $j\in\{2, \dots, n-1\}$, 
\[
q((j-1)k/\tau)\leq \stu_{jk} \leq q((j+1)k/\tau).
\]
By \cref{lem:quantileapprox}, $\P(B_\tau) \geq 1 - 2ne^{-2k^2/\tau}$. 
Recall that $k = \lfloor \tau^\beta\rfloor$ and $n = \lfloor \tau/k\rfloor$; as long as $\beta \in (0.5, 1)$, 
$\P(B_\tau) \to 1$ as $\tau \to \infty$.
On the event $B_\tau \cap \{\stu_{jk}\leq b_\tau, \stu_{(j+1)k}\geq a_\tau\}$, 
\[
|\stu_{(j+1)k} - \stu_{jk}| &\leq |q((j-1)k/\tau) - q((j+2)k/\tau)|\\
b_\tau \geq \stu_{jk} \geq \stu_{(j+1)k} - (\stu_{(j+1)k}-\stu_{jk}) &\geq a_\tau - |q(j+2)k/\tau) - q((j-1)k/\tau)|\\
a_\tau \leq \stu_{(j+1)k} \leq \stu_{jk} + (\stu_{(j+1)k}-\stu_{jk}) &\leq b_\tau + |q(j+2)k/\tau) - q((j-1)k/\tau)|.
\]
Since the target has contiguous slices, $q$ is continuous,
and by assumption $\E[h(\lambda)|u]$ is continuous, and so both are uniformly continuous on compacta.
Using these results above, there exists a sequence of $(a_\tau, b_\tau)$ expanding slowly enough that
on the event $B_\tau \cap \{\stu_{jk}\leq b_\tau, \stu_{(j+1)k}\geq a_\tau\}$, for some $\xi\in[0,1)$,
as $t\to\infty$
\[
\max_{w\in\scW} L_w(\stu_{jk}, \stu_{(j+1)k}, \stu_{(j+1)k}-\stu_{jk}) &\to 0\\
 \max_{w\in\scW} M_w(\stu_{jk}, \stu_{(j+1)k}) = O(k^\xi).
\]
Therefore
\[
&\P\lt(\sup_{a_\tau \leq u \leq b_\tau, w\in\scW} \lt|\E\lt[h\lt(\frac{\lambda}{w}\rt)\m|u\rt] - \shmu\lt(h\lt(\frac{\lambda}{w}\rt), u\rt) \rt| > \epsilon\rt)\\
&= O\lt(2|\scW| n e^{-k^{1-\xi}\epsilon/2}\rt) = o(1).
\]
 The result follows.
\eprfof

%% file: pseudocode.tex
\section{Pseudocode}\label{sec:pseudocode}

\balg[h]
\caption{\texttt{SliceSampleStep}: One slice sampling transition}\label{alg:slicesampling}
\balgc[1]
\Require state $x$, density value $\pi_x$, initial width function $w$
\LineComment Draw slice value $u$, direction $\rho$
\State $u \dist \Unif[0, \pi_x]$
\State $\rho \dist m(\d\rho; u)$
\LineComment Initialize inner slice end bounds $b_\ell, a_r$, and outer slice value cache $\scL$
\State $b_\ell \gets 0$
\State $a_r \gets 0$
\State $\scL \gets [\,]$
\LineComment If tuning $w(\cdot)$, in all subsequent code within this function let $\pi(y)$ for $y\in\reals$ represent calling \texttt{Eval}$(y,x,\rho,u,b_\ell,a_r,\scL)$ and 
keeping track of the values of $b_\ell,a_r,\scL$ across calls. This is left implicit to keep the pseudocode clear.
If not tuning $w(\cdot)$, then $\pi(y)$ for $y\in\reals$ simply represents evaluating $\pi(x+\rho y)$, and the call to \texttt{GetBounds} below can be removed.
\State $\shell, \shr, \pi_\ell, \pi_r, \scC \gets$ (\texttt{SteppingOut} or \texttt{Doubling} or \texttt{LazyDoubling})$(u, w(u))$
\State $y, \pi_y \gets$ \texttt{Shrinkage}$(u, \shell, \shr, \pi_\ell, \pi_r, \scC, w(u))$
\State $a_\ell, b_\ell, a_r, b_r \gets$\texttt{GetBounds}$(b_\ell, a_r, \scL)$
\State \Return $x + \rho y, \pi_y, u, \rho a_\ell, \rho b_\ell, \rho a_r, \rho b_r$
\State
\ealgc
\ealg

\balg[h]
\caption{\texttt{Eval}: evaluation of slice density with slice bounds tracking}\label{alg:eval}
\balgc[1]
\Require query $y$, original state $x$, direction $\rho$, slice value $u$, inner slice bounds $b_\ell \leq a_r$, outer bounds cache $\scL$
\LineComment Evaluate the density
\State $\pi_y = \pi(x + \rho y)$
\LineComment The remainder of the code keeps track of the best available upper/lower bounds on slice edges
\If{$\pi_y \geq u$}
	\LineComment If the draw is in the slice, move the inner bounds
	\State $b_\ell \gets b_\ell \wedge y$
	\State $a_r \gets a_r \vee y$
\Else
	\LineComment If the draw is beyond the slice, just store it for now
	\State \texttt{append}$(\scL, y)$
\EndIf
\State \Return $\pi_y, b_\ell, a_r, \scL$
\ealgc
\ealg

\balg[h]
\caption{\texttt{GetBounds}: evaluate the final upper/lower bounds on slice edges}\label{alg:getbounds}
\balgc[1]
\Require inner slice bounds $b_\ell \leq a_r$, outer bounds cache $\scL$
\State $a_\ell \gets \max\{y \in \scL : y \leq b_\ell\}$
\State $b_r \gets \min\{y \in \scL : y \geq a_r\}$
\State \Return $a_\ell, b_\ell, a_r, b_r$
\ealgc
\ealg

\balg[h]
\caption{\texttt{SteppingOut}: approximate slice finding via stepping out}\label{alg:steppingout}
\balgc[1]
\Require slice variable $u$, window size $w$
\LineComment Set the initial window
\State $V \dist \Unif[0,1]$
\State $\shell \gets - V w$
\State $\shr \gets (1-V)w$
\LineComment Expand window leftward 
\While{$\pi(\shell) \geq u$} 
\State $\shell \gets \shell - w$ 
\EndWhile
\LineComment Expand window rightward 
\While{$\pi(\shr) \geq u$} 
\State $\shr \gets \shr + w$ 
\EndWhile
\State \Return $\shell, \shr, 0, 0, [\,]$
\ealgc
\ealg

\balg[h]
\caption{\texttt{Doubling}: approximate slice finding via cached doubling}\label{alg:doubling}
\balgc[1]
\Require slice variable $u$, window size $w$
\LineComment Set the initial window
\State $V \dist \Unif[0,1]$
\State $\shell \gets - V w$
\State $\shr \gets (1-V)w$
\LineComment Initialize cache, density values, and slice edge bounds
\State $i \gets 0$
\State $\scC \gets [\, ]$
\State $\pi_\ell \gets \pi(\shell)$
\State $\pi_r \gets \pi(\shr)$
\While{$\pi_\ell \geq u$ or $\pi_r \geq u$}
\State $Z_i \dist \Bern(0.5)$
\If{$Z_i = 1$} 
\LineComment Leftward expansion. Cache the previous density value and expand the window
\State \texttt{append}$(\scC, \pi_\ell)$
\State $\shell \gets \shell - (\shr-\shell)$
\State $\pi_\ell \gets \pi(\shell)$
\Else
\LineComment Rightward expansion. Cache the previous density value and expand the window
\State \texttt{append}$(\scC, \pi_r)$
\State $\shr \gets \shr + (\shr-\shell)$
\State $\pi_r \gets \pi(\shr)$
\EndIf
\State $i\gets i+1$
\EndWhile
\State \Return $\shell, \shr, \pi_\ell, \pi_r, \scC$
\ealgc
\ealg

\balg[h]
\caption{\texttt{Shrinkage}: shrinkage algorithm for drawing the next state}\label{alg:shrinkage}
\balgc[1]
\Require slice variable $u$, approximate slice $\shell, \shr$, density values $\pi_\ell, \pi_r$, cache $\scC$, window size $w$
\State $\ell\gets \shell$
\State $r\gets \shr$
\While{true}
\State $V \dist \Unif[0,1]$
\State $y \gets (1-V)\ell + Vr$
\State  $\pi_y \gets \pi(y)$ 
\If{$\pi_y \geq u$ and (\texttt{Accept} or \texttt{LazyAccept})$(u, y, \shell, \shr, \pi_\ell, \pi_r, \scC, w)$}
\State\Return $y, \pi_y$
\ElsIf{$y < 0$}
 \State $\ell \gets y$
\Else
\State $r \gets y$
\EndIf
\EndWhile
\ealgc
\ealg

\algnewcommand\algorithmiccontinue{\textbf{continue}}
\algnewcommand\Continue{\State \algorithmiccontinue}

\balg[h]
\caption{\texttt{Accept}: accept algorithm to check proposal validity}\label{alg:accept}
\balgc[1]
\Require slice variable $u$, proposal $y$, approximate slice $\shell, \shr$, density values $\pi_\ell,\pi_r$, cache $\scC$, window size $w > 0$
\If{\texttt{SteppingOut}} 
\State \Return\texttt{true}
\EndIf
\State $i \gets $\texttt{length}$(\scC)$
\While{$i>0$}
\State $m\gets (\shr+\shell)/2$
\State $\pi' \gets \scC[i]$
\State $i\gets i-1$
\LineComment{Move the appropriate boundary and update the density value using the cache if possible}
\If{$y < m$}
\If{$\shell \leq 0 \leq \shr$}
\State $\shr \gets m$
\State $\pi_r \gets \pi'$
\Else
\State $\shr \gets m$
\State $\pi_r \gets\pi(\shr)$
\EndIf
\Else
\If{$\shell \leq 0 \leq \shr$}
\State $\shell \gets m$
\State $\pi_\ell \gets \pi'$
\Else
\State $\shell \gets m$
\State $\pi_\ell \gets\pi(\shell)$
\EndIf
\EndIf
\If{$\pi_\ell < u$ and $\pi_r < u$}
\State \Return \texttt{false}
\EndIf
\EndWhile
\State \Return \texttt{true}
\ealgc
\ealg

\balg[h]
\caption{\texttt{TunedSliceSampling}}\label{alg:tunedslicing}
\balgc[1]
\Require state initialization $x_0$, number of draws $T$
\State $\pi_0 \gets \pi(x_0)$
\State $w(\cdot) \gets 1$ (the constant function with value 1)
\For{$t\in\{1,2,\dots,2T\}$}
\State $x_t, \pi_t, u_t, a_{\ell t}, b_{\ell t}, a_{r t}, b_{r t} \gets$\texttt{SliceSampleStep}$(x_{t-1},\pi_{t-1},w(\cdot))$
\If{\texttt{ispow2}$(t)$}
\State $w(\cdot) \gets $\texttt{Tune}$\lt(\lt(u_j, \lambda_j=\frac{b_{rj} + a_{rj} - b_{\ell j}-a_{\ell j}}{2}\rt)_{j=\lfloor t/2\rfloor +1}^t\rt)$ (see \cref{sec:tuning})
\EndIf
\EndFor
\State \Return $(x_t)_{t=T+1}^{2T}$
\ealgc
\ealg

\balg[h]
\caption{\texttt{LazyDoubling}: approximate slice finding via lazy cached doubling}\label{alg:lazydoubling}
\balgc[1]
\Require slice variable $u$, window size $w$
\LineComment Set the initial window
\State $V \dist \Unif[0,1]$
\State $\shell \gets - V w$
\State $\shr \gets (1-V)w$
\LineComment Initialize cache, density values, and slice edge bounds
\State $i \gets 0$
\State $\scC \gets [\, ]$
\State $\pi_\ell \gets$\texttt{null}
\State $\pi_r \gets$\texttt{null}
\While{\texttt{true}}
\State $Z_i \dist \Bern(0.5)$
\LineComment Check if we should continue doubling with lazy evaluation. 
\LineComment{Note: in the below checks, $a<b$ is assumed to return \texttt{false} if either $a$ or $b$ are \texttt{null}}
\If{$\pi_\ell=$\texttt{null} and $\pi_r=$\texttt{null}}
\LineComment Evaluate the end opposite to the one that is about to move
\If{$Z_i=1$}
\State $\pi_r \gets \pi(\shr)$
\Else
\State $\pi_\ell \gets \pi(\shell)$
\EndIf
\EndIf
\If{$\pi_\ell < u$ and $\pi_r=$\texttt{null}}
\State $\pi_r \gets \pi(\shr)$
\EndIf
\If{$\pi_\ell=$\texttt{null} and $\pi_r < u$}
\State $\pi_\ell \gets \pi(\shell)$
\EndIf
\If{$\pi_\ell < u$ and $\pi_r < u$}
\State \Return $\shell, \shr, \pi_\ell, \pi_r, \scC$
\EndIf
\If{$Z_i = 1$} 
\LineComment Leftward expansion. Cache the previous density value (or null) and expand the window
\State \texttt{append}$(\scC, \pi_\ell)$
\State $\shell \gets \shell - (\shr-\shell)$
\State $\pi_\ell \gets$\texttt{null}
\Else
\LineComment Rightward expansion. Cache the previous density value (or null) and expand the window
\State \texttt{append}$(\scC, \pi_r)$
\State $\shr \gets \shr + (\shr-\shell)$
\State $\pi_r \gets$\texttt{null}
\EndIf
\State $i\gets i+1$
\EndWhile
\ealgc
\ealg

\balg[h]
\caption{\texttt{LazyAccept}: lazy cached accept algorithm to check proposal validity}\label{alg:lazyaccept}
\balgc[1]
\Require slice variable $u$, proposal $y$, approximate slice $\shell, \shr$, density values $\pi_\ell,\pi_r$, cache $\scC$, window size $w > 0$
\If{\texttt{SteppingOut}} 
\State \Return\texttt{true}
\EndIf
\State $i \gets $\texttt{length}$(\scC)$
\While{$i>0$}
\State $m\gets (\shr+\shell)/2$
\State $\pi' \gets \scC[i]$
\State $i\gets i-1$
\LineComment{Move the appropriate boundary and update the density value using the cache if possible}
\If{$y < m$}
\State $\shr \gets m$
\If{$\shell\leq 0 \leq \shr$}
\State $\pi_r \gets \pi'$
\Else
\State $\pi_r \gets$\texttt{null}
\EndIf
\Else
\State $\shell \gets m$
\If{$\shell\leq 0 \leq \shr$}
\State $\pi_\ell \gets \pi'$
\Else
\State $\pi_\ell \gets$\texttt{null}
\EndIf
\EndIf
\LineComment{Try to skip evaluations if one edge is known to be in the slice} 
\LineComment{Note: in the below checks, $a>b$ is assumed to return \texttt{false} if either $a$ or $b$ are \texttt{null}}
\If{$\pi_\ell \geq u$ or $\pi_r \geq u$}
\Continue
\EndIf
\If{$\pi_\ell=$\texttt{null} and $\pi_r=$\texttt{null}}
\LineComment{Evaluate left edge first if the right boundary moved and vice versa}
\If{$y<m$}
\State $\pi_\ell \gets \pi(\shell)$
\Else
\State $\pi_r \gets \pi(\shr)$
\EndIf
\EndIf
\If{$u>\pi_\ell$ and $\pi_r=$\texttt{null}}
\State $\pi_r \gets \pi(\shr)$
\EndIf
\If{$u>\pi_r$ and $\pi_\ell=$\texttt{null}}
\State $\pi_\ell\gets \pi(\shell)$
\EndIf
\If{$u>\pi_r$ and $u>\pi_\ell$}
\State \Return\texttt{false}
\EndIf
\EndWhile
\State \Return \texttt{true}
\ealgc
\ealg